\documentclass[12pt]{iopart}
\usepackage{iopams}

\expandafter\let\csname equation*\endcsname\relax

\expandafter\let\csname endequation*\endcsname\relax

\usepackage{amsmath}
\usepackage{bm}
\usepackage{graphicx}
\usepackage{xcolor}
\usepackage{cite}

\usepackage{tikz}
\usetikzlibrary{arrows}
\usetikzlibrary{decorations, decorations.text,backgrounds}

\usepackage[utf8]{inputenc} 
\usepackage[colorlinks=true,linkcolor=blue,citecolor=blue,urlcolor=blue]{hyperref}
\usepackage[all]{hypcap} 

\newcommand{\dd}[0]{d} 
\newcommand{\ee}[0]{\mathrm{e}}

\newcommand{\hyp}{\mathrm{_{2}F_{1}}}

\newcommand{\FDJ}[4]{~{_{2\,}\! \mathrm{F} \! _{\,1}} \left(#1, #2; #3; #4  \right)}

\newcommand{\etaC}[0]{\eta_{C}} 
\newcommand{\Th}[0]{T_{h}} 
\newcommand{\Tc}[0]{T_{c}}

\newcommand{\wout}[0]{w_{out}} 
\newcommand{\qin}[0]{q_{in}} 
 
\begin{document} 
\title[Fluctuations in heat engines]{Fluctuations in heat engines}

\author{Viktor Holubec$^1$ and Artem Ryabov$^2$}

\address{Charles University, 
Faculty of Mathematics and Physics, 
Department of Macromolecular Physics, 
V~Hole\v{s}ovi\v{c}k\'{a}ch~2, 
CZ-180~00~Praha, Czech Republic}

\eads{$^1$ \mailto{viktor.holubec@mff.cuni.cz}, $^2$ \mailto{artem.ryabov@mff.cuni.cz}}
\vspace{10pt}
\begin{indented}
\item[] \today 
\end{indented}

\begin{abstract} \textcolor{black}{At the dawn of thermodynamics, Carnot's constraint on efficiency of heat engines stimulated the formulation of one of the most universal physical principles, the second law of thermodynamics. In recent years, the field of heat engines acquired a new twist due to enormous efforts to develop and describe microscopic machines based on systems as small as single atoms. At microscales, fluctuations are an inherent part of dynamics and thermodynamic variables such as work and heat fluctuate. Novel probabilistic formulations of the second law imply general symmetries and limitations for the fluctuating output power and efficiency of the small heat engines.} Will their complete understanding ignite a similar revolution as the discovery of the second law? Here, we review the known general results concerning fluctuations in the performance of small heat engines. To make the discussion more transparent, we illustrate the main abstract findings on exactly solvable models and provide a thorough theoretical introduction for newcomers to the field.
\end{abstract}
%
%
\submitto{\JPA}
%
\maketitle
%

\section{Introduction} 
Noises and fluctuations are often regarded as negative effects, which devaluate signals and make our lives more unpredictable.  
However, they are at the very heart of thermodynamics~\cite{Carnot:1824, Clausius:1850, Clausius:1854, Thomson:1852, Kubo:book1968,callen1998thermodynamics}. Indeed, the fundamental notions of heat and entropy are closely related to thermal fluctuations. Only due to these fluctuations, we are able to use heat engines to transform the easily accessible thermal energy into the useful mechanical work, and thus to propel our industrial society. 

During the last decades, we have witnessed an enormous effort to explore energy transformations in living organisms, miniaturize electronic circuits necessary for development of computers, and advance experimental micromanipulation techniques. With  the  current  experimental  resolution,  one  can  observe how thermal fluctuations transfer working medium of a heat engine between its microstates. A periodic driving of energies of the microstates and temperature of the environment then allows us to systematically convert energy of the thermal fluctuations into mechanical work. We call such operating  machines as microscopic {\em cyclic heat engines}. Another option is to couple the working medium to several thermal environments at different constant temperatures and to convert part of the resulting heat flux into work. Such microscopic machines operate autonomously under stationary conditions, and we refer to them as {\em steady-state heat engines}. 

If one can neglect fluctuations, e.g.\ at macroscopic scales, fundamental limitations on performance of both these types of heat engines are identical and dictated by the second law of classical thermodynamics~\cite{callen1998thermodynamics}. However, at the microscale, the fluctuating output work in cyclic and steady-state heat engines become fundamentally different stochastic processes subjected to distinct restrictions~\cite{Holubec/Ryabov:2018}. Mathematical description of these processes and intuitive understanding of their physical behavior are the main aims of this review. We will address them using exact solutions to specific models and generalizations of the second law, which take into account fluctuations.

The consistent theoretical framework describing thermodynamics of small systems is known as the stochastic thermodynamics. Its development up to 2012 is comprehensibly reviewed in Ref.~\cite{Seifert:2012}. Concerning microscopic machines, this early period of development was mainly devoted to generalizations of classical problems: With a few exceptions \cite{Chvosta/etal:2010a}, performance of stochastic heat engines was studied on the level of mean values with a particular focus on the efficiency at maximum power~\cite{Schmiedl2007,CalvoHernandez/etal:2015,Tu:2021}. Only recently, a systematic attention has been devoted to fluctuations in performance of small heat engines. Our main goal here is to present a coherent and intuitive overview of main achievements in this rapidly growing field.

Concretely, Sec.~\ref{sec:dynamics} of this review presents mathematical tools necessary for analysis of dynamics of small systems under the broadly used Markovian description. Section~\ref{sec:TD} contains definitions of basic thermodynamic quantities for cyclic and steady-state heat engines and provides further mathematical tools for studying their fluctuations. These two sections introduce basic concepts of stochastic thermodynamics, which are employed in the rest of the review. 
A reader familiar with these basic notions can jump directly to Sec.~\ref{sec:general_results}, where we review the known general results on fluctuating thermodynamic performance of classical heat engines (with a few quantum side remarks). Namely, Sec.~\ref{sec:propertiesWPD} describes general properties of probability density functions (PDFs) for stochastic work and heat. Section~\ref{sec:fluct_teor} discusses generalizations of the Crooks and the Jarzynski fluctuation theorems for heat engines. In Sec.~\ref{sec:FluctEff}, we discuss universal properties of PDF for stochastic efficiency. And Sec.~\ref{sec:TUR} is devoted to implications of thermodynamic uncertainty relations for the performance of heat engines. In particular, we focus on the imposed bounds on power and its fluctuations close to the reversible efficiency.

All these concepts and results are illustrated on specific models which are to a great extent exactly solvable. These models and the corresponding solutions are reviewed in Sec.~\ref{sec:Examples}. We believe that such a collection of exactly solvable models can help the readers to gain a better intuitive understanding of how to apply the general results in specific situations.

This review is conceptually self contained. However, for readers willing to explore other achievements of stochastic thermodynamics or to get a deeper understanding of its theoretical tools, we can recommend a number of excellent reviews focusing on various major topics in this filed. First, the book~\cite{Sekimoto:2010} contains a pedagogical introduction into stochastic energetics defining heat and work for individual stochastic trajectories. It also covers several models of microscopic machines not discussed here like the Feynman and the B{\" u}ttiker-Landauer ratchets, and information engines. Then, there are several comprehensible reviews of fluctuation theorems (FTs) explaining their properties from different perspectives~\cite{Evans/Searles:2002, Harris/Schuetz:2007, Kurchan:2007, Jarzynski:2011, Bochkov/Kuzovlev:2013}, review articles focusing on thermodynamic uncertainty relations~\cite{Marsland/England:2018, Seifert:2019, Horowitz/Gingrich:2020}, stochastic thermodynamics of information~\cite{Parrondo/etal:2015, Lutz/Ciliberto:2015}, and experimental applications of the theory~\cite{Bustamante/etal:2005, Ritort:2006, Ciliberto/etal:2013, Ciliberto:2017}. Properties of steady-state heat engines (with the main focus on quantum systems) are discussed in Ref.~\cite{Benenti/etal:2017}. We are aware of the fact that this list is highly subjective and incomplete. We have selected its entries based on their relation to the fluctuations in heat engines. 

In the whole review, we will measure the energy in units of temperature and thus we set $k_B = 1$. To avoid introducing too many symbols, we will often distinguish between different functions by their variables. For example $L$ will denote matrix of transition rates and $L(s_{in})$ will be the corresponding tilted rate matrix. We will also omit writing independent variables if their values will be clear from the context.

\section{Dynamics}
\label{sec:dynamics}

In this section, we review dynamic equations describing stochastic time evolution of small systems. If not specified otherwise, we assume the classical Markovian dynamics~\cite{Gillespie:1992,vanKampen2007}, which is in general described by the Generalized master equation (GME)~\footnote{Equations of the form~\eref{eq:GME} can also describe some non-Markovian systems, in which case 
$\mathcal{L} p(m,t)$ depends on the history of the function $p(m,t)$, rendering the dynamic equation non-local in time. For quantum systems, the function $p(m,t)$ in general describes not only PDF for $m$ but also coherences in the system~\cite{breuer2002theory}, which are neglected in our description.}
\begin{equation}
\frac{\partial }{\partial t} p(m,t) 
=  \mathcal{L}(t)  p(m,t) 
\label{eq:GME}
\end{equation}
for the probability density $p(m,t)$ to find the system at microstate $m$ at time $t$. The time-dependence in the operator $\mathcal{L}(t)$ comes from time dependence of temperature and driving protocol necessary to describe cyclically operating heat engines.

\subsection{Discrete state space}
\label{sec:discrete_Dyn}

For systems with a discrete state space, such as models of molecular machines or even incoherent quantum models, the microstate label $m$ assumes a (typically finite) number of discrete values, say $m = 1,\dots,N$. The dynamic equation can then be written in the matrix form 
\begin{equation}
 \frac{\dd }{\dd t} {\mathbf p}(t) = L(t) {\mathbf p} (t) 
 \label{eq:discrete}
\end{equation}
for the vector ${\mathbf p}(t) = [p_1(t),\dots,p_N(t)]^\intercal$ of occupation probabilities $p_m(t)$ of the individual microstates. The non-diagonal elements $L_{m n}(t)$ of the transition rate matrix $L(t)$ describe probabilities (per unit time) for transitions from state $n$ to state $m$. The diagonal element $L_{n n}(t)$ is given by the sum over all transitions from the state~$n$: $L_{nn}(t) = -\sum_{m\neq n} L_{m n}(t)$\textcolor{black}{, which secures that the normalization,  $\sum_{m=1}^N p_m(t)$, is constant over time}.

To ensure thermodynamic consistency of the model, the transition rates must obey the so-called (local) detailed balance condition~\cite{Maes2003} 
\begin{equation}
\label{eq:detailed_balance}
\frac{ L_{mn}(t)}{L_{nm}(t)} 
= \ee^{- \beta(t) [U(m,t)-U(n,t) ]},
\end{equation}
where $\beta(t) = 1/T(t)$ \textcolor{black}{is the inverse temperature of the environment} and $U(m,t)$ denotes energy of microstate $m$ at time $t$. The condition ensures that the probabilities $p_m(t)$ are described by the Boltzmann equilibrium distribution
\begin{equation}
p_m(t) =  \exp[- \beta(t) U(m,t)]/Z(t), 
  \label{eq:Boltzmann}
\end{equation}
if the driving is much slower than all system's relaxation processes and if the system is connected to a single heat bath at a time. The normalization  $Z(t)$ is given by $Z(t)=\sum_m \exp[- \beta(t) U(m,t)]$.

In case of time-independent transition rates, the solution to the Master equation~ \eref{eq:discrete} is given by the matrix exponential 
\begin{equation}
   {\mathbf p}(t) = \exp[(t-t') L ] {\mathbf p}(t'). 
   \label{eq:MatrixExp1}
\end{equation}
For problems with time-dependent rates, the solution instead involves the so-called time-ordered matrix exponential $\exp_{\rightarrow}[\int_{t'}^t dt'' \, L(t'')]$. While its exact evaluation is difficult, it can be always calculated approximately, as described in Sec.~\ref{sec:2level_PC} on a specific example.

The matrix exponential yields complete information about dynamics of the system. Moreover, when using the method of moment generating functions described in Sec.~\ref{sec:static_HE_TD}, it can also yield information about functionals of the stochastic process such as heat and work. Alternatively, the complete information about the dynamics and energetics of Markov processes with discrete state space can be obtained using suitable simulation algorithms~\cite{Gillespie:1992,Holubec2011}.

\subsection{Continuous state space} 
\label{sec:continuous_Dyn}

For systems where $m$ labels a continuum of microstates, the GME~\eref{eq:GME} is a partial differential equation. For example, for an underdamped Brownian particle, the microstate $m$ specifies position $\bf x$ and momentum $\bf p$ of the particle and the operator $\mathcal{L}$ assumes the form 
\begin{equation}
\mathcal{L}(t)
= \nabla_{\bf p} \cdot \left[\gamma^2 D(t)  \nabla_{\bf p} + \gamma {\bf p} + \nabla_{\bf x} U({\bf x},t)   \right]
- \nabla_{\bf x} {\bf p}/m_g,
\label{eq:underdamped}
\end{equation}
where $m_g$ is the particle mass and $U({\bf x},t)$ is potential. The nabla operators $\nabla_{\bf x}$ and $\nabla_{\bf p}$ denote gradients with respect to position and momentum, respectively. The thermodynamic consistency~\eref{eq:detailed_balance} in this case requires that the friction coefficient, $\gamma$, diffusion coefficient, $D$, and temperature $T$ are interconnected by the fluctuation-dissipation theorem \begin{equation}
D(t) = T(t)/\gamma.
\label{eq:FDT}
\end{equation}
The underdamped dynamics describes particles in diluted gasses so that the relaxation time of their momentum is comparable or larger than that of the position~\cite{vanKampen2007, Risken1996}.

For particles immersed in environments with a larger friction, such as water, the underdamped description can be significantly simplified by assuming that momentum degrees of freedom are at all times in equilibrium. Then the microstate $m$ is specified by the system position $\bf x$ and the operator $\mathcal{L}$ reads
\begin{equation}
\mathcal{L}(t)
= \nabla \cdot \left[D(t) \nabla  + \frac{1}{\gamma} \nabla U({\bf x},t)    \right],
\label{eq:overdamped}
\end{equation}
where $\nabla \equiv \nabla_{\bf x}$. This case corresponds to the most frequently used description of the so-called Brownian heat engines with the working medium composed of Brownian particles immersed in water, as witnessed by the available experimental~\cite{Blickle2012,Martinez2016,Martinez2017} and theoretical~\cite{Schmiedl2007,Speck:2011,Holubec2014,Holubec2015,Ryabov/etal:2013} results. Note that these systems can also be driven by non-conservative forces. In such case, the potential force $- \nabla_{\bf x} U({\bf x},t)$ in Eqs.~\eref{eq:underdamped} and \eref{eq:overdamped} has to be replaced by the total force ${\bf F}({\bf x}, t)$ containing also the non-potential component.

If the operator $\mathcal L$ assumes either of the two forms \eref{eq:underdamped} and \eref{eq:overdamped}, the GME~\eref{eq:GME} is
called the Fokker-Planck equation. Even for time-independent driving, this equation can be solved exactly in few special cases only ~\cite{Risken1996}. The most notable case is the Brownian motion in a parabolic potential~\cite{Schmiedl2007,Speck:2011,Kwon/etal:2013,Ryabov/etal:2013}, which yields forces linear in the position variable. In the overdamped regime~\eref{eq:overdamped}, we discuss this solution in detail in Sec.~\ref{sec:harmonic}. 

Besides various approximate techniques~\cite{Risken1996},
the universally applicable approach to problems where exact solutions are out of reach is provided by so called Brownian dynamics simulations~\cite{Gillespie:1992,Risken1996}. They are based on the stochastic differential \textcolor{black}{(Langevin)} equations
\begin{eqnarray}
\dot{\mathbf x}(t) &=& \frac{{\mathbf p}(t)}{m_g},
\label{eq:LExu}\\
\dot{\mathbf p}(t) &=& - \gamma \frac{{\mathbf p}(t)}{m_g} - \nabla_{\bf x} U({\bf x},t) + \sqrt{2D(t)\gamma^2}\, {\bm \eta}(t),
\label{eq:LEvu}
\end{eqnarray}
and 
\begin{equation}
 \dot{\mathbf x}(t) = - \frac{1}{\gamma}\nabla_{\bf x} U({\bf x},t) + \sqrt{2D(t)}\, {\bm \eta}(t), 
 \label{eq:LExo}
\end{equation}
corresponding to the underdamped and overdamped cases, respectively. Above, the vector ${\bm \eta}(t)$ is the Gaussian white noise fulfilling $\left< {\bm \eta}(t)\right> = {\bm 0}$ and $\left< \eta_i(t)\eta_j(t')\right> =\delta_{ij} \delta(t-t')$. Numerical integration of these equations yields trajectories of the stochastic process \textcolor{black}{$\{\mathbf{x}(t), \mathbf{p}(t) \}$, which bare complete information about its dynamics and energetics.}

Furthermore, for overdamped systems, the continuous state-space can be discretized in a thermodynamically consistent way~\cite{Holubec/etal:2019}. Then the Fokker-Planck equation transforms into a matrix Master equation, which can be solved by numerical evaluation of the time-ordered matrix exponential. 

\subsection{Steady-state and cyclic heat engines}

Traditionally, heat engines are cyclically operating machines transforming energy accepted from a thermal environment (heat) into mechanical work. Examples of such \emph{cyclic heat engines} can be found in cars and other devices with combustion engines. Another way to transform heat to work is to connect the working medium simultaneously to two reservoirs at different temperatures and transform part of the resulting heat flux into work. \textcolor{black}{Turbines in thermal plants, solar cells, but also wind turbines operate this way (the wind is propelled by temperature gradients).} In the rest of this work, we will call these machines \emph{steady-state heat engines}. 

\textcolor{black}{On the level of average dynamics and thermodynamics, there exist formal mappings between the periodic and steady-state heat engines~\cite{Esposito/Parrondo:PRE2015, Raz/etal:PRX2016, Brandner/etal:PRX2015, Barato/Seifert:NJP2017}. However, from a practical perspective, these devices are fundamentally different.} Due to a higher level of required control, cyclic heat engines are usually harder to realize and maintain in practice than steady-state ones. But the higher level of control also has advantages since cyclic heat engines can fundamentally outperform their steady-state counterparts. Readers experienced in stochastic thermodynamics can see directly Sec.~\ref{sec:TUR} for more details. Beginners are advised to first go through the current section explaining details of regimes of operation of the two classes of heat engines.

\subsection{Steady-state heat engines}

\begin{figure}
\centerline{
     \begin{tikzpicture}[
      scale=0.5,
      level/.style={thick},
      transU/.style={thick,->,shorten >=2pt,shorten <=2pt,>=stealth},
      transD/.style={thick,<-,shorten >=2pt,shorten <=2pt,>=stealth},
    ]
     \draw[level,red] (0cm,-11em) -- (-2cm,-11em) node[left] {hot source $\left|h\right>$ at $T_h$};
    \draw[level] (2cm,-2em) -- (0cm,-2em) node[left] {quantum dot $\left|d\right>$};
    \draw[level,blue] (4cm,7em) -- (2cm,7em) node[left] {cold drain $\left|c\right>$ at $T_c$};
    \draw[transU,blue] (0.8cm,-2em) -- (2.8cm,7em) node[midway,left] {$L_{cd}$};
    \draw[transU,red] (-1.2cm,-11em) -- (0.8cm,-2em) node[midway,left] {$L_{dh}$};
    \draw[transD,blue] (1.2cm,-2em) -- (3.2cm,7em) node[midway,right] {$L_{dc}$};
    \draw[transD,red] (-0.8cm,-11em) -- (1.2cm,-2em) node[midway,right] {$L_{hd}$};
    \draw[->] (7cm,-11em) -- (7cm,11em) node[right] {$U(m,t)$};
    \draw[dashed] (4cm,7em) -- (7.1cm,7em) node[right] {\color{blue}{$\mu_c$}};
    \draw[dashed] (2cm,-2em) -- (7.1cm,-2em) node[right] {0};
    \draw[dashed] (0cm,-11em) -- (7.1cm,-11em) node[right] {\color{red}{$\mu_h$}};
		 \draw[dashed]  (8.5cm,7em) -- (15.cm,7em);
    \draw[dashed] (8.5cm,-2em) -- (12.cm,-2em);
    \draw[dashed]  (9.0cm,-11em) -- (15.cm,-11em);
		\draw[<->,shorten >=2pt,shorten <=2pt,blue] (12.cm,-2em) -- (12.cm,7em) node[midway,right] {$-\dot{q}_{out}$};
		\draw[<->,shorten >=2pt,shorten <=2pt,red] (12.cm,-11em) -- (12.cm,-2em) node[midway,right] {$\dot{q}_{in}$};
		\draw[<->,shorten >=2pt,shorten <=2pt,black!60!green] (15.cm,-11em) -- (15.cm,7em) node[midway,right] {$\dot{w}_{out}$};
    \end{tikzpicture}
} 
\caption{The simplest model of a steady-state heat engine: A single-level quantum dot connected to two leads at different temperatures and chemical potentials. \textcolor{black}{The dynamics of this system is assumed to be a hopping process where electrons jump from the energy level $\left|i\right>$ to $\left|j\right>$ with the transition rate $L_{ji}$.} In the steady state, the engine utilizes the average heat flux $\dot{q}_{in}$ due to the temperature gradient $T_h-T_c$ to drive electrons against the difference $\mu_c-\mu_h$ of chemical potential, and generates power \textcolor{black}{$\dot{w}_{out} = \dot{q}_{in} - \dot{q}_{out}$.} The model can be interpreted as a simple microscopic realization of a thermoelectric device.}
\label{Fig:SSHE}
\end{figure}
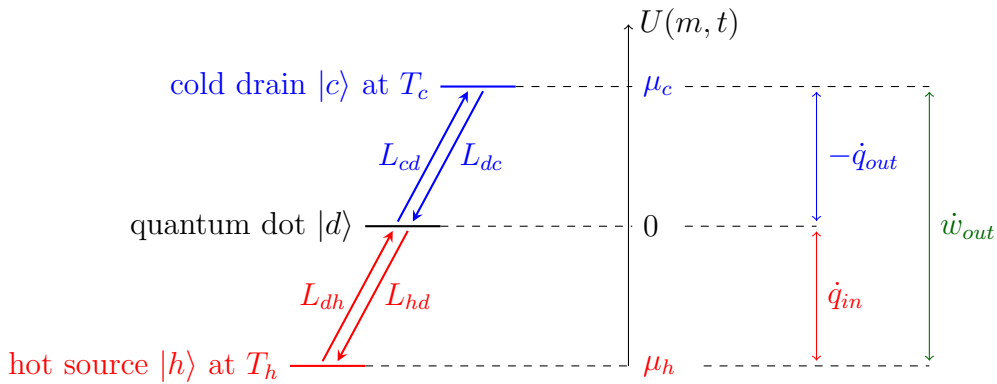

Steady-state heat engines require simultaneous coupling of their stochastic dynamics to at least two heat reservoirs at different temperatures. For discrete models, individual transitions can be induced by different reservoirs, as shown in Fig.~\ref{Fig:SSHE} for a simple model of a thermoelectric device. The different temperatures will then appear in the detailed balance condition~\eref{eq:detailed_balance} and hence in the hopping rates $L_{ji}$. 
In the continuous models, the system-reservoir couplings can be realized in various ways: Having a system of interacting particles, one can couple each of them to a distinct heat bath. More generally, one can connect individual degrees of freedom of a multidimensional system to different noise sources, which is a common situation in thermal ratchet models~\cite{Ryabov/etal:2016, Holubec/etal:2017}. Another option is to create a spatially-dependent temperature profile~\cite{Sekimoto:2010, Seifert:2012}.  



Due to the constant conditions, dynamics of steady-state engines eventually reaches a stationary regime where the microstate distribution $p(m)$ does not depend on time:
\begin{equation}
 p(m)  =   \lim_{t\to \infty} p(m,t). 
\end{equation}
Here, $p(m,t)$ is the solution~\eref{eq:MatrixExp1} to Eq.~\eref{eq:GME} for an arbitrary initial condition. The stationary distribution $p(m)$ can be also found as a time-independent solution of the GME~\eref{eq:GME}, 
\begin{equation}
\mathcal{L} p(m) =0,
\end{equation}
i.e., it is given by the eigenvector of the operator $\mathcal{L}$ to the eigenvalue zero.
In models presented in this review, we assume the dynamics to be ergodic and thus the stationary distribution is unique.

\subsection{Cyclic heat engines}

\begin{figure}
\centerline{
     \begin{tikzpicture}[
      level/.style={thick},
      transU/.style={thick,->,shorten >=2pt,shorten <=2pt,>=stealth},
      transD/.style={thick,<-,shorten >=2pt,shorten <=2pt,>=stealth},
    ]
    \node[red!50!blue] at (7.0,3.5) {$T(t)$};
    \node[red!50!blue] at (-0.5,3.5) {$T(t)$};
    \begin{scope}[scale=0.6]
    \draw[rounded corners,red!50!blue] (-2, -2) rectangle (9.5, 6.5) {};
    \draw[level] (2cm,-2em) -- (0cm,-2em) node[left] {$\left|d\right>$};
    \draw[level] (2cm,9em) -- (0cm,9em) node[left] {$\left|u\right>$};
    \draw[transU] (0.8cm,-2em) -- (0.8cm,9em) node[midway,left] {$L_{ud}(t)$};
    \draw[transD] (1.2cm,-2em) -- (1.2cm,9em) node[midway,right] {$L_{du}(t)$};
    \draw[->] (5cm,-2em) -- (5cm,13em) node[right] {$U(m,t)$};
    \draw[dashed] (2cm,9em) -- (5.1cm,9em) node[right] {$E_u(t)$};
    \draw[dashed] (2cm,-2em) -- (5.1cm,-2em) node[right] {0};
	\draw[dashed]  (7.2cm,9em) -- (8.cm,9em);
    \draw[dashed] (6.0cm,-2em) -- (8.cm,-2em);
		\draw[<->,shorten >=2pt,shorten <=2pt,red!50!blue] (8.cm,-2em) -- (8.cm,9em) node[midway,right,red!50!blue] {$q$};
	\draw[<->,shorten >=2pt,shorten <=2pt,thick,black!60!green] (5.cm,12em) -- (5.cm,6em) node[right] {$w_{out}$};
	\end{scope}
\begin{scope}[scale=1.5,xshift=7.5cm]
\draw[rounded corners,red!50!blue] (-3.3, -0.8) rectangle (1.5, 2.6) {};
\tikzset{shift={(-1.2,0)}}
\draw[bend left,<->,black!60!green]  (-1.8,1.2) to (-1,1.6);
\draw[bend left,<->,black!60!green]  (1,1.6) to (1.8,1.2) node[right,black!60!green] {$w_{out}$};
\draw[bend left,<->,red!50!blue]  (-0.56,0.3) to (0.44,0.3) node[above] {$q$};
\draw[very thick,fill={rgb:black,1;white,4}] (-0.06,-0.05) circle (0.24cm);
\tikzset{shift={(0,-0.55)}}
\draw[scale=1,dashed,domain=-1.2:1.2,smooth,variable=\x] plot ({\x},{1.4*\x*\x});
\draw[scale=1,domain=-1.4:1.4,smooth,variable=\x] plot ({\x},{1.0*\x*\x});
\draw[scale=1,dashed,domain=-1.55:1.55,smooth,variable=\x] plot ({\x},{0.8*\x*\x});
\node at (0.6,1.8) {$U(x,t)$};
\end{scope}
    \end{tikzpicture}
}
\caption{Paradigmatic examples of cyclic heat engines are the two-level cyclic heat engine (left) and the Brownian heat engine (right). \textcolor{black}{The dynamics of the two-level system is governed by the master equation~\eqref{eq:discrete} with jumps from the state $\left|i\right>$ to $\left|j\right>$ occurring with the rates $L_{ji}(t)$. The dynamics of the Brownian heat engine obeys the overdamped Fokker-Planck equation with the evolution operator~\eqref{eq:overdamped}. 
Cyclic heat engines are driven by a periodic variation of the heat bath temperature $T(t)$ and the system potential $U(m,t)$ or microstate energies $E_i(t)$. During parts of the cycle when temperature is large, the system is typically excited to higher values of the potential or higher energy levels. Subsequently, the engine performs work, $w_{out}$, by decreasing the energy of the excited microstates. 
}
}
\label{Fig:CyclicHE}
\end{figure}
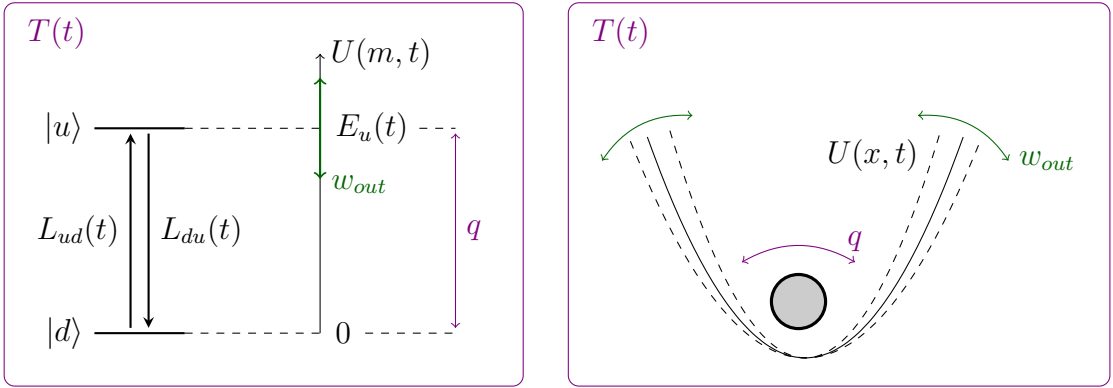

Figure~\ref{Fig:CyclicHE} shows two paradigmatic examples of cyclic heat engines: The two-state discrete  model with time-dependent state energies and the continuous model of a Brownian particle moving in the time-dependent parabolic potential. Both these models are to a large extent exactly solvable, i.e., most of their properties can be derived analytically. For more details, see Secs.~\ref{sec:2level} and~\ref{sec:harmonic}. 


The cyclic heat engines are driven by a $t_p$-periodic protocol $\pi(t) = \left[ U(m,t), T(t)\right]$, $\pi(t)=\pi(t+t_p)$. Due to the periodicity of the driving, the engine microstate distribution $p(m,t)$ also becomes a $t_p$-periodic function in the long-time limit. This limit, where $p(m,t) = p(m,t+t_p)$ holds, is frequently referred to as the \emph{time-periodic steady state} or  the \emph{limit cycle}. In the following, we assume that the cyclic heat engines operate in this limit cycle.

The distribution $p(m,t)$ within the limit cycle can be calculated by two complementary approaches. Both of them are conveniently formulated using the Green function $G(m,t|m',t')$ of the dynamic equation~\eref{eq:GME}, i.e.\ its solution satisfying $G(m,t'|m',t') = \delta(m-m')$, with $\delta(.)$ being the Dirac delta function for continuous systems and the Kronecker delta for discrete ones. The Green function thus describes the time evolution of a system that departed with certainty from the microstate $m'$ at time $t'$. Heaving the Green function, we can evolve in time any initial distribution $p(m',t')$. Within the limit cycle, this distribution will not change after one period. Hence we can derive the limit-cycle distribution by solving the so called Chapman-Kolmogorov equation
\begin{equation} 
p(m,t+t_p) =  \int \dd m\,' G(m,t+t_p|m',t) p(m',t). 
\label{eq:PGtp}
\end{equation}

In the discrete case, the Chapman-Kolmogorov equation \eqref{eq:PGtp} assumes the vector form 
\begin{equation}
    {\mathbf p}(t + t_p) = G(t+t_p,t) {\mathbf p}(t), 
    \label{eq:limitCyclePDF}
\end{equation}
where the element $mn$ of the matrix $G(t,t')$ is equal to $G(m,t|n,t') \equiv G_{mn}(t,t')$. The state of the system ${\mathbf p}(t)$ during the limit cycle is thus the eigenvector of the Green function corresponding to the eigenvalue 1. This insight can be used to solve the equation~\eqref{eq:limitCyclePDF} numerically. If we introduce a suitable discretization of the state variable $m$, the same method can also be applied to solve Eq.~\eqref{eq:PGtp} for the continuous state space~\cite{Holubec/etal:2019}.


As the second alternative approach yielding the limit-cycle distribution, one can perform the limit 
\begin{equation}
    p(m,t) =  \lim_{t'\to -\infty} G(m,t|m',t'),
    \label{eq:limitCyclePDF2}
\end{equation}
which is nothing but the long-time limit of a solution for an arbitrary initial condition. 


\section{Thermodynamics}
\label{sec:TD}

{\color{black} 
The dynamics of small systems is inherently stochastic and so is the thermodynamic performance of machines based on these systems. To identify work and heat for a given trajectory $\{ m(t)\}$, a convenient starting point is an expression for the change of system's potential energy, $\dd  U(m(t),t)$, during the infinitesimal time interval $[t,t+dt]$. For overdamped systems and Markov jump processes, this change can be interpreted as the first law of thermodynamics (conservation of energy) and expressed in the form~\cite{Sekimoto:2010,Seifert:2012}
\begin{equation} 
\dd u(t) \equiv \dd  U(m(t),t) = 
\delta w(t) + \delta q(t) =  
\left. \left[ \frac{\partial U(m,t)}{\partial t} \dd t + \frac{\partial U(m,t)}{\partial m} \dd m(t) \right] \right|_{m=m(t)} .
\label{eq:1st_law_overdamped}
\end{equation}
Here, $\dd u$ denotes the increase in the system internal energy during the infinitesimal time interval of duration $dt$, and $\delta w$ and $\delta q$ are the corresponding amounts of work and heat transferred \emph{into} the system. In the underdamped case, $\dd u(t) = d H(m(t),t)$ holds, where the system Hamiltonian $H(m,t) =K(m,t)+ U(m,t)$ includes also the kinetic energy $K(m,t)$. 

Which part of $\dd  U(m(t),t)$ [or $dH(m(t),t)$] is associated with $\delta w$ and $\delta q$ depends on the situation at hand. The definitions differ significantly for cyclic and steady-state heat engines. Let us first discuss the cyclic ones.
}

\subsection{Thermodynamics of cyclic heat engines}
\subsubsection{Work, heat, and efficiency.} 
As illustrated in Fig.~\ref{Fig:CyclicHE}, microscopic cyclic heat engines are driven by altering the potential $U(m,t)$ in a similar way as the macroscopic combustion engines are driven by a piston. Accordingly, the variation of $U(m,t)$ represents the experimental tool for extracting work from the system. 

{\color{black} This intuitive picture is in accord with the formal definition of the output work used in stochastic thermodynamics for cyclic heat engines based on overdamped Brownian systems or on Markov systems with discrete energy levels. For them, $\delta w$ is associated with the first term on the right-hand side of~\eqref{eq:1st_law_overdamped}, $\delta w = (\partial_t U) dt$, and the remaining part of $dU$ is attributed to heat: $\delta q =(\partial_m U) dm$~\cite{Sekimoto:2010,Seifert:2012}. For cyclic heat engines based on underdamped Brownian systems, one should substitute the potential in these definitions by the full Hamiltonian, $H(m,t) =K(m,t)+ U(m,t)$, including the kinetic energy $K(m,t)$.
Such defined work obeys the Jarzynski equality and the Crooks fluctuation theorem~\cite{JarzynskiCompariosn2007}. These definitions also agree with notions of work and heat from the classical statistical thermodynamics~\cite{Reif:1965}. 

However, this work differs from the one introduced in textbooks on mechanics~\cite{Goldstein/etal:2002} as ``work = force $\times$ displacement'. Actually, both these work definitions are reasonable and valid for different situations\cite{JarzynskiCompariosn2007, HorowitzJarzynskiJSTAT, VilarRubi2008, Peliti2008JSTAT, HorowitzJarzynski2008, ZimanyiSilbey2009}. Specifically, the work defined as $\delta w = (\partial_t U) dt$, which we will use in the rest of this review, corresponds to the situation when the time dependent driving changes the immediate internal energy of the system. On the other hand, the mechanistic definition of work suits to the situation when the time dependent part of the Hamiltonian is not understood as part of the internal energy but rather as a source of an external force.

In the quantum regime, the problem of defining the fluctuating work and heat is even less settled. A key role here is played by the measurement procedure and by the coherences (offdiagonal elements of the density matrix operator). If coherences do not emerge during the dynamics, the so-called two-point measurement scheme~\cite{Talkner/etal:PRE2007, Campisi/etal:RMP2011} yields the stochastic work, which is consistent with both the first law of thermodynamics and the quantum fluctuation theorems. For a general recent analysis of various definitions of the fluctuating work, see Refs.~\cite{Perarnau-Llobet/etal:PRL2017, Hovhannisyan/Imparato:arxiv2021} and references therein. 

Returning back to the Markovian cyclic heat engines subjected to the first law in the form~\eqref{eq:1st_law_overdamped},} the overall work per cycle is given by the stochastic integral
\begin{equation} 
w_{out}  = -\int_0^{t_p}\dd t \left.\frac{\partial U(m,t)}{\partial t}\right|_{m=m(t)} .
\label{eq:work_cyclic}
\end{equation}
This expression is most easily understood for discrete systems such as the two-level system in Fig.~\ref{Fig:CyclicHE}. If the system resides in microstate $m$ at time $t$, the work flux (power) 
\begin{equation}
\dot{w}(t) = \frac{\delta{w}(t)}{\dd t} = \left.\frac{\partial U(m,t)}{\partial t}\right|_{m=m(t)} 
\label{eq:dw_cyclic}
\end{equation}
flows \emph{into}/\emph{from} the system if the energy of microstate $m$ at time $t$, $U(m,t)$, increases/decreases. A further intuition can be gained by comparing the energy levels to occupied/empty massless ``elevators''. \textcolor{black}{If an elevator is occupied by a heavy passenger and goes down in a gravity field, it can lift a weight connected to it by a pulley, and thus produce work. An empty elevator can produce no work. To lift an occupied elevator we are required to supply work.}

Heat flux into the system is given by the second contribution in Eq.~\eref{eq:1st_law_overdamped}:
\begin{equation}
\dot q(t) = \frac{\delta{q}(t)}{\dd t} = \left.\frac{\partial U(m,t)}{\partial m}\right|_{m=m(t)} \dot{m}(t).
\label{eq:dq_cyclic}
\end{equation}
The heat flows into the system whenever it changes its microstate and thus it is closely related to the probability current. Using again the elevator analogy, the heat can be understood as energy associated with jumping between elevators (microstates) with different \textcolor{black}{heights and thus potential energies}. Jumping to a level with a larger energy requires heat from the bath. The energy gained by falling onto lower energies is transferred into the bath.

For assessing thermodynamic efficiency of cyclic heat engines, it is necessary to identify which parts of the heat contribute to the input, $q_{in}$, and output, $q_{out}$, heats. While it always depends on skills of experimentalists which part of the heat they can use as a resource~\cite{Holubec2020,Gronchi2021}, reasonable definitions are~\footnote{{\color{black} Different definitions of input and output heat apply, e.g., in heat engines with heat regenerators, such as the Stirling engine. These regenerators allow reusing part of the heat flowing out of the system and thus significantly enhance the engine efficiency. The corresponding input heat is then typically larger than~\eqref{eq:qin_cyclic}, and output heat is smaller than~\eqref{eq:qout_cyclic}.}}
\begin{eqnarray}
q_{in} &=& \int_0^{t_p} dt\, \dot{q}(t) \theta[\dot{Q}(t)],
\label{eq:qin_cyclic}\\
q_{out} &=& - \int_0^{t_p} dt\, \dot{q}(t) \theta[-\dot{Q}(t)].
\label{eq:qout_cyclic}
\end{eqnarray}
Here,
\begin{equation}
\dot{Q}(t) \equiv \left< \dot{q}(t)\right> 
= \int dm\, U(m,t) \dot{p}(m,t)
\end{equation}
is the average heat flux into the system at time $t$. Similarly as other average thermodynamic quantities for cyclic heat engines, it can be calculated based on the limit cycle PDF~\eref{eq:limitCyclePDF2}. 
The Heaviside theta functions $\theta(.)$ in Eqs.~\eqref{eq:qin_cyclic} and~\eqref{eq:qout_cyclic} ensure that if the heat on average flows into the system, we count it into the total input heat and vice versa.



Having defined the stochastic work per cycle and input heat, it is natural to introduce the stochastic power $w_{out}/t_p$ and the stochastic efficiency 
\begin{equation} 
    \tilde{\eta} = \frac{w_{out}}{q_{in}}.
    \label{eq:effSt}
\end{equation}
{\color{black} 
However, such defined efficiency have in general rather peculiar properties. In particular, its moments usually diverge and thus the average stochastic efficiency $\left< \tilde{\eta} \right>$ is rarely equal to the traditional  efficiency 
\begin{equation}
    \eta = \frac{W_{out}}{Q_{in}} , 
    \label{eq:eff}
\end{equation}
defined as the ratio of average performed work $W_{out} = \left<w_{out}\right>$ to the average input heat $Q_{in} = \left<q_{in}\right>$. One exception are models which yield $w_{out} \propto q_{in}$, see Secs.~\ref{sec:W_twolevel} and \ref{sec:CHE}. For more details, see the discussion of Eq.~\eqref{eq:tails_rho_eta}. 
}

\textcolor{black}{So far, we defined the averages $W_{out}$ and $Q_{in}$ of stochastic work and heat $w_{out}$ and $q_{in}$ as ensemble averages, i.e., averages taken over many realizations of the process $m(t)$. However, in experiments and simulations, it is often advantageous to measure a single long trajectory composed of many engine cycles. Then the average work and heat are approximated by averaging $w_{out}$ and $q_{in}$ over the $n$ cycles forming the long trajectory: $W_{out} \approx -\frac{1}{n}\int_0^{n t_p} \dot{w}(t) dt$ and $Q_{in} \approx \frac{1}{n}\int_0^{n t_p} \dot{q}(t) \theta[\dot{Q}(t)] dt$. The traditional efficiency~\eqref{eq:eff} is approximated by their ratio. On the other hand, the PDF for the stochastic efficiency~\eqref{eq:effSt} follows by calculating the ratios $w_{out}/q_{in}$ separately during the individual cycles. Notably, if the analysed trajectory is finite, such obtained approximation for $\eta$ is a random variable with similar properties as $\tilde{\eta}$, as discussed in Sec.~\ref{sec:FluctEff}.}
 
\subsubsection{Fluctuations of stochastic work and heat.}
\label{sec:WHC}
 To investigate fluctuations of work and heat in cyclic heat engines on analytical grounds, it is enough to consider the dynamic equation for the joint process $\{m(t), w(t)\}$, which reads~\cite{Imparato2005,Subrt/Chvosta:2007}
\begin{equation}
\frac{\partial }{\partial t} p(m,w,t) 
= \left[ \mathcal{L}(t)
- \frac{\partial U(m,t)}{\partial t} \frac{\partial }{\partial w} \right] p(m,w,t).
\label{eq:PDFworkCyclic}
\end{equation}
The second term on the right-hand side accounts for the work $\partial U(m,t)/\partial t$ done on the system per unit time while it resides in state $m$. For diffusive systems, Eq.~\eref{eq:PDFworkCyclic} is just the Fokker-Planck equation~\cite{Risken1996} corresponding to the stochastic process $\{m(t), w(t)\}$ described by the Langevin equations~\eref{eq:LExu}--\eref{eq:LExo} and \eref{eq:dw_cyclic}.

Both the PDF for stochastic power $w_{out}/t_p$ and the PDF for the stochastic efficiency $\tilde \eta$, Eq.~\eref{eq:effSt}, can be obtained from the solution $G(w_1,m_1,t_1|w_0,m_0,t_0)$ to Eq.~\eref{eq:PDFworkCyclic} for the initial condition $G(w_1,m_1,t_0|w_0,m_0,t_0) = \delta(w_1-w_0)\delta(m_1-m_0)$, describing the system that at time $t_0$ resides with certainty in the state $m_0$ and the work done on it is $w_0$. This Green function can be obtained explicitly only in very few situations most of which are reviewed in Ref.~\cite{holubec2014non}, also see Sec.~\ref{sec:Examples} for examples with direct applications to heat engines. Alternatively, the PDFs can be obtained by numerical techniques mentioned at the ends of Secs.~\ref{sec:discrete_Dyn} and \ref{sec:continuous_Dyn}.

Having the Green function, the PDF for the output work reads
\begin{equation}
\rho_{w_{out}} = \rho_{w_{out}}(t_p) = \int \dd m_1 \int \dd m_0\,
G(-w_{out},m_1,t_p|0,m_0,0)\, p(m_0,0),
\label{eq:PDFwoutC}
\end{equation}
where $p(m_0,0)$ is the state of the system within the limit cycle~\eref{eq:PGtp}.
To calculate the PDF for the stochastic efficiency, we need to first obtain the joint PDF $\rho(w_{out},q_{in})$ for $w_{out}$ and $q_{in}$. Assuming that the heat on average flows into the system from $t=0$ to $t=t_h$ during the cycle, the Markovian property of the dynamics can be used to obtain the PDF $\rho(w_{out},q_{in})$ from the Green function as
\begin{multline}
\rho(w_{out},q_{in}) = 
\int \dd w \int \dd m_2 \int \dd m_1 \int \dd m_0\,
\delta\left\{q_{in} - \left[U(m_1,t_h) - U(m_0,0) - w\right]\right\}\\
\times
G(-w_{out},m_2,t_p|w,m_1,t_h)\,
G(w,m_1,t_1|0,m_0,0)\,
p(m_0,0).
\label{eq:PDFwoutqinC}
\end{multline}
In words, this equation integrates over all trajectories which yield the heat $q_{in}$ during the initial part of the of the cycle of duration $t_h$ and provide the total output work $w_{out}$. The PDF $\rho(w_{out},q_{in})$ then gives the PDFs $\rho_{q_{in}}$ and $\rho_{\tilde{\eta}}$ for input heat and stochastic efficiency by the marginalizations
\begin{eqnarray}
\rho_{q_{in}} &=& \rho_{q_{in}}(t_p) = \int dw_{out}\, \rho(w_{out},q_{in}),\\
\rho_{\tilde{\eta}} &=& \int\int dw_{out}\, dq_{in}\, \delta\left(\tilde{\eta} - \frac{w_{out}}{q_{in}}\right) \rho(w_{out},q_{in}).
\label{eq:eff_PDF}
\end{eqnarray}
These PDFs yield all moments of stochastic thermodynamic variables describing the heat engine in question. The price for this complete information is the difficulty to determine them analytically. We leave a more detailed discussion of properties of the PDFs $\rho_{w_{out}}$, $\rho_{q_{in}}$, and $\rho_{\tilde{\eta}}$ to Secs.~\ref{sec:propertiesWPD}--\ref{sec:FluctEff}. In the rest of this section, we discuss an alternative method for evaluation of the second moment of output work based on the solution of the dynamic equation~\eref{eq:GME}. 

In fact, the Green function $G(m,t\,|\,m',t')$ of Eq.~\eref{eq:GME} and the limit-cycle probability density $p(m,t)$ render all time-correlation functions of the underlying Markov process $m(t)$ and hence also all moments of the thermodynamic variables involved~\cite{Holubec2014,holubec2014non}. For example, the two-time correlation function 
\begin{equation}
\left< h[m(t)]\,f[m(t')] \right>_{c} = \left\{ 
  \begin{array}{l l}
    \,\int\int dm\, dm'\,
h(m)\,f(m') G(m,t|m',t')p(m',t')\,\,, & \quad t\geq t'\\
    \,\int\int dm\,dm'\,
f(m)\,h(m') G(m,t'|m',t)p(m',t)\,\,, & \quad t\leq t'
  \end{array} \right.
\label{eq:time_cor_fun_continuous}
\end{equation}
yields the second moment of the random work~\eref{eq:dw_cyclic} done per cycle via the formula
\begin{equation}
\left<w_{\rm out}^2\right> = \int_{0}^{t_p}{d t'} \int_{0}^{t_p}d t''\, 
\left< \frac{{\partial}{U}[m(t'),t']}{{\partial}t'}\frac{{\partial}
{U}[m(t''),t'']}{{\partial}t''} \right>_{c}\,\,.
\label{eq:work_second_moment}
\end{equation}
Higher moments can be obtained using a similar procedure. The resulting equations, however, contain $2 n$ integrations, where $n$ denotes the degree of the moment, and thus quickly become practically useless. Let us now turn our attention to fluctuations of work and heat in steady-state heat engines.

\subsection{Heat and work in steady-state heat engines}
\label{sec:static_HE_TD}

Steady-state heat engines operate under time-independent conditions in contact with several heat baths at different temperatures and thus the work definition~\eref{eq:dw_cyclic} used for cyclic heat engines is not applicable as it gives 0.  This means that all energetic quantities defined for steady-state heat engines, including work, are of the form \eref{eq:dq_cyclic} related to the motion of particles/excitations in fixed energy landscapes $U(m)$.
 
 Basic definitions can be most easily understood using the specific example in Fig.~\ref{Fig:SSHE}, showing an engine based on a single-level quantum dot connected to two leads at different chemical potentials $\mu_i$ and temperatures $T_i$. The red transitions are assumed to be caused by the hot reservoir and thus the corresponding energy intake $q_{in} = -\mu_h > 0$ (in case the system jumps from the hot lead to the quantum dot) is identified as input heat. Similarly, the energy needed for jumps from the quantum dot to the cold lead, $\mu_c$, is identified as heat obtained from the cold bath. The corresponding output heat (positive if energy is transferred into the cold bath) is thus given by $q_{out} = -\mu_c$. The work $w_{out} = \mu_c - \mu_h = q_{in} - q_{out}$ is then done by the engine if an electron is transferred from the hot to the cold lead against the gradient of chemical potential.
 
Analytical calculation of PDFs for work and heat in this case can be based on so-called tilted evolution operators. While this technique can be also used for cyclic heat engines, the time-dependent evolution operators used therein require some extra care as sketched briefly in Sec.~\ref{sec:2level_PC}. Here, we explain the technique for the specific example of the quantum dot in Fig.~\ref{Fig:SSHE}. Its general exposition is given in Ref.~\cite{Holubec/etal:2019}.

Specifically, we show how to construct the joint PDF $p(w_{out},q_{in},t)$ for output work and input heat
\begin{equation}
    w_{out}(t) = \int_0^t dt' \dot{w}_{out}, \quad  q_{in}(t) = \int_0^t dt' \dot{q}_{in}
    \label{eq:integrated_work_heat}
\end{equation}
integrated over the time interval $(0,t)$ during the operation of the heat engine in the steady state. Probability that the system residing in microstate $m$ will during an infinitesimal time interval jump into microstate $n$ is given by the transition rate $L_{nm}$. Describing the state of the quantum dot in the steady state by the vector ${\bf p} = [p_0, p_d]^\intercal$ of probabilities that no/one electron occupies the quantum dot, the rate matrix in the GME~\eref{eq:GME} reads
\begin{equation}
L = \begin{pmatrix}
- L_{dh} - L_{dc} &  \phantom{-}L_{hd} + L_{cd}\\
\phantom{-}L_{dh} + L_{dc} & - L_{hd} - L_{cd}\\
\end{pmatrix}.
\label{eq:rateMatrixd1}
\end{equation}
As discussed above, the amount of heat $-\mu_h$ is transferred into the system if the electron jumps from $\left|h \right>$ to $\left|d\right>$ and vice versa. 
Further, $w_{out} = \mu_c$ when the electron jumps from $\left|d\right>$ to $\left|c\right>$ and $w_{out} =  \mu_h$ for jumps from $\left|d\right>$ to $\left|h\right>$. Probability that these jumps occur during an infinitesimal time interval \textcolor{black}{of duration $dt$} are described by elements of the matrix $\exp(dtL) \approx 1+dt L$, where $1$ denotes the unity matrix. Multiplying the transition rates by PDFs for the corresponding increments to $w_{out}$ and $q_{in}$, we find that the matrix $1+dtL(w_{out},q_{in})$, with
\begin{equation}
L(w_{out},q_{in}) = \begin{pmatrix}
- L_{dh} - L_{dc} & \delta(w_{out} - \mu_h) \delta(q_{in}+ \mu_h) L_{hd} + \delta(w_{out} - \mu_c)L_{cd}\\
\delta(q_{in} + \mu_h) L_{dh} + L_{dc} & - L_{hd} - L_{cd}\\
\end{pmatrix}
\label{eq:deltaMatrix}
\end{equation}
describes the Green function for work and heat transferred during the infinitesimal time interval. \textcolor{black}{Namely, the PDF that system found in microstate $n$ at time 0 and in microstate $m$ at time $dt$ performed work $w_{out}$ and accepted heat $q_{in}$ is given by the matrix element $[1+dtL(w_{out},q_{in})]_{mn}$.}

The unconditioned PDF that the system in steady state ${\bf p}$ performs during the time $dt$ work $w_{out}(dt)$ and accepts heat $q_{in}(dt)$ follows by the summation $p(w_{out},q_{in},dt) = \left<+\right|[1+dtL(w_{out},q_{in})] {\bf p}$. For discrete state spaces, the operator $\left<+\right|$ is just a row vector of ones, for continuous state spaces, it imposes integration over final microstates.  The Markovianity of the stochastic process implies that PDFs for longer integration times can be obtained by convolutions of the form $p(w_{out},q_{in},dt)  = \left<+\right|[1+dtL(w,q)]\star[1+dtL(w,q)](w_{out},q_{in}) {\bf p}$ over the work and heat variables. For long times, it is thus advantageous to employ the Laplace transform and instead of the PDF $p[w_{out},q_{in},t]$ calculate its moment generating function. Denoting the Laplace variables for $q_{in}$ and $w_{out}$ as $s_{in}$ and $s_w$, respectively, we find that $p(s_w,s_{in},t+dt) = \left<+\right|[1+dtL(s_w,s_{in})] p(s_w,s_{in},t)$ and thus
\begin{equation}
    \dot{{\bf p}}(s_w,s_{in},t) = L(s_w,s_{in}){\bf p}(s_w,s_{in},t),
    \label{eq:tiltedRM}
\end{equation}
where the so-called tilted rate matrix $L(s_w,s_{in})$ is Laplace transformed matrix~\eref{eq:deltaMatrix}:
\begin{equation}
L(s_w,s_{in}) =  \begin{pmatrix}
- L_{dh} - L_{dc} & \exp[(s_{in}-s_{w})\mu_h]L_{hd} +  \exp(-s_{w}\mu_c)L_{cd}\\
\exp(s_{in}\mu_h) L_{dh} + L_{dc} & - L_{hd} - L_{cd}\\
\end{pmatrix}.
\label{eq:TiltedrateMatrix}
\end{equation} 
The equation~\eref{eq:tiltedRM} can be again solved by the matrix exponential. Taking into account that at time 0 no work and heat were transferred and that the system operates in the time independent steady state, we get
\begin{equation}
 {\mathbf p}(s_w,s_{in},t) = \exp\left[L(s_w,s_{in}) t\right] {\mathbf p}.
\label{eq:mexp}
\end{equation}
The generating function for the input and output heat accepted up to time $t$ then follows by summing over final microstates:
\begin{equation}
\chi(s_w,s_{in},t) = \left<+\right|\exp\left[L(s_w,s_{in}) t\right] {\mathbf p}.
\label{eq:generatingFun}
\end{equation}
After the inverse Laplace transform, $\chi(s_w,s_{in})$ yields the joint PDF for $w_{out}(t)$ and $q_{in}(t)$. Alternatively, the generating function can be directly applied for evaluation of raw and central moments of these variables. Namely the raw moments of input heat read
\begin{equation}
\left<[q_{in}(t)]^n\right> =
(-1)^n \left.\frac{\partial^n}{\partial s_{in}^n}
\chi(s_{w},s_{in},t) \right|_{s_{in}=s_{w}=0}
\label{eq:raw_moments_qin}
\end{equation}
and similarly for $w_{out}(t)$.

At long times the PDF $\rho(w_{out},q_{in})$ can be approximated using the so-called large deviation theory~\cite{Touchette:2009,Holubec/etal:2019}. Specifically, this theory postulates that up to sub-exponential contributions the PDF is given by
\begin{equation}
\rho(w_{out},q_{in},t) \sim \exp\left[t I\left(\frac{w_{out}}{t},\frac{q_{in}}{t}\right)\right],
\label{eq:LDFqinqout}
\end{equation}
where $I(x,y)$ is the so-called large deviation or rate function. Let us now show how it follows from the moment generating function~\eref{eq:generatingFun}. Using Eq.~\eref{eq:LDFqinqout} in the definition of $\chi(s_w,s_{in},t)$, we find
\begin{align} \nonumber
    \log \chi(s_w,s_{in},t)
    & = \log \iint dw_{out}\, dq_{in}\,
    \exp(-s_{w}w_{out} - s_{in}q_{in})\rho(w_{out},q_{in},t) \\ \nonumber 
   & \sim \log \iint dw_{out}\, dq_{in}\,
    \exp\left\{t\left[I\left(\frac{w_{out}}{t},\frac{q_{in}}{t}\right) - s_w\frac{w_{out}}{t} - s_{in}\frac{q_{in}}{t}\right]\right\} \\
   & \approx t \max_{w_{out},q_{in}} \left[I\left(w_{out},q_{in}\right) - s_{w} w_{out} - s_{in}q_{in} \right].
    \label{eq:SCGF}
\end{align}
The last expression follows for large $t$ by the Laplace approximation of the integral. On the other hand, writing the matrix exponential in the moment generating function~\eref{eq:generatingFun} using eigenvalues $\lambda_i(s_w,s_{in})$ and the corresponding eigenvectors of the tilted rate matrix, we find
\begin{multline}
    \frac{1}{t} \log \chi(s_w,s_{in},t)
    = \frac{1}{t} \log \left\{ \left<+\right|\exp\left[L(s_w,s_{in}) t\right] {\mathbf p} \right\} = \\
    \frac{1}{t} \log \left\{ \sum_i c_i(s_w,s_{in}) \exp\left[t\lambda_i(s_w,s_{in})\right] \right\} \approx 
    \lambda_{max}(s_w,s_{in}).
    \label{eq:LDF_help}
\end{multline}
The coefficients $c_i$ follow from products of the eigenvectors and vectors ${\mathbf p}$ and $\left< +\right|$. In the last step, we assumed large time $t$ and approximated the sum by the term corresponding to the largest eigenvalue of the tilted matrix, $\lambda_{max}(s_w,s_{in})$.

Combining these two results and assuming differentiability of the generating function, we find using the Legendre-Fenchel transform of Eq.~\eref{eq:LDF_help} that
\begin{equation}
    I(w_{out},q_{in}) = \min_{s_w,s_{in}}
    \left[\lambda_{max}(s_w,s_{in}) + s_{w}w_{out} + s_{in}q_{in} \right].
    \label{eq:rateqqSS}
\end{equation}
The fluctuations of the integrated work and heat are thus for large times controlled by the largest eigenvalue of the tilted rate matrix.

\textcolor{black}{Let us now again consider the experimentally relevant situation of observing the stochastic steady-state heat engine for a long time. Similarly as for cyclic heat engines, the standard efficiency is then defined as the ratio $\eta = w_{out}(t)/q_{in}(t)$ of work and heat measured over the whole trajectory.  Unless the trajectory is infinite ($t\to \infty$), this $\eta$ is again a stochastic quantity with properties of the stochastic efficiency $\tilde{\eta}$. Since the work and heat are given by the integrated work and heat~\eqref{eq:integrated_work_heat} involved in the above calculation, the properties of such defined stochastic efficiency can be deduced from the generating function~\eqref{eq:rateqqSS} (see Sec.~\ref{sec:FluctEff}).}

This section concludes our presentation of general mathematical background needed for theoretical analysis of cyclic and steady-state heat engines. In the next section, we review known general results about the behavior of the above defined PDFs. In Sec.~\ref{sec:Examples}, we then present several exactly solvable models which demonstrate these findings.

\section{General results}
\label{sec:general_results}

Here, we sum up known general results concerning heat and work fluctuations in cyclic and steady-state heat engines. Sec.~\ref{sec:propertiesWPD} describes general properties of PDFs for work and heat in cyclic heat engines. Sec.~\ref{sec:fluct_teor} shows how the known detailed and integral fluctuation theorems apply to heat engines. Properties of stochastic efficiency $\tilde{\eta}$ are reviewed in Sec.~\ref{sec:FluctEff}. And the last Sec.~\ref{sec:TUR} of this chapter is devoted to application of thermodynamic uncertainty relations to heat engines, especially to their implications on trade-offs between power, power fluctuations, and efficiency close to reversible efficiency.

\subsection{Properties of work and heat PDFs in cyclic heat engines} 
\label{sec:propertiesWPD}

The PDFs for work and heat in cyclic heat engines bear several general properties~\cite{Subrt/Chvosta:2007,holubec2014non,Chvosta/etal:2010a}. First, for quasi-static cycles, the work PDF is given by  $\delta$ function located at the value of the average work~\cite{Speck2004,holubec2014non}. Noteworthy, the PDF for input heat can be non-trivial even in the quasi-static limit. To understand this, consider the two-level system in Fig.~\ref{Fig:CyclicHE}. Quasi-static limit means that the energy levels move so slowly that the system is at all times in thermodynamic equilibrium~\eref{eq:Boltzmann}. From the point of view of motion of energy levels, the equilibrium is established immediately after the beginning of the process as the average \textcolor{black}{waiting times between jumps among the energy levels are vanishingly small compared to the time needed to change the microstate energies. The work process that measures times spent in the individual microstates is hence self-averaging, and the resulting PDF does not depend on the initial and final states of the system and parametrization of the protocol.} Heat, on the other hand, corresponds to jumps between the microstates and thus it strongly depends on the initial and final microstate. For example, for a time-independent protocol \textcolor{black}{where no work can be done}, zero amount of heat is transferred if the system ends up in the initial state, while ending up in a state with a different energy than the initial one corresponds to a nonzero transferred heat. As described in Sec.~\ref{sec:TD}, work in steady-state heat engines has properties of heat in cyclic heat engines. And it is the described fundamental difference between the self-averaging work in cyclic heat engines and the boundary-states dependent heat-like work in steady-state heat engines that implies different fluctuations of the two quantities reviewed in Sec.~\ref{sec:TUR}. Second, the work PDF for slowly driven but not quasi-static cycles is Gaussian near its maximum located at the average work~\cite{Speck2004}~\footnote{Interestingly, it turns out that in slowly driven quantum systems this property is lost if the driving induces coherence in the system~\cite{Miller2019,Scandi2020}.}. This can be viewed as a manifestation of the central limit theorem. Noteworthy, also this result does not hold for heat PDFs. Finally, for infinitely fast cycles, the work and heat PDFs are given just by $\delta$ functions located at 0 as the system has no time to leave its microstate during the cycle and the energy of all microstates at the end and at the beginning of the cycle equal. 

For systems with finite changes in microstate energies over the cycle, the work and heat PDFs in addition acquire finite supports. For work PDF, its boundaries are determined by maximum and minimum (negative maximum) changes in energies of the microstates over the cycle. For heat PDF, the support is bounded by maximum and minimum energy differences between the microstates. Further, if the microstates are discrete and thus there is a nonzero chance that the system will spend the whole cycle in the same microstate (no jumps during the cycle), the PDFs for work and heat have a singular component represented by  $\delta$ function located at 0 with the weight given by the probability that no jump occurs during the cycle. Work and heat PDFs for non-cyclic processes have similar properties. Just the  $\delta$ functions in work PDFs do not need to be located at zero and there can be more of them -- one for each microstate with non-zero probability of staying during the whole time evolution. Concerning heat PDF, the $\delta$ function is always located at zero as no jumps always yield no heat transfer. For more details on shapes of work and heat PDFs, see Refs.~\cite{Subrt/Chvosta:2007,holubec2014non,Chvosta/etal:2010a}.

\begin{figure}
    \centering
    \includegraphics[width=1.0\textwidth]{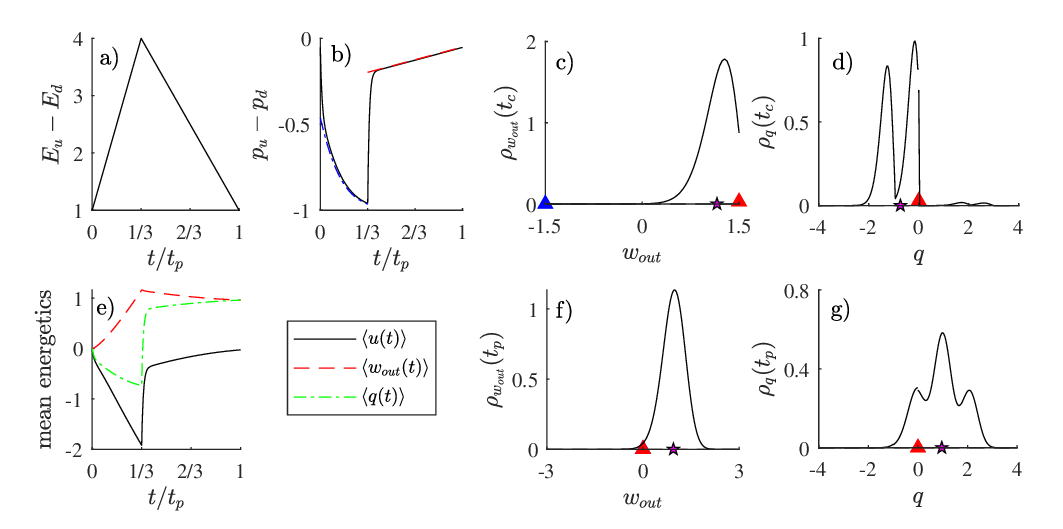}
    \caption{Dynamics and thermodynamics of the two-level system depicted in Fig.~\ref{Fig:CyclicHE} slowly driven by \textcolor{black}{the piece-wise linear protocol for energy levels and piece-wise constant protocol for the bath temperature} analyzed in Sec.~\ref{sec:W_twolevel}. a) The piece-wise linear protocol for difference of energies of the two levels $E_u(t)-E_d(t)$. \textcolor{black}{The blue 
    part of the protocol corresponds to the cold bath temperature, $T/T_c = 1$, and the red part to hot temperature, $T/T_c = 10$.} b) The difference $p_u(t)-p_d(t)$ of occupation probabilities during the limit cycle (solid line) is for the assumed slow driving close to the equilibrium Boltzmann PDF~\eref{eq:Boltzmann} corresponding to the immediate energies and temperatures (blue dashed line for cold isotherm and red for the hot one). A notable difference occurs only after the sudden changes in temperate. e) Mean output work, $\left<w_
    {out}(t)\right>$, heat, $\left<q(t)\right>$, and internal energy of the system, $\left<u(t)\right> = \left<U(m,t)\right>$, during the cycle. (c-d) PDFs for output work and heat into the system integrated over the cold isotherm. (f-g) PDFs for output work and heat integrated over the whole cycle. The stars show mean values of the individual PDFs. The triangles (vertical arrows) depict singular parts of the individual PDFs which are for the slow driving negligible. The PDFs are plotted over their whole supports. Beyond the supports, they vanish. Parameters taken
    $h_1 =  1/2$, $h_2 =  2$, $t_c = 1$, $t_h = 2$, and $\nu = 30$.
    }
    \label{fig:2levelSlow}
\end{figure}

\begin{figure}
    \centering
    \includegraphics[width=1.0\textwidth]{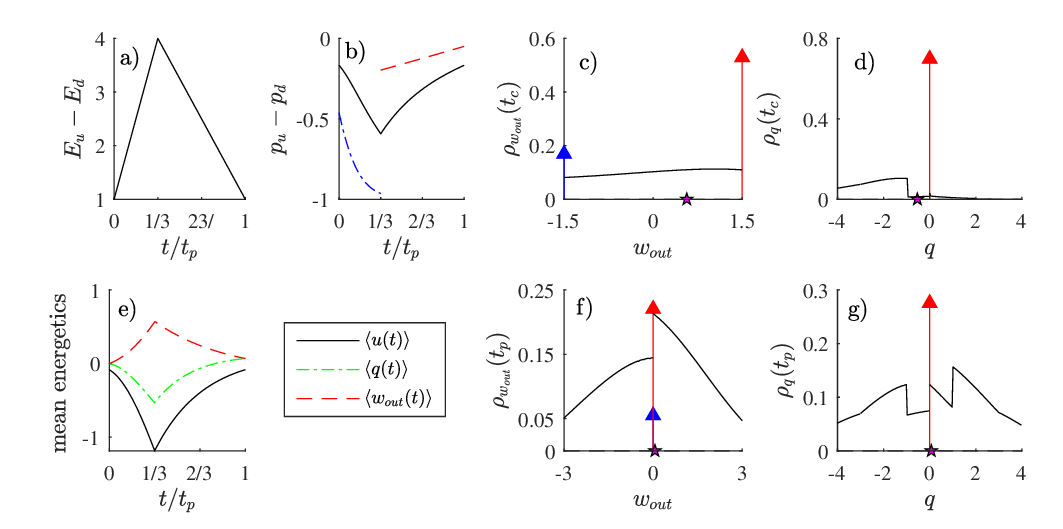}
    \caption{Dynamics and thermodynamics of the two-level system depicted in Fig.~\ref{Fig:CyclicHE} driven by \textcolor{black}{the piece-wise linear protocol for energy levels and piece-wise constant protocol for the bath temperature} analyzed in Sec.~\ref{sec:W_twolevel}. We took $\nu = 1$ corresponding to a fast driving. Other parameters and meanings of the individual curves and symbols are the same as in Fig.~\ref{fig:2levelSlow}. Note that the system is far from equilibrium during the whole cycle (b) and shapes of all the PDFs are far from a Gaussian PDF (c-d and f-g). In particular, the weights of the singular parts of the PDFs, depicted by the arrows, are quite large.
    }
    \label{fig:2levelFast}
\end{figure}

Most of the described features of work and heat PDFs are demonstrated in Figs.~\ref{fig:2levelSlow} and \ref{fig:2levelFast} using the two-level model described in Sec.~\ref{sec:W_twolevel}. Specifically, we consider the system driven by \textcolor{black}{the piece-wise linear protocol for energy levels and piece-wise constant protocol for the bath temperature}~\eref{eq:k1}-\eref{eq:k2} and calculate PDFs for work $w_{out}(t) = \int_0^t dt'\, \dot{w}$ and heat $q(t) = \int_0^t dt'\, \dot{q}$ at the end of the cold isotherm ($t = t_c$) and end of the whole cycle ($t=t_p$). 

In Fig.~\ref{fig:2levelSlow}, we \textcolor{black}{show a cycle which duration is much longer than the system relaxation time, controlled by the parameter $\nu$.}
The resulting system dynamics is indeed close to equilirbium as witnessed by both the occupation probabilities converging to the Boltzmann PDF (b) and Gaussian shapes of the work PDFs at $t_c$ (c) and $t_p$ (f). Due to the finite supports of the work PDFs, they are Gaussian only near their maximums. Heat PDFs, on the other hand, develop even for this quasistatic process several Gaussian-like peaks, each corresponding to a different combination of initial and final states of the system. Due to the frequent jumps, the singular parts of the heat and work PDFs, depicted by the arrows (triangles), are in this case negligible. Figure~\ref{fig:2levelFast} corresponds to the same protocol as Fig.~\ref{fig:2levelSlow} \textcolor{black}{but with a large system relaxation time (small parameter $\nu$).} The resulting dynamics is thus far from equilirbium (b) and the PDFs (c-d and f-g) are strongly non-Gaussian with significant singular parts. Note that the PDF for work per cycle exhibits a single and the corresponding heat PDF several discontinuities.

\subsection{Fluctuation theorem for heat engines}
\label{sec:fluct_teor}

One of the cornerstones of stochastic dynamics is the local detailed balance condition
\begin{equation}
    \frac{p_F(\Gamma|m_i)}{p_R(\Gamma_R|m_f)} = \exp\left[\Delta s_R(\Gamma)\right],
    \label{eq:LDB}
\end{equation}
which is the stochastic counterpart of time reversal invariance of microscopic dynamics~\cite{Maes2003}. It relates the conditional probability $p_F(\Gamma|m_i)$ to observe a stochastic forward trajectory $\Gamma$ from microstate $m_i$ to microstate $m_f$, the entropy $\Delta s_R(\Gamma)$ transferred into the thermal environment during this process, and the conditional probability $p_R(\Gamma_R|m_f)$ to observe the time-reversed trajectory $\Gamma_R$ under the time-reversed dynamics. The probabilities are conditioned on the initial microstates $m_i$ and $m_f$ of the two trajectories. The detailed balance condition implies the famous Jarzynski equality~\cite{Jarzynski1997} and Crooks' fluctuation theorem~\cite{Crooks1999}.

Let us now derive counterparts of these results for cyclic heat engines operating between two reservoirs at constant temperatures $T_h$ and $T_c$, such as the one investigated in Figs.~\ref{fig:2levelSlow} and \ref{fig:2levelFast}. Due to the cyclic (or steady-state) operation regime of the engine, the probabilities of the initial and final states, $m_i$ and $m_f$, of the process $\Gamma$ in Eq.~\eref{eq:LDB} are drawn from the same probability distribution $p(m)$. The ratio of unconditioned probabilities for the forward and reversed path thus reads
\begin{equation}
 \frac{p_F(\Gamma)}{p_R(\Gamma_R)} = \frac{p(\Gamma|m_i)}{p_R(\Gamma_R|m_f)} \frac{p(m_i)}{p(m_f)} = \exp\left[\Delta s_R(\Gamma) + \Delta s(\Gamma)\right] = \exp[\Delta s_{tot}(\Gamma)],
 \label{eq:DBUC}
\end{equation}
where $\Delta s(\Gamma) = - \log [p(m_f)/p(m_i)]$ is the entropy change in the system during the forward process and $\Delta s_{tot}(\Gamma)$ is the corresponding total entropy change. For heat engines with two reservoirs, 
$\Delta s_R(\Gamma) =  -{q_{in}(\Gamma)}/{T_h} + {q_{out}(\Gamma)}/{T_c}$.
Using the standard procedure~\cite{Crooks1999,Sinitsyn:2011}, i.e., multiplying Eq.~\eref{eq:DBUC} by $\delta[\Delta s - \Delta s(\Gamma)]\delta[q_{in} - q_{in}(\Gamma)]\delta[q_{out} - q_{out}(\Gamma)]$, integrating over the trajectories $\Gamma$, employing the one to one correspondence between $\Gamma$ and $\Gamma_R$, and the antisymmetry of all the involved thermodynamic variables with respect to time reversal, we obtain the fluctuation theorem for heat engines in the form~\cite{Garcia-Garcia/etal:2010, Verley/etal:2014} 
\begin{equation}
    \frac{\rho(\Delta s, q_{out}, q_{in})}{\rho_R(-\Delta s, -q_{out}, -q_{in})}
    = \exp\left(\Delta s_{tot}\right).
    \label{eq:detailFTHE}
\end{equation}

If the working medium is at the beginning of the cycle at thermal equilibrium with the cold bath, $p(m) \propto \exp[- U(m,0)/T_c]$ and $\Delta s (\Gamma) =  \Delta u(\Gamma)/T_c$. Then the same procedure yields the detailed fluctuation theorem for joint PDF for work, input heat, and increase of internal energy per cycle, $\Delta u$~\cite{Sinitsyn:2011,Campisi2014}:
\begin{equation}
    \frac{\rho(\Delta u, w_{out}, q_{in})}{\rho_R(-\Delta u, -w_{out}, -q_{in})} =  \frac{\rho(w_{out}, q_{in})}{\rho_R(-w_{out}, -q_{in})}
    = \exp\left(\Delta s_{tot}\right) = \exp\left[\frac{1}{T_c}\left(\eta_C q_{in} - w_{out}\right)\right],
    \label{eq:detailFTHE2}
\end{equation}
where $\eta_C = 1 - T_c/T_h$ denotes the Carnot efficiency. The second equality in Eq.~\eqref{eq:detailFTHE2} follows by integrating over $\Delta u$. For a model where these results can be derived from an explicit joint PDF, see Sec.~\ref{sec:CHE}.

Multiplying by $p_R$, dividing by the right-hand side, and integrating over all the variables, this detailed fluctuation theorem implies the integral fluctuation theorem
\begin{equation}
\left<  \exp\left(-\Delta s_{tot}\right) \right> = 
\left<  \exp\left[-\Delta s - \frac{q_{in}}{T_c}(\eta_C - \tilde{\eta})\right] \right>
= 1.
\label{eq:IFT}
\end{equation}
These results measure probability that engine cyclic operation decreases entropy of the universe and thus they highlight the probabilistic character of the second law. Namely, probability of $\Delta s_{tot} < 0$ decays exponentially with $|\Delta s_{tot}|$. Note, however, that stochastic efficiencies $\tilde{\eta} > \eta_C$ are not necessary unlikely for cycles where change in the system entropy is negative. This effect diminishes if heat and work are measured over multiple cycles -- while heat and work are extensive in the number of cycles, the entropy change is not.

The fluctuation theorems thus generalize the Carnot result
\begin{equation}
 \eta \le \eta_C,
 \label{eq:CarnotT}
\end{equation}
which follows from Eq.~\eref{eq:IFT} after applying the Jensen's inequality $\exp(\left< x\right>) \le \left< \exp( x) \right>$ and noticing that for cyclic processes $\left<\Delta s\right> = 0$. Furthermore, the detailed fluctuation theorem~\eref{eq:detailFTHE} implies under quite general conditions a remarkable symmetry of the PDF for stochastic efficiency discussed in the following section~\ref{sec:FluctEff}. And it can be used as one of the starting points for the derivation of thermodynamic uncertainty relations reviewed in Sec.~\ref{sec:TUR}.

\subsection{Stochastic efficiency}
\label{sec:FluctEff}


\textcolor{black}{The previous section shows that the fluctuation theorem for entropy production~\eqref{eq:DBUC} represents a generalization of the Carnot theorem~\eqref{eq:CarnotT}. In this section, we ask if there is a generalization of the Carnot theorem based on the properties of the stochastic efficiency $\tilde{\eta}$, defined as the ratio of stochastic work and heat.
}

Consider a heat engine (either cyclic or steady-state) operating with two heat reservoirs at constant temperatures $\Th$ and $\Tc$. The classical and stochastic efficiencies $\eta$ and $\tilde{\eta}$ per cycle of operation of the cyclic heat engine are given by Eqs.~\eqref{eq:effSt} and \eqref{eq:eff}. Counterparts of these definitions for steady-state heat engines should be calculated using heat and work fluxes in the steady state: $\eta = \left<\dot{w}_{out}\right>/\left<\dot{q}_{in}\right>$ and $\tilde{\eta} = \dot{w}_{out}(t)/\dot{q}_{in}(t)$. Regardless details of definitions of work and heat in the definition of $\tilde{\eta}$, the joint PDF $\rho(\wout,\qin)$ yields the efficiency PDF according to~\cite{Gingrich/etal:2014,Proesmans2015}
\begin{multline}
\label{eq:rho_eta_definition}
\rho(\tilde\eta,t)= \int \int d\wout d\qin \, 
\delta\left(\tilde\eta-\frac{\wout}{\qin} \right) \rho(\wout,\qin) = \\
\frac{1}{\tilde\eta^2} 
\int d\wout |\wout| \rho\left(\wout, \frac{ \wout}{\tilde \eta} \right).
\end{multline} 
This result suggests that if the integral on the right-hand side does not vanish for $\tilde \eta\to\pm\infty$, $\rho(\tilde\eta,t)$ exhibits heavy tails and its moments $\left<\tilde{\eta}^k\right>$ diverge for $k\ge 1$. An example where the integral can be evaluated explicitly are models with Gaussian fluctuations of all thermodynamic currents~\cite{Polettini/etal:2015, Jiang/etal:2015, Proesmans/etal:2016b, Park/etal:PRE:2016, Vroylandt/etal:PRE:2016, Proesmans/VanDenBroeck:Chaos:2017, Gupta/Sabhapandit:PRE:2017, Gupta:JSTAT:2018}.  Then, it indeed holds that 
\begin{equation}
\label{eq:tails_rho_eta}
\rho(\tilde\eta) \sim \frac{1}{\tilde\eta^2}, 
\qquad |\tilde\eta|\to\infty,
\end{equation} 
yielding diverging moments of the efficiency, including the mean value, i.e. $\left<\tilde{\eta}\right> \neq \eta$. From experimentalist's perspective, this means that fluctuations of the empirical average $\left<\tilde\eta \right>$ diverge with the number of experimental trajectories \textcolor{black}{over which the average is calculated}. The power-law tails~(\ref{eq:tails_rho_eta}) arise because the stochastic input heat in the denominator of $\tilde\eta$ can be very close to zero even for relatively large values of $|w_{out}|$.

{\color{black}
Beyond the nonexistence of moments, the stochastic efficiency computed from the quantities per cycle (cyclic heat engines) or instantaneous energy fluxes (steady-state heat engines) bares no special properties. 
However, this changes if one considers stochastic efficiencies evaluated from work and heat measured over many cycles or over long operational times:
\begin{equation}
\tilde\eta(t) = \frac{\wout(t)}{\qin(t)}.
\label{eq:etaFLDF}
\end{equation}
Here, $\wout(t)$ and $q_{in}(t)$ are defined as integrals \eqref{eq:integrated_work_heat} over $\dot{w}_{out}(t)$ and $\dot{q}_{in}(t)$ from 0 to $t$. For cyclic heat engines, the integration time is a whole number multiple of the cycle period, $t = n t_p$.
These quantities are often obtained from experiments where stable experimental conditions are most easily achieved for single long trajectories. Having such a long trajectory, the average energy fluxes in steady-state heat engines are most reasonably approximated by the time averages $\wout(t)/t$ and $q_{in}(t)/t$, obtained from Eqs.~\eqref{eq:integrated_work_heat}, and similarly for cyclic heat engines. In experiments, the stochastic efficiency~\eqref{eq:etaFLDF} thus often approximates the standard efficiency $\eta$.
While the PDF for this definition of $\tilde{\eta}$ still obeys the properties described by Eqs.~\eqref{eq:rho_eta_definition} and \eqref{eq:tails_rho_eta}, it acquires several interesting new features.
}

They can be derived from the specialized version of the fluctuation theorem for heat engines~\eqref{eq:detailFTHE2}. Let us now assume that (i) the joint PDF for the integrated work and heat can be for large times written using the large deviation form $\rho(w_{out},q_{in},t) \sim \exp[t I(w_{out}/t,q_{in}/t)]$, (ii) $\rho(w_{out},q_{in},t)$ obeys Eq.~\eqref{eq:detailFTHE2}, and (iii) the long-time limit of the corresponding scaled cumulant generating function~\eqref{eq:SCGF}, 
\begin{equation}
\lim_{t\to \infty} \frac{1}{t} \log \xi(s_w,s_{in},t) = \lim_{t\to \infty} \frac{1}{t} \left<\exp(-s_w w_{out} - s_{in}q_{in}) \right>,
\label{eq:CGFFE}
\end{equation}
is a smooth function of its arguments and has a unique maximum. Then the PDF for the stochastic efficiency fulfills the large deviation principle, $\rho_{\tilde{\eta}} \sim \exp(t I_{\tilde{\eta}})$, and its rate function
\begin{equation}
\label{eq:contraction_LDFeta}
I_{\tilde{\eta}} = \max_{\qin} I\left(\tilde\eta\frac{\qin}{t},\frac{\qin}{t}\right)
\end{equation}
shows interesting properties. Namely, for cyclic heat engines driven by time-symmetric protocols and steady-state heat engines the position of its maximum is located at the standard thermodynamic efficiency $\eta$, and the position of its minimum at the reversible efficiency $\eta_C$~\cite{Verley2014,Verley/etal:2014b, Verley/etal:2014}.
For time-symmetric protocols, $\eta_C$ ($\eta$) thus represent the least (most) likely value of efficiency $\tilde{\eta}(t)$ for large $t$. For non-symmetric protocols, the position of the maximum of $I_{\tilde{\eta}}$ is still located at $\eta$, but the minimum has no special meaning anymore. Instead, the reversible efficiency $\eta_C$ mark the intersection point of rate functions for the forward and for the time reversed protocols for the engine~\cite{Gingrich/etal:2014}.

\begin{figure}
    \centering
    \includegraphics[width=1.0\textwidth]{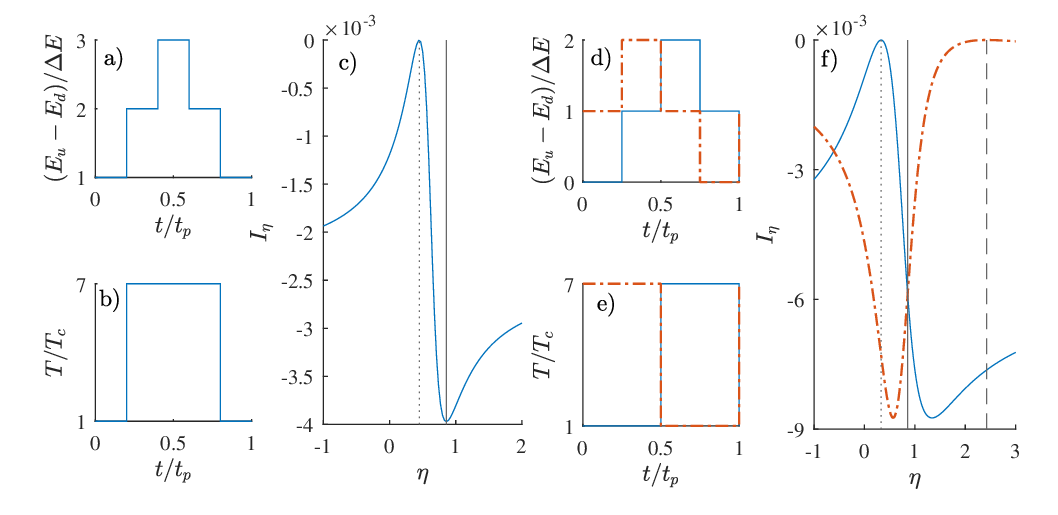}
    \caption{Large deviation functions for stochastic efficiency for the two-level system depicted in Fig.~\ref{Fig:CyclicHE} driven by the piece-wise constant protocol analyzed in Sec.~\ref{sec:2level_PC}. The symmetric protocol (a-b) implies the large deviation function with minimum at the reversible efficiency $\eta_C$ (solid line in c) and maximum at the standard efficiency~\eref{eq:eff} (dotted line). Large deviation functions for the time-asymmetric (solid blue line) and the corresponding time-reversed (\textcolor{black}{thick} dot dashed red line) protocols (d-e) intersect at the reversible efficiency (solid line in f). Maximums of these large deviation functions (dotted and dashed lines in f) mark the corresponding standard efficiencies $\eta$. We use the same parameters as Ref.~\cite{Gingrich/etal:2014}:
    $\Delta E = 2.375$, $T_c = 2$, $T_h = 14$, $B = 0.05$, and $t_p = 10$.
    }
    \label{fig:LD}
\end{figure}

As an illustration, we plot in Fig.~\ref{fig:LD} rate functions for the stochastic efficiency~\eqref{eq:etaFLDF} for a two-level system driven by a time symmetric (a-c) and a time-asymmetric (d-e) protocols. Details of the calculations are described in Sec.~\ref{sec:2level_PC}. The rate function corresponding to the symmetric driving (c) exhibits a global maximum at the corresponding standard efficiency $\eta$, similarly as the rate functions for the forward and time-reversed protocols in d. The global minimum of the rate function for the symmetric driving and the intersection of the rate functions for the forward and backward protocols are then indeed located at $\tilde{\eta}=\eta_C$. This example is also interesting from the point of view of breaking the assumptions of the presented theory. Namely, all the time-asymmetric protocols composed of only 4 constant segments yield in this case stochastic output work proportional to the input heat. The resulting stochastic efficiencies for all trajectories thus equal the standard efficiency and the corresponding PDF $\rho_{\tilde{\eta}} = \delta(\tilde{\eta} - \eta)$ does not obey the large deviation principle. A similar situation occurs in the Brownian heat engine driven by a piece-wise constant protocol discussed in Sec.~\ref{sec:CHE}.

The position of the maximum of $I_{\tilde{\eta}}$ at $\tilde{\eta} = \eta$ 
follows from the assumption that the quantities $\wout(t)$, $\qin(t)$, and $\tilde\eta(t)$ obey the large deviation principle. Since $\wout(t)/t$ and $\qin(t)/t$ converge with increasing $t$ towards their stationary mean values, the PDF $\rho(\tilde\eta,t)$ concentrates around the point 
$\eta = \langle \wout \rangle/ \langle \qin \rangle$ as $t\to\infty$. As follows from its dependence on the symmetry of the protocol, the value of $I(\eta_C)$ corresponding to the reversible efficiency requires in addition to the validity of the large deviation principle also the symmetry imposed by the specialized fluctuation theorem~\eqref{eq:detailFTHE2}~\cite{Verley/etal:2014,Gingrich/etal:2014,Verley2014,Verley/etal:2014b}.

As an example, consider the situation with a time-symmetric protocol~\cite{Verley2014,Verley/etal:2014b, Verley/etal:2014}, when the forward and backward PDFs in Eq.~(\ref{eq:detailFTHE}) are identical. The stochastic efficiency is given by the reversible efficiency, $\tilde\eta=\etaC$, for reversible trajectories with $\Delta s_{tot}(t)=0$. In this case, Eq.~(\ref{eq:detailFTHE}) implies the symmetry $\rho(\wout,\qin,t)=\rho(-\wout,-\qin,t)$ of the PDF for work and heat and thus also of the corresponding rate function. The rate function $I(\etaC\qin/t,\qin/t)$ is thus an even function in $q_{in}$ with an extreme at $q_{in}=0$. Equation~\eqref{eq:contraction_LDFeta} further implies that $I_{\tilde{\eta}}(\tilde\eta) \geq I(0,0)$ and thus the rate function for the stochastic efficiency has a global minimum at
$\tilde\eta = \eta_C$. This proves that $\etaC$ is the least likely efficiency in the long-time limit~\cite{Verley/etal:2014,Verley2014,Verley/etal:2014b}. For a similar argument for time-asymmetric protocols, see Ref.~\cite{Gingrich/etal:2014}.

At first glance, Eq.~\eqref{eq:detailFTHE2} universally follows from the exact fluctuation theorem~\eqref{eq:detailFTHE} for large integration times $t$, when the time-nonextensive entropy change in the system, $\Delta s$, and the internal energy change, $\Delta u$, become negligible compared to the time-extensive integrated work and heat. However, Ref.~\cite{Manikandan/etal:PRL:2019} explains that the reality is more subtle and the boundary terms $\Delta u$ and $\Delta s$ can limit the region of convergence of the scaled cumulant generating function~\eqref{eq:CGFFE}. This can induce discontinuities in the derivatives of the rate function $I(w_{out}/t,q_{in}/t)$ and break its symmetry properties strongly enough that the value $I(\eta_C)$ looses its above described special meaning~\cite{Manikandan/etal:PRL:2019}.

Besides the presented general conclusions, the efficiency statistics has been evaluated for several model situations~\cite{Rana/etal:2014, Proesmans/VanDenBroeck:NJP:2015, Rana/etal:PhysicaA:2016}, addressing e.g.\ the role of coupling between system's internal degrees of freedom~\cite{Sune/Imparato:JPhysA:2019}, many-particle gas models of a cylinder enclosed by a piston and coupled to a thermostat~\cite{Cerino/etal:PRE:2015}, and a system of interacting unicyclic machines undergoing a phase transition~\cite{Vroylandt/etal:PRL:2020}. For feedback cooling techniques, the concept of stochastic efficiency was reformulated in terms of fluctuating information flows~\cite{Rosinberg/Horowitz:EPL:2016, Paneru/etal:NatCommun:2020, Paneru/Pak:APX:2020}. The stochastic efficiency was also introduced for quantum machines, see e.g.\ Refs.~\cite{Esposito/etal:2015, Agarwalla/etal:2015, Cuetara/Esposito:NJP:2015, Crepieux/Michelini:JSTAT:2016, Tang/etal:PRB:2018} for models of quantum dots and/or thermoelectric junctions, and Ref.~\cite{Denzler/Lutz:PRR:2020} for an example of a quantum periodic cycle.  

As for experiments, to the best of our knowledge, there exist only a few works reporting measurements of efficiency fluctuations~\cite{ Paneru/etal:NatCommun:2020, Paneru/Pak:APX:2020, Martinez2016,Proesmans/etal:2016b}.

\subsection{Thermodynamic uncertainty relations for heat engines}
\label{sec:TUR}

Thermodynamic uncertainty relations (TURs) are
inequalities that relate relative fluctuations of a certain observable and the corresponding total entropy production. They are thus frequently interpreted as thermodynamic upper bounds on precision of measurements of the observable. 
Since their discovery~\cite{Barato/Seifert:2015}, TURs have been derived for several Markovian models~\cite{Gingrich/etal:2016, Dechant/Sasa:2018, Liu/etal:2019} and from fluctuation theorems~\cite{Zhang:2019, Hasegawa/VanVu:2019, Timpanaro/etal:2019}. For recent reviews on TURs, see Refs.~\cite{Marsland/England:2018, Seifert:2019, Horowitz/Gingrich:2020}. 

In their original form~\cite{Barato/Seifert:2015, Pietzonka/Seifert:2018}
\begin{equation}
\label{eq:TUR_J}
\frac{J^2}{D} \leq \sigma,
\end{equation} 
TURs relate the average current $J = \lim_{t\to\infty} o(t)/t$ of a time-extensive observable $o(t)$, its dispersion $D =\lim_{t\to\infty} \langle [o(t)/t-J]^2 \rangle t/2$, and the total average entropy production rate $\sigma$.
Examples of the observable $o(t)$ are the integrated output work $w_{out}(t)$ and input heat $q_{in}(t) = w_{out}(t) + q_{out}(t)$~\eqref{eq:integrated_work_heat}, which invites application of the TUR~\eqref{eq:TUR_J} to steady-state heat engines. 

This application~\cite{Pietzonka/Seifert:2018} uncovered a trade-off between power, power fluctuations, and efficiency of these machines. Namely, introducing output power of a steady-state heat engine by $P = \lim_{t\to \infty }w_{out}(t)/t$, defining the so called constancy~\footnote{\textcolor{black}{The constancy is the long-time limit of variance of the stochastic power, $w_{out}(t)/t$, that decreases as $1/t$, multiplied by $t$. As such, it represents a non-zero number that characterises fluctuations of the power measured from long trajectories.}}
\begin{equation}
\Delta_P = \lim_{t\to\infty} \left<[w_{out}(t)/t-P]^2\right> t,
\label{eq:constancy}
\end{equation}
and noticing that the entropy production rate for a steady-sate heat engine in contact with two reservoirs at temperatures $T_c$ and $T_h$ obeys 
$\sigma= \lim_{t\to \infty} [q_{out}(t)/(t\Tc)-q_{in}(t)/(t\Th)] = P (\eta_C-\eta)/(T_c \eta)$, the TUR~\eqref{eq:TUR_J} can be written in the form~\cite{Pietzonka/etal:2016,Pietzonka/Seifert:2018}
\begin{equation}
\label{eq:eta_bound_TUR}
\eta \leq \frac{\etaC}{1+2 P \Tc /\Delta_P}.
\end{equation}
It implies that the efficiency of a steady-state heat engine producing a well defined output power (non-negligible $P/\Delta_P$) is always smaller than $\etaC$. The engine can work close to $\eta_C$ only if either $P$ vanishes, power fluctuations $\Delta_P$ diverge, or the engine operates at an extremely low $T_c$. 

As already noted in Ref.~\cite{Pietzonka/Seifert:2018} based on the analysis of Brownian clocks~\cite{Barato/Sefiert:2016}, this result does not imply the same behavior for cyclic heat engines. The reason is that the TUR~\eqref{eq:TUR_J} applies only to observables related to energy/mass transport and it does not hold for output work~\eqref{eq:work_cyclic} in cyclic heat engines. As discussed in Secs.~\ref{sec:static_HE_TD} and \ref{sec:propertiesWPD}, work in steady-state heat engines has properties of heat rather than those of the work in cyclic heat engines. Therefore, the two works have a much different statistics. \textcolor{black}{Indeed, the cyclic heat engines driven quasi-statically~\cite{Holubec/Ryabov:2018} by tuning the relaxation time of the system can operate arbitrarily close to the Carnot efficiency and deliver a well defined positive output power.}

\begin{figure}
    \centering
    \includegraphics[width=1.0\textwidth]{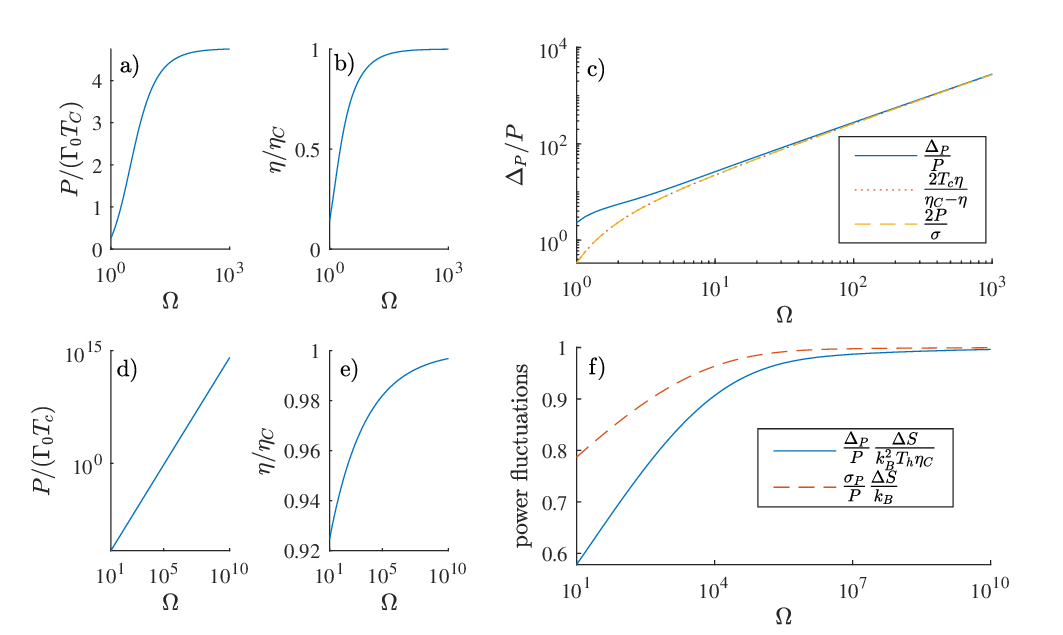}
    \caption{Power and power fluctuations close to the reversible efficiency $\eta_C$ for steady-state and cyclic heat engines. (a-c) Performance of a simple model solar cell operating as a steady-state heat engine, depicted in Fig.~\ref{Fig:colarCell} and described in Sec.~\ref{sec:solar_cell} and Ref.~\cite{Pietzonka/Seifert:2018}. As the efficiency reaches $\eta_C$ (b), the power saturates (a) and its fluctuations diverge according to Eq.~\eqref{eq:TUR26}.   Parameters taken: $x_c=10$, $x_h = 0.2$, $x_{ru} = 1$, and $x_{ru} = 1.2 + 7/\Omega$. (d-f) Performance of the cyclic Brownian heat engine depicted in Fig.~\ref{Fig:CyclicHE} and detailed in  Sec.~\ref{sec:W_parabolic} and Ref.~\cite{Holubec/Ryabov:2018}, including the used parameters. With $\eta \to \eta_C$ (e), its power increases (d) and power fluctuations converge to the values given by Eqs.~\eqref{eq:PRLSATRU} (f).}
    \label{fig:TURALL}
\end{figure}

Figure~\ref{fig:TURALL} illustrates striking differences between the performance of steady-state and cyclic heat engines showing power, efficiency and power fluctuations for paradigmatic examples of the two types of machines. The quantities are plotted as functions of a scaling parameter $\Omega$ yielding reversible efficiency $\eta_C$ in the limit $\Omega \to \infty$. Panels~(a-c) demonstrate performance of the three-level model of steady-state solar cell considered in Ref.~\cite{Pietzonka/Seifert:2018} and reviewed in Sec.~\ref{sec:solar_cell}. For an illustration using a steady-state thermoelectric heat engine, see Ref.~\cite{Kheradsoud/eta:2019}. Panels~(d-f) correspond to the Brownian heat engine based on the breathing parabola model discussed in Ref.~\cite{Holubec/Ryabov:2018} and reviewed in Sec.~\ref{sec:W_parabolic}.

Panels (a-c) verify that, while reaching the reversible efficiency $\eta_C$ (b), the output power of the solar cell saturates (a) and its fluctuations diverge (c). Furthermore, \textcolor{black}{the overlapping solid, dotted, and dashed lines show} that for large values of $\Omega$ this model saturates the inequality $\eqref{eq:eta_bound_TUR}$, which can be written as
\begin{equation}
\frac{\Delta_P}{P} \ge \frac{2 T_c \eta}{\eta_C-\eta} = \frac{2 P}{\sigma}.  
\label{eq:TUR26}
\end{equation}
On the other hand, as the efficiency of the Brownian heat engine converges to $\eta_C$ (e), its power increases (d) and the power fluctuations saturate (f). The saturation values can be evaluated analytically using calculations similar to those performed in Sec.~\ref{sec:CHE}. The result is
\begin{equation}
  \frac{\Delta_P}{P}  = \frac{\Delta S}{k_B^2 T_h \eta_C}, \quad
  \frac{\sigma_P}{P}  = \frac{\Delta S}{k_B},
  \label{eq:PRLSATRU}
\end{equation}
where $\sigma_P^2 = \left<w_{out}^2 - W_{out}^2\right>/t_p^2$ denotes the variance of the output power, and $\Delta S$ the entropy change in the system during the hot isotherm. For details, see Sec.~\ref{sec:W_parabolic} and Ref.~\cite{Holubec/Ryabov:2018}.

Counterparts of the bound~\eqref{eq:eta_bound_TUR} valid for periodic processes were derived by introducing a set of variational parameters formally reminiscent of occupation probabilities for the individual microstates~\cite{Koyuk/etal:2018} and for slowly driven systems~\cite{Miller/etal:PRL2021, Miller/etal:PRE2021}. Another family of inequalities for entropy production providing bounds on efficiency of cyclic heat engines was derived in Ref.~\cite{Koyuk/Seifert:2019}. 
A recent comparative study of TURs for periodic systems performed for a specific model of driven electron pump can be found in Ref.~\cite{Harunari/etal:2020}. 
All these results are in agreement with the conclusions of Ref.~\cite{Holubec/Ryabov:2018}, which rely on two broadly fulfilled assumptions: (i) Dynamics of the system is described by the GME~\eqref{eq:GME}. (ii) For a constant protocol, the GME has a unique (equilibrium) stationary solution. The same assumptions are required for the validity of the Jarzynski identity~\cite{Speck2007}, which can thus be used as a check whether given system can be used as a basis of a cyclic heat engine operating close to reversible efficiency with well defined nonzero output power.

To the best of our knowledge, TURs have not yet been tested experimentally in realizations of microscopic machines. However, several utilizations of TURs to infer entropy production have been proposed recently~\cite{Seifert:2019, Li/etal:2019, Manikandan/etal:2020, Skinner/Dunkel:2021}. 

\section{Exactly solvable models of cyclic heat engines}
\label{sec:Examples}

In addition to being interesting by themselves, exactly solvable models played a fundamental role in the development of stochastic thermodynamics. The general results such as fluctuation theorems for entropy production and thermodynamic uncertainty relations were first found for a particular model. Simple model systems also help us to establish an intuitive understanding of complex physics out of equilibrium. Moreover, actual applications of theory to experiments are frequently tested by means of experimental realizations of such simple and in theory solvable models.

As in classical statistical physics, the paradigmatic examples of discrete and continuous systems, where analytical solutions in stochastic thermodynamics can be obtained, are two-level systems and harmonic oscillators.
Known results for the former are discussed in the next section. Sec.~\ref{sec:harmonic} is devoted to the latter. Finally, in Sec.~\ref{sec:solar_cell} we review a simple steady-state model of solar cell used to demonstrate validity of TURs in the previous section.

\subsection{Two-level system}
\label{sec:2level}

The two-state Markov process has a prominent position both in theory and experiment~\cite{Schuler/etal:2005, Tietz/etal:2006} as a minimal non-trivial model with discrete energy spectrum that describes well more complicated setups for low enough temperatures. In practice, such processes can represent dynamics of a defect in crystal~\cite{Schuler/etal:2005, Tietz/etal:2006} and a single-level quantum dot or a tunnel junction~\cite{Kung/etal:2012, Koski/etal:2013}. They also serve as a general coarse-grained description of transition dynamics in systems having time-dependent free-energy landscapes with two sharp minima, such as in studies of stochastic resonance~\cite{Jop/etal:2008} or macromolecules with two conformational states~\cite{Ritort:2006, Ritort:2007}.

The dynamics~\eref{eq:GME} of the two-level model is exactly solvable for an arbitrary protocol~\cite{Chvosta/etal:2010a} and thus one can evaluate analytically first two moments of work and heat using the approach suggested in Eqs.~\eref{eq:time_cor_fun_continuous} and
\eref{eq:work_second_moment}. However, exact solutions for work and heat PDFs are known only for two special classes of protocols: piece-wise constant and piece-wise linear. Although the solutions for piece-wise linear protocols for microstate energies are the more involved from the two, they were obtained earlier in a series of papers by Chvosta et al.~\cite{Subrt/Chvosta:2007,Chvosta2007,Chvosta/etal:2010a, Chvosta/etal:2010b}. These solutions are remarkable especially because all other known solutions were obtained for piece-wise constant driving.

Exact solutions for specific piece-wise constant protocols were obtained in Refs.~\cite{Verley/etal:2013, Verley/etal:2014} and applied in study of stochastic efficiency~\cite{Verley/etal:2014b,Gingrich/etal:2014},
current fluctuations~\cite{Barato/Chetrite:2018}, time-reversal symmetric Crooks and Gallavotti–Cohen fluctuation relations~\cite{Mandaiya/Khaymovich:2019}, and
thermodynamic uncertainty relations~\cite{Harunari/etal:2020}. Moment-generating function for work was also discussed recently in~\cite{Salazar:2020}. 

In the next section, we review the general approach to piece-wise constant protocols in discrete systems. Review of results for piece-wise linear protocols is given in Sec.~\ref{sec:W_twolevel}.

\subsubsection{Piece-wise constant protocol.}
\label{sec:2level_PC}
Consider the two-level system depicted in Fig.~\ref{Fig:CyclicHE} with energy levels and temperature varying according to a periodic piece-wise constant protocol. Several examples of such protocols are given in Fig.~\ref{fig:LD}(a-b) and (d-e). However, it is important to note that an arbitrary protocol can be approximated by a piece-wise constant driving~\cite{Holubec/etal:2019,Chvosta/etal:2020} and thus the following approach can be applied for extracting information on work and heat fluctuations in an arbitrarily driven two-level system described by the GME~\eref{eq:GME}. Moreover, the key steps of the presented technique do not change with increasing number of microstates and thus it applies for arbitrary discrete systems~\cite{Mandaiya/Khaymovich:2019}. Finally, Ref.~\cite{Holubec/etal:2019} shows that the overdamped continuous systems~\eqref{eq:overdamped} can be well approximated by discrete systems and thus the presented method yields work and heat fluctuations also for overdamped Brownian heat engines.

During each constant segment of the protocol, the transition rate matrix is constant. We employ the transition rates presented in Ref.~\cite{Gingrich/etal:2014}, where the rate matrix for the $i$th segment reads
\begin{equation}
    L_i = \begin{pmatrix}
- \exp(B/T_i) & \exp[-(B - E_i)/T_i]\\
\exp(B/T_i) & - \exp[-(B - E_i)/T_i]\\
\end{pmatrix}. 
\label{eq:rate2level}
\end{equation}
Here, $T_i = T(t)$, $B - E_i = U(u,t) = E_u(t)$, $t \in ] t_i, t_{i+1} [$. The jumps in the protocol occur at times $t_i$, $i=1,\dots,n+1$, and the energy $U(d,t) = E_d(t)$ of the down microstate $d$ is set to zero during the whole cycle. We assume that duration of all the $n$ constant segments of the protocol is equal, i.e. $t_{i+1} - t_{i} = t_p/n$, $i=1,\dots,n$. Finally, the protocol is time-periodic and thus $T_0 = T_{n}$ and similarly for $E_i$.

We will now show how to calculate the generating function for output work~\eref{eq:work_cyclic} and input heat~\eref{eq:qin_cyclic} and also the corresponding rate function. We will employ a strategy analogous to that with tilted rate matrices described for steady-state heat engines in Sec.~\ref{sec:static_HE_TD}.

If the system dwells at the time $t_i$ of the $i$th jump on the upper level, it performs work $-(E_i-E_{i-1})$. Since $E_d(t) = 0$, these are the only occasions when work is done. The propagators during the jumps are unity matrices because the system has no time to change its state. The Laplace transform of PDF for work done during the jumps conditioned on initial and final states is thus described by the matrix
\begin{equation}
    G_i(s_w) = \begin{pmatrix}
1 & 0\\
0 & - \exp[(E_i-E_{i-1})s_w]\\
\end{pmatrix},
\label{eq:Ptiltedw}
\end{equation}
where $s_w$ denotes the Laplace variable corresponding to $w
_{out}$. Specifically, the matrix element $[L_i(s_w)]_{kl}$ gives the generating function for the case when the system was before the jump at microstate $l$ and after the jump at microstate $k$.

Heat is absorbed from reservoirs whenever the system jumps between the microstates $d$ and $u$. It equals $E_i$ if the jump is from $d$ to $u$ at time $t$ and $- E_i$ for the opposite jump. The heat transferred during the constant segments of the protocol equals the difference in the internal energy $\Delta u$ of the system. Therefore, we do not need to tilt the rate matrix \eref{eq:rate2level}, but, similarly as above for work, we can tilt directly the propagator $G_i = \exp(L_i t_p/n)$ for the whole $i$th segment. The corresponding conditioned generating function for the input heat reads
\begin{equation}
    G_i(s_{in}) = \begin{pmatrix}
[G_i]_{dd} & \exp(- E_i s_{in}) [G_i]_{du}\\
\exp(E_i s_{in}) [G_i]_{ud} & [G_i]_{uu}\\
\end{pmatrix}.
\label{eq:Ptiltedq}
\end{equation}
The Chapman-Kolmogorov equation implies that the propagator $G$ for the whole cycle follows from the propagators for the individual segments $G_i$,  $i=1,\dots,n$, as the matrix product
\begin{equation}
    G = G_{n} \dots G_1,
\end{equation}
which is thus equal to the ordered matrix exponential $G = \exp_{\rightarrow}[\int_0^{t_p} dt' L(t')]$. In a similar manner, the tilted propagators \eref{eq:Ptiltedw} and \eref{eq:Ptiltedq} yield the conditioned moment generating function $G(s_w,s_{in})$ for output work $w_{out}$ and input heat $q_{in}$ per cycle: 
\begin{equation}
    G(s_w,s_{in}) = G_{n+1}(s_w)\tilde{G}_{n}(s_{in})G_{n}(s_w)
    \dots
    \tilde{G}_1(s_{in})G_1(s_w).
\end{equation}
To take into account just the heat flowing into the system from the hot bath, $\tilde{G}(s_k)$ is given by $G(s_k)$ if $T_k = T_h$ and $G_k$ otherwise.

The matrix $G(s_w,s_{in})$ can be used for calculation of generating function for heat and work measured over $k\ge 1$ cycles as
\begin{equation}
    \chi_k(s_w,s_{in}) = \left<+\right|[G(s_w,s_{in})]^k {\mathbf p},
\end{equation}
where the state of the system at the beginning of the cycle, $\mathbf p$, is determined by the periodicity condition ${\mathbf p} = G {\mathbf p}$. This generating function can be used either to calculate moments of work and heat as in Eq.~\eref{eq:raw_moments_qin} or for calculation of the PDF for work and heat using the inverse Laplace transform. Instead, we will use it for derivation of the rate function for work and heat, which we then apply to calculation of the rate function for the stochastic efficiency.

The rate function $I(w_{out},q_{in})$ can be obtained using a similar procedure as Eq.~\eref{eq:rateqqSS}. The only difference in the derivation is that now the tilted propagator has the form $[G(s_w,s_{in})]^k$ instead of $\exp[t L(s_w,s_{in})]$. However, writing  $[G(s_w,s_{in})]^k = \exp\left[k \log G(s_w,s_{in}) \right]$ implies
\begin{equation}
    I(w_{out},q_{in}) = \min_{s_w,s_{in}}
    \left[\log \lambda_{max}(w_{in},s_{out}) + s_{in}q_{in} + s_w w_{out} \right].
    \label{eq:ratewqC}
\end{equation}
If the stochastic output work and input heat are not proportional to each other, the generating function $I(w_{out},q_{in})$ can be used for calculation of generating function for the stochastic efficiency~\eref{eq:effSt} through the transformation
\begin{equation}
I_{\tilde{\eta}} = \max_{q_{in}} I(\tilde{\eta}q_{in}, q
_{in}),
\end{equation}
which follows from applying the Laplace approximation in the marginalization~\eref{eq:eff_PDF} of the two-dimensional PDF.

To demonstrate the general features of the rate function for efficiency described in Sec.~\ref{sec:FluctEff}, we calculated rate functions for the symmetric protocol composed of 5 segments and forward and time-reversed 4 segments protocols depicted in Fig.~\ref{fig:LD}. Interestingly, for an arbitrary symmetric 4 branches protocol, $w_{out}$ and $q_{in}$ are proportional to each other, $\tilde{\eta}$ is $\delta$-distributed, and the rate function $I_{\tilde{\eta}}$ does not exist.

\subsubsection{Piece-wise linear protocol.}
\label{sec:W_twolevel}
The PDF for work done on a two-level system by linearly varying \textcolor{black}{energies of} its energy levels in time was derived for two types of transition rates in Refs.~\cite{Subrt/Chvosta:2007,Chvosta2007}. Subsequently, the method has been generalized to piece-wise constant protocols and applied to derivation of heat and work PDFs for a two-stroke cyclic heat engine~\cite{Chvosta/etal:2010a, Chvosta/etal:2010b}. Below, we review the main results for the heat engine. For a more detailed review, see Ref.~\cite{holubec2014non}.

The engine was studied for the Glauber transition rates  
\begin{eqnarray}
\label{eq:k21_twolevel}
L_{du}(t) &=& \frac{\nu}{1+\exp\left\{-\beta \left[E_{u}(t)-E_{d}(t) \right]\right\}}, \\ 
L_{ud}(t) &=& \frac{\nu}{1+\exp\left\{\beta \left[E_{u}(t)-E_{d}(t) \right]\right\}},
\end{eqnarray} 
where the parameter $\nu$ controls relaxation time of the two-level system and $E_m(t) = U(m,t)$, $m = u,d$ are energies of the two levels. Let us now consider a purely linear protocol for the microstate energies and a constant protocol for the bath temperature:
\begin{equation}
\label{eq:E1_linear}
E_u(t)=h+b(t-t'), \quad E_d(t)=-E_u(t), \quad T(t) = T,
\end{equation}
where $h$ and $b$ are constants, and $t\geq t'$. For $b>0$, the matrix elements $G_{mn}(w,t|w',t') = G(w,m,t|w',n,t')$ of the Green function for the partial differential  equation~\eref{eq:PDFworkCyclic} are given by~\cite{Chvosta/etal:2010a} 
\begin{eqnarray}
\fl \frac{G_{uu}}{2\beta}&=& 
\left[\frac{1+u}{u+ \ee^{\widetilde \tau} }\right]^{a}\,
\delta(\widetilde{\tau}-\widetilde{\eta}) - \frac{\widetilde{\theta}\,aux^{a}(1-x)y}{2}  \\ \nonumber
\fl &\times&  
\left[\frac{\FDJ{1+a}{-a}{1}{\phi}}{(1+ux)^{1+a}(1+uy)^{1-a}}
- (1+a)(1+u) 
(1+uxy)\frac{\FDJ{2+a}{1-a}{2}{\phi}}
{(1+ux)^{2+a}(1+uy)^{2-a}}\right],
\label{g11}
\end{eqnarray}
\begin{eqnarray}
\fl \frac{G_{ud}}{2\beta} = \frac{\widetilde{\theta}\,
aux^{a}y}{2} \frac{\FDJ{a}{1-a}{1}{\phi}}{(1+ux)^{a}(1+uy)^{1-a}},
\label{g12}
\end{eqnarray}
\begin{eqnarray} 
\fl \frac{G_{du}}{2\beta} = \frac{\widetilde{\theta}\,
ax^{a}}{2} \frac{\FDJ{1+a}{-a}{1}{\phi}}{(1+ux)^{1+a}(1+uy)^{-a}},
\label{g21}
\end{eqnarray} 
\begin{eqnarray} 
\fl \frac{G_{dd}}{2\beta}&=&
\left[\frac{1+u\exp(-\widetilde{\tau})}{1+u}\right]^{a}\,
\delta(\widetilde{\tau}+\widetilde{\eta})
+ \frac{\widetilde{\theta}\,aux^{a}(1-y)}{2}  \\
\fl \nonumber
&\times& \bigg[ \frac{\FDJ{a}{1-a}{1}{\phi}}{(1+ux)^{1+a}(1+uy)^{1-a}}
- (1-a)(1+u) 
(1+uxy)\frac{\FDJ{1+a}{2-a}{2}{\phi}}{(1+ux)^{2+a}(1+uy)^{2-a}}\bigg].
\label{g22}
\end{eqnarray}
To shorten the notation, we have defined the reduced work and time variables $\eta=2\beta w$ and
$\tau=\Omega t$, with $\Omega = 2 \beta |b|$. The dimensionless combination $a=\nu/\Omega$ measures the degree of irreversibility of the process. We have also introduced the abbreviations $\widetilde{\tau} = \tau - \tau'$, $\widetilde{\eta} = \eta - \eta'$,
$\widetilde{\theta}=\theta(\widetilde{\tau}+\widetilde{\eta})\theta(\widetilde{\tau}-\widetilde{\eta})$, where
$\theta(z)$ is the unit step function,
$u=\exp(-2\beta h|b|/b-\Omega t')$,
$x=\exp\left[-(\widetilde{\tau}+\widetilde{\eta})/2\right]$,
$y=\exp\left[-(\widetilde{\tau}-\widetilde{\eta})/2\right]$, and
\begin{equation}
\label{abbrphi}
\phi=-u\,\frac{1-x}{1+ux}\,\frac{1-y}{1+uy}\,\,.
\end{equation} 
Finally, $\hyp(\alpha,\beta;\gamma;z)$ is the Gauss hypergeometric function \cite{Slater:1960}. The Green function for $b < 0$ follows after interchanging indexes $u$ and $d$ in Eqs.~\eref{g11}--\eref{g22}.  

For the engine, we assume the piecewise linear periodic protocol: 
\begin{eqnarray}
     E_{u}(t) = h_{1}+\frac{\displaystyle h_{2}-h_{1}}{\displaystyle t_c} t, \quad T(t) =T_c, \quad t\in[0,t_c], 
     \label{eq:k1}\\
      E_{u}(t) = h_{2}-\frac{\displaystyle h_{2}-h_{1}}{\displaystyle t_h}\,(t-t_c), \quad T(t) =T_h, \quad t\in [t_c,t_p),
\label{eq:k2}
\end{eqnarray}
and $E_{d}(t)=-E_{u}(t)$, where $t_p = t_c + t_h$. Invoking the Chapman-Kolmogorov equation, the Green function for work during the whole cycle reads 
\begin{equation}
G(w,t|w',t') =
\begin{cases} 
      G^c(w,t|w',t'), & t' < t < t_c,\\
\displaystyle  \int dw' G^h(w,t|w',t_c)
 G^c(w',t_c|0,0), & t'<t_c<t<t_p,\\
 G^h(w,t|w',t'), & t_c< t' < t < t_p,
   \end{cases}
\label{eq:GlinC}
\end{equation}
where $G^c$ and $G^h$ are Green functions for the first (cold) and second (hot) branch of the cycle,
respectively. Having the Green function for work, the PDFs for heat and work per cycle follow from Eqs.~\eref{eq:PDFwoutC} and \eref{eq:PDFwoutqinC} in Sec.~\ref{sec:WHC}. Similar approach can be applied for linear protocols composed of more branches.

The Green function~\eref{g11}--\eref{g22} at first glance possesses a bounded support, imposed by the unit step functions, and a singular part, proportional to the $\delta$ functions. In Sec.~\ref{sec:propertiesWPD}, we discuss these and other general properties of work PDFs for systems with discrete state space. The formulas~\eref{g11}--\eref{eq:GlinC} are applied in Figs.~\ref{fig:2levelSlow} and~\ref{fig:2levelFast} to illustrate these general properties. Namely, we show PDFs for work and heat after the individual branches. While the PDFs at the end of the cycle follow directly from Eqs.~\eref{eq:PDFwoutC} and \eref{eq:PDFwoutqinC}, those after the first branch are obtained using the formulas
\begin{align}
\rho_{w_{out}}(t_c) &= \left<+\right|
G(-w_{out},t_c|0,0)\, {\mathbf p}(0),\\
\rho_{q_{in}}(t_c) &= 
\int \dd w \sum_{mn} \delta\left\{q_{in} - \left[E_m(t_c) - E_n(0) - w\right]\right\}
G_{mn}(w,t_c|0,0)\,
p_n(0)
\end{align}
where ${\mathbf p}(t)$ denotes the state of the system during the limit cycle. Interestingly, also ${\mathbf p}(t)$ can be calculated from the propagator~\eref{eq:GlinC}. For example as ${\mathbf p}(t) =  \lim_{t'\to -\infty}\int \int dw\, dw'\, G(w,t|w',t') {\mathbf p}_0$, where ${\mathbf p}_0$ is an arbitrary normalized initial condition.


\subsection{Overdamped harmonic oscillator}
\label{sec:harmonic}

For continuous models, diffusion in parabolic potential depicted in Fig.~\ref{Fig:CyclicHE} plays an analogously paradigmatic role as the two-level systems for discrete ones. Indeed, a broad class of situations with weak enough noise involve stochastic dynamics near a single minimum of the (free) energy landscape that can be approximated by a parabolic potential. For example in  experiments, forces applied on single molecules, e.g., by optical tweezers, are often approximately linear in particle's displacement from the center of laser beam~\cite{Trepagnier/etal:2004, Carberry/etal:2004, Carberry/etal:2007, Andrieux/etal:2007, Khan/Sood:2011, Mestres/etal:2014}. 

There are two qualitatively distinct scenarios how the parabolic potential $U(x,t)$ can vary in time. In the so-called ``sliding parabola'' model, the position of the potential minimum $\lambda(t)$ changes with time, i.e.,  $U(x,t)=k[x-\lambda(t)]^2/2$. The so-called ``breathing parabola'' model then involves time-dependent stiffness $k(t)$ of the potential, 
\begin{equation}
\label{eq:parabola_breathing}
U(x,t)=\frac{k(t)}{2}x^2.
\end{equation} 
For both these situations, the Langevin equations \eref{eq:LExu}--\eref{eq:LExo} for velocity and position are linear. \textcolor{black}{The resulting processes $x(t)$} are thus linear functionals of the Gaussian white noise and their PDFs are also Gaussian~\cite{Risken1996}. The first two moments of work and heat can then be evaluated analytically using the approach suggested in Eqs.~\eref{eq:time_cor_fun_continuous} and
\eref{eq:work_second_moment}.

In the sliding parabola model, also the Langevin equation~\eref{eq:dw_cyclic} for work is linear and thus the work PDF is Gaussian~\cite{vanZon/Cohen:2003, vanZon/Cohen:2004, Cohen:2008, Nickelsen/Engel:2011, Subasi/Jarzynski:2013, Kim/etal:2014}. Unfortunately, the sliding parabola model can not be used as a basis of a heat engine that would provide a positive output power because the mean position $\left<x(t)\right>$, coupled to $\lambda(t)$ in the expressions for work~\eref{eq:dw_cyclic} and heat~\eref{eq:dq_cyclic}, is independent of temperature~\cite{Holubec2015}.

On the other hand, the breathing parabola model can be used as a basis of a heat engine but does not allow for a general solution for heat and work PDFs for an arbitrary stiffness $k(t)$. The task to find the Green function~\eref{eq:PDFworkCyclic} for the partial differential equation for work can be reduced to solving a Riccati equation~\cite{Ryabov/etal:2013}, or an equivalent problem of coupled ordinary differential equations~\cite{Speck:2011, Kwon/etal:2011, Kwon/etal:2013}. Exact solutions of these problems are known only for a few specific functions $k(t)$. Namely, for a piece-wise constant protocol, where $k(t)$ and $T(t)$ involve a single jump \cite{Kwon/etal:2011} or two jumps \cite{Holubec/Ryabov:2017, Holubec/Ryabov:2018}; in the slow-driving limit [$\dot k(t)$ small compared to the relaxation time of the system] where the work PDF is Gaussian for any $k(t)$~\cite{Speck:2011, Hoppenau/Engel:2013}; and if $k(t)$ is a rational function of time~\cite{Ryabov/etal:2013, Kwon/etal:2013}. In addition to exact approaches, theories were developed to predict asymptotics of work PDFs for small and large values of work~\cite{Engel:2009, Nickelsen/Engel:2011, Noh/etal:PRL2013, Ryabov/etal:2013, Holubec/etal:2015, Manikandan/Krishnamurthy:2017}. Also an Onsager-Machlup-type theory has been applied to obtain approximate solutions of the problem~\cite{Deza/etal:2009}. Finally, the method of approximating arbitrary protocol by a piece-wise constant driving, described in Sec.~\ref{sec:2level_PC}, has been for the breathing parabola model solved analytically for an arbitrary finite number of steps~\cite{Chvosta/etal:2020}. It thus allows to analytically approximate work PDF for an arbitrary protocol with an arbitrary precision. 

Most of the known solutions correspond to the overdamped limit~\eref{eq:overdamped} with the notable exception of Ref.~\cite{Kwon/etal:2013}. This is because the overdamped approximation offers a simpler description but also due to its high experimental relevance for Brownian heat engines~\cite{Blickle2012,Martinez2016,Martinez2017}. Let us now review the most frequently used results for the overdamped case. The next section is devoted to the derivation of PDFs for work and heat for an engine driven by a piece-wise constant protocol with two jumps. In Sec.~\ref{sec:W_parabolic}, we review derivation of the Green function for position, which allows evaluation of second moments of work and heat for an arbitrary driving.

\subsubsection{Piece-wise constant protocol.}
\label{sec:CHE}
\begin{figure}
    \centering
    \includegraphics[width=1.0\textwidth]{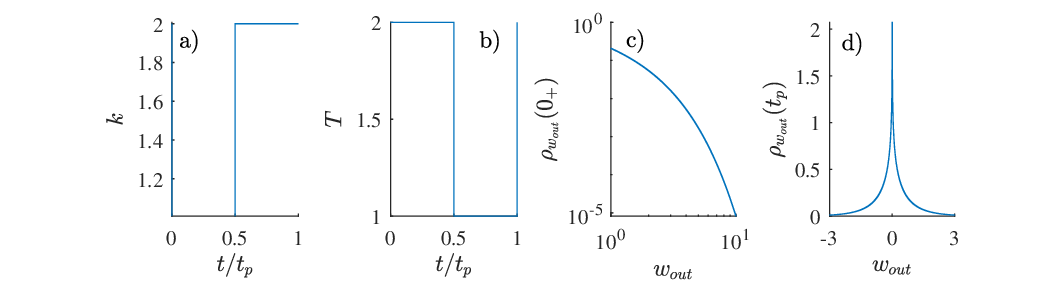}
    \caption{(a-b) Piece-wise constant protocol for stiffens $k$ and temperature $T$ for the Brownian heat engine based on the parabolic potential depicted in Fig.~\ref{Fig:CyclicHE} and discussed in Sec.~\ref{sec:CHE}. c) PDF for work done by the engine during the adiabatic expansion (first branch). d) PDF for work done per cycle.
    }
    \label{fig:critical}
\end{figure}

In this section, we provide analytical derivations of work, heat, and efficiency PDFs in the arguably most simple stochastic heat engine with continuous state-space. We consider an overdamped Brownian particle in a parabolic potential with the protocol composed of two quasi-static isothermal-isochoric branches interconnected by two infinitely fast adiabats~\cite{Holubec/etal:2017}. During the whole cycle, the state of the system is thus described by the Boltzmann-like PDF
\begin{equation}
    p(z,k/T) = \sqrt{\frac{k}{2 \pi T}} \exp\left(- \frac{k}{2 T} z^2\right),
    \label{eq:BoltzmannCritical}
\end{equation}
even though the fast adiabatic branches bring it far from equilibrium. An example of the protocol is depicted in Figs.~\ref{fig:critical}(a) and (b). In the discussion below, we will keep the depicted ordering of the individual branches: adiabatic expansion, hot isotherm, adiabatic compression, cold isotherm.

The engine performs work only during the adiabatic branches. During the adiabatic expansion, the system is in thermal equilibrium with reservoir at temperature $T_h$ and the stiffness abruptly decreases from $k_1$ to $k_2$. The branch is so fast that the system dwells in a fixed microstate, say $x$, and the corresponding stochastic work done equals to the decrease in potential energy $(k_1-k_2) x^2/2$. The PDF for position during the adiabatic branch is given by Eq.~\eref{eq:BoltzmannCritical} with $z=x$ and $k/T=k_1/T_h$ and thus the work PDF after the first branch reads
\begin{multline}
  \rho_{w_{out}}(0_+) =  \int dx\,\delta\left[w_{out} - (k_1-k_2) x^2/2\right] p(x,k_1/T_h)
    =\\ 
  \sqrt{\frac{\eta}{\pi T_h} \frac{1}{|w_{out}|}} \exp\left(-\frac{\eta}{T_h} w_{out}\right) \theta(w_{out}).
  \label{eq:criticalWPD0}
\end{multline}
The unit step function $\theta$ in the expression originates from the fact that decreasing the stiffness leads to an energy decrease regardless of $x$. Similarly, during the adiabatic compression the stiffness increases from $k_2$ to $k_1$ and the work done reads $(k_2-k_1) y^2/2$, where $y$ is the miscrostate occupied during the second adiabat. The total stochastic work per cycle 
\begin{equation}
    w_{out}(x,y) = \frac{1}{2}(k_1-k_2)(x^2 - y^2)
    \label{eq:workCHE}
\end{equation}
thus depends just on the microstates $x$ and $y$ occupied during the adiabatic branches. Due to the quasi-static isotherms mediating the two adiabats, the random variables $x$ and $y$ are independent and distributed according to the PDF~\eqref{eq:BoltzmannCritical} with $k/T = k_1/T_h$ for $z=x$ and $k_2/T_c$ for $z=y$. The PDF for the output work per cycle thus reads
\begin{multline}
    \rho_{w_{out}}(t_p) = \int dx\int dy\,\delta(w_{out} - w_{out}(x,y)) p(x,k_1/T_h) p(y,k_2/T_c)
    =\\ 
    \sqrt{\frac{k_1 k_2}{T_c T_h}} \frac{1}{\pi  (k_1-k_2)} \exp\left(-\frac{k_1 T_c -k_2 T_h }{k_1 - k_2}\frac{w}{2 T_c T_h}\right) K_0{\left(\frac{k_1 T_c+k_2 T_h }{k_1 -k_2} \frac{|w|}{2T_c T_h}\right)},
    \label{eq:eref}
\end{multline}
where $K_0$ denotes the modified Bessel function of the second kind. The PDFs for work after the first adiabat~\eqref{eq:criticalWPD0} and the whole cycle~\eqref{eq:eref} are plotted in Figs.~\ref{fig:critical}(c) and (d). The shown shapes of the PDFs are characteristic for piece-wise constant protocols with a single and two jumps, respectively.

The heat flows into the system during the hot isotherm. This branch is isochoric, no work is done, and the input heat is given just by the increase in the internal energy,
\begin{equation}
    q_{in}(x,y) = \frac{1}{2} k_1 (x^2 - y^2).
    \label{eq:heatCHE}
\end{equation}
The stochastic input heat is thus proportional to the stochastic output work and also their PDFs are proportional. Namely, $\rho_q(q_{in}) = \eta \rho_w(\eta q_{in})$, where $\eta = (k_1-k_2)/k_1 < \eta_C$ is the efficiency of the engine.

Since stochastic output work and input heat per cycle are proportional to each other, stochastic efficiency $\tilde{\eta} = w_{out}/q_{in}$ is in this model deterministic and given by $\eta$, i.e., $\rho_{\tilde{\eta}} = \delta(\tilde{\eta} - \eta)$. Therefore, the present model represents a simple example where the large deviation theory of stochastic efficiency does not apply. The reason is the failure of the large deviation principle: even though the work and heat by themselves obey the large deviation principle, the stochastic efficiency is independent of the number of cycles over which they are measured. The model yields a nontrivial PDF for $\tilde{\eta}$ if considered in the underdamped regime, where the stochastic work is still given by Eq.~\eqref{eq:workCHE} and the stochastic input heat~\eqref{eq:heatCHE} acquires another contribution $(p_x^2-p_y^2)/(2m)$ from momentum degrees of freedom. The PDFs for $p_x$ and $p_y$ are still of the form~\eqref{eq:BoltzmannCritical} with $k/T = 1/(m T_h)$ for $z=p_x$ and $k/T = 1/(m T_c)$ for $z=p_y$. The rest of the calculation follows along similar lines as in Eqs.~\eqref{eq:criticalWPD0} and \eqref{eq:eref} and we leave it as an exercise for an interested reader.

The PDF~\eref{eq:eref} yields all moments of $w_{out}$, $q_{in}$, and also the internal energy difference per cycle, $\Delta u = q_{in}$. Due to the close relation of work and heat to changes in internal energy in this model, it is not surprising that the second moment of work is determined by heat capacity of the working medium~\cite{Holubec/etal:2017}. This relation of fluctuations to heat capacity was used in Ref.~\cite{Holubec/etal:2017} as an argument for diverging work and power fluctuations in the critical heat engine proposed in Ref.~\cite{Campisi2016}. However, this negative effect can be circumvented by employing adiabatic branches where the system is not disconnected from the bath~\cite{Holubec/Ryabov:2018}.

The simplicity of the PDF~\eref{eq:eref} allows to directly demonstrate validity of the detailed fluctuation theorem~\eref{eq:detailFTHE} for heat engines for this model. Due to the proportionality of $w_{out}$, $q_{in}$, and $\Delta u$, the joint PDF is given by
\begin{equation}
    \rho(w_{out},q_{in},\Delta u) = 
    \delta(w_{out} - \eta \Delta u)
    \delta(w_{out} - \eta q_{in})
    \rho_{w_{out}}(t_p).
    \label{eq:PDF_FORWARD_CHE}
\end{equation}
While this might not be obvious at first glance, the used protocol is time-reversal symmetric. To see this, it is enough to realized that the reversed protocol can be plotted on top of the forward one after a suitable shift in time. We conclude that the joint PDF for the reversed process is again given by Eq.~\eref{eq:PDF_FORWARD_CHE}. This allows us to explicitly evaluate the ratio in Eq.~\eref{eq:detailFTHE}:
\begin{equation}
    \frac{\rho(w_{out},q_{in},\Delta u)}{\rho(-w_{out},-q_{in},-\Delta u)}
    = \exp\left(-\frac{k_1 T_c -k_2 T_h }{k_1 - k_2}\frac{w}{2 T_c T_h}\right)
    = \exp\left[\frac{1}{T_c} q_{in} (\eta_C-\eta)\right].
\end{equation}
The second equality follows from symmetry of the  $\delta$ function with respect to sign reversal and the third one is obtained after the substitution $w_{out} = \eta q_{in}$. 
    
\subsubsection{Arbitrary protocol.}
\label{sec:W_parabolic}
In the breathing parabola problem, one can calculate analytically the variance of output work for an arbitrary protocol for the stiffness and temperature on the basis of Eq.~\eref{eq:work_second_moment}. This is possible because the Green function for the Fokker-Planck equation 
\textcolor{black}{
\begin{equation}
\frac{\partial }{\partial t} p(x,t) 
=  \left[D \frac{\partial^2}{\partial x^2} + \frac{k(t)}{\gamma} \frac{\partial}{\partial x} x \right]  p(x,t),
\label{eq:GMEBP}
\end{equation}
describing the PDF of the} particle position, $x$,
can be obtained analytically~\cite{Risken1996}. It is given be the Gaussian PDF
\begin{equation}
    G(x,t|x',t') = \frac{1}{\sqrt{2\pi \sigma^2(t|x',t')}}
    \exp\left[- \frac{(x-m(t|x',t'))^2}{2\sigma^2(t|x',t')}\right],
\end{equation}
where $m(t|x',t') = \left<x(t)\right>$ and $\sigma^2(t|x',t') = \left<[x(t)]^2\right> - \left<x(t)\right>^2$ are average and variance of the particle position at time $t$ conditioned on its initial position $x'$ at time $t'$. These functions obey the ordinary differential equations
\begin{eqnarray}
\frac{d}{dt} m(t|x',t') &=& - \frac{1}{\gamma} k(t) m(t|x',t'),
\label{eq:m}\\
\frac{d}{dt} \sigma^2(t|x',t') &=& - \frac{2}{\gamma} k(t) \sigma^2(t|x',t') + \frac{2 k_B}{\gamma}T(t),
\label{eq:sigma}
\end{eqnarray}
with the initial conditions $m(t'|x',t') = x'$ and $\sigma(t'|x',t') = 0$, and the well-known solution reads 
\begin{eqnarray}
m(t|x',t') &=& x' \exp\left[-\gamma^{-1}\int_{t'}^t dt''\,k(t'')\right],\\
\sigma^2(t|x',t') &=&  2 k_B \gamma^{-1}
\int_{t'}^t dt'' T(t'') \exp\left[-2\gamma^{-1}\int_{t''}^t dt'''\,k(t''')\right].
\label{eq:sigma2maxeta}
\end{eqnarray}
The PDF for $x$ during the limit cycle then follows by $p(x,t) = \lim_{t'\to -\infty} G(x,t|x',t')$. The equations \eref{eq:m} and \eref{eq:sigma} are most easily derived from the Langevin equation~\eqref{eq:LExo} corresponding to the Fokker-Planck equation for $p(x,t)$.

In Fig.~\ref{fig:TURALL}, we show average power, power fluctuations, and efficiency calculated using this Green function for the protocol 
\begin{eqnarray}
      k(t)&=&\displaystyle{\frac{1}{\sigma_0^2}\frac{k_B T_h}{(1+b_h t)^2} - \frac{\gamma b_h}{1+b_h t}}, \quad T(t) = T_h,\quad t\in [0,t_h), \\
      k(t)&=&\displaystyle \frac{1}{\sigma_f^2}\frac{k_B T_c}{[1+b_c (t-t_h)]^2} - \frac{\gamma b_c}{1+b_c (t-t_h)},\quad T(t) = T_c \quad t\in [t_h,t_p),
      \label{eq:k}
\end{eqnarray}
that maximizes $W_{out}$ for fixed entropy change in the system during the hot isotherm, $\Delta S = k_B\log{\sigma_f/\sigma_0}$, and durations $t_h$ and $t_c$ of the isotherms~\cite{Schmiedl2007,Holubec2014}.  The durations of the adiabatic branches are assumed to be negligible compared to $t_c$ and $t_h$ and thus $t_p = t_c+t_h$. The parameters $\sigma_0^2$ and $\sigma_f^2$ stand for the variances of the particle position at the beginning and end of the hot isotherm, and $b_h = \left(\sigma_f/\sigma_0 - 1\right)/t_h$ and $b_c = \left(\sigma_0/\sigma_f - 1\right)/t_c$. In the figure, we plot the output power and its fluctuations as functions of the scaling parameter $\Omega$ defined by the equations
\begin{equation}
\sigma^2_f \propto \Omega^{-\xi}, \quad t_p \propto  \Omega^{(\chi - 1)\xi},
\label{eq:scaling}
\end{equation}
with $\sigma_f^2/\sigma_0^2 > 0$, $\xi>0$ and $\chi \in (0,1]$. In the limit $\Omega \to \infty$, the cyclic Brownian heat engine can provide diverging output power with limited fluctuations and reversible efficiency~\cite{Holubec/Ryabov:2018}. For steady-state heat engines, such a performance is forbidden by thermodynamic uncertainty relations as discussed in Sec.~\ref{sec:TUR}.

Note that in Eqs.~\eqref{eq:m}--\eqref{eq:k} we reintroduced the Boltzmann constant $k_B$. This is because we use in the figure the experimentally relevant parameters $\gamma \approx 1.89 \times 10^{-8}$ kg/s, $T_c = 293.15$ K and $T_h = 5273.15$ K~\cite{Holubec/Ryabov:2018}. Further, we took $\sigma^2_f = \Omega^{-\xi}$, $\sigma_f^2/\sigma_0^2 = 3$, $\xi = 3$, $\chi = 0.05$, $t_c=t_h=t_p/2$, and
\begin{equation}
    t_p = \frac{2\gamma^{1-\chi}}{T_h \eta_C \Delta S} 
    (\sigma_f - \sigma_0)^{2(1-\chi)}.
\end{equation}

\subsection{Solar cell as steady-state heat engine}
\label{sec:solar_cell}

\begin{figure}
\centerline{
     \begin{tikzpicture}[
      scale=0.5,
      level/.style={thick},
      transU/.style={thick,->,shorten >=2pt,shorten <=2pt,>=stealth},
      transD/.style={thick,<-,shorten >=2pt,shorten <=2pt,>=stealth},
    ]
    \draw[level,blue] (-8cm,-11em) -- (-12cm,-11em)  node[left] {$\mu_l$};
     \draw[level,black] (-4cm,-11em) -- (0cm,-11em) node[right] {$\left|d\right>$};
    \draw[level] (2cm,2em) -- (-2cm,2em) node[left] {$\left|u\right>$};
    \draw[level,blue] (6cm,2em) -- (10cm,2em) node[right] {$\mu_r$};
    \draw[transU,red,dashed] (-3.2cm,-11em) -- (-1.2cm,2em) node[midway,left] {\textcolor{red}{$L_{ud}^h$} \textcolor{black}{+} \textcolor{black}{$L_{ud}^c $}};
    \draw[transU,blue] (-2.8cm,-11em) -- (-0.8cm,2em);
    \draw[transD,red,dashed] (-1.2cm,-11em) -- (0.8cm,2em);
    \draw[transD,blue] (-0.8cm,-11em) -- (1.2cm,2em) node[midway,right] {\textcolor{red}{$L_{du}^h$} \textcolor{black}{+} \textcolor{black}{$L_{du}^c $}};
    \draw[transU,blue] (2.5cm,2.3em) -- (5.5cm,2.3em) node[midway,above] {$L_{ru}$};
    \draw[transD,blue] (2.5cm,1.7em) -- (5.5cm,1.7em) node[midway,below] {$L_{ur}$};
    \draw[transU,blue] (-7.5cm,-10.7em) -- (-4.5cm,-10.7em) node[midway,above] {$L_{dl}$};
    \draw[transD,blue] (-7.5cm,-11.3em) -- (-4.5cm,-11.3em) node[midway,below] {$L_{ld}$};
    \end{tikzpicture}
}
\caption{A simple model of solar cell: the two-level quantum dot (black energy levels in the middle) is connected to two cold leads at different chemical potentials $\mu_l<\mu_r$ (blue energy levels). The transitions of electrons between the leads and the dot are induced solely by thermal fluctuations of the substrate (solid blue transitions) \textcolor{black}{and described by the transition rates $L_{ij}$}. When occupied, the quantum dot can be excited/de-excited both by fluctuations from the substrate (solid blue transitions \textcolor{black}{with the rates $L_{ij}^c$}) and by photons arriving from the hot sun (red dashed transitions \textcolor{black}{with the rates $L_{ij}^h$}). The cell performs work by pumping electrons against the gradient $\mu_r-\mu_l$ of chemical potential.}
\label{Fig:colarCell}

\end{figure}
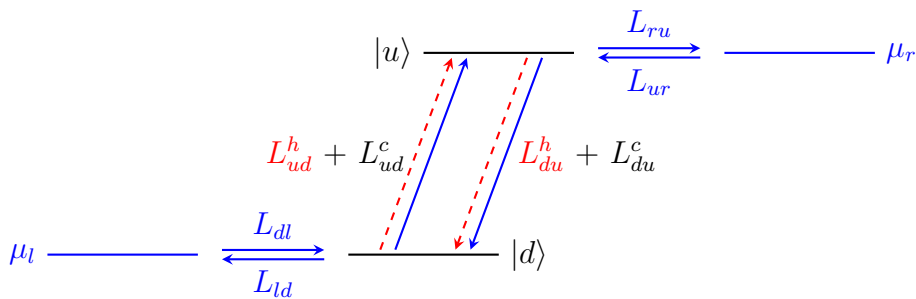

In this section, we review the derivation of power and its fluctuations for the simple model of solar cell depicted in Fig.~\ref{Fig:colarCell} and used in Sec.~\ref{sec:TUR} to demonstrate limitations imposed on the performance of steady-state heat engines by the thermodynamic uncertainty relations.

The considered solar cell has three microstates. Denoting them as 0 (no electron in quantum dot), $d$ (electron in lower state of the dot), and  $u$ (excited dot), the rate matrix in the GME~\eqref{eq:GME} for the probability vector ${\mathbf p}(t) = [p_0(t), p_d(t), p_u(t)]^\intercal$ reads
\begin{equation}
L=  \begin{pmatrix}
- L_{dl} - L_{ur} & L_{ld}  & L_{ru}\\
 L_{dl} & - L_{ld} -  L^c_{ud} - L^h_{ud} &  L^c_{du} + L^h_{du}\\
L_{ur} & L^c_{ud} + L^h_{ud} & - L_{ru} -  L^c_{du} - L^h_{du}
\end{pmatrix}.
\label{eq:solar_cell}
\end{equation}
For Fig.~\ref{fig:TURALL}, we employ the frequently used transition rates following from the Fermi-Dirac and Bose-Einstein statistics for electrons and phonons, respectively~\cite{Pietzonka/Seifert:2018}:
\begin{equation}
L_{sr} = \Gamma/[1 + \exp(x_{sr})],\quad  L_{rs} = \Gamma/[1 + \exp(-x_{sr})],
\end{equation}
with $x_{rs} = (E_s - \mu_r)/T_c$, $r = l,r$, $s = u,d$, and
\begin{equation}
L^r_{ud} = \Gamma_r/[\exp(x_{r})-1], \quad L^r_{du} = \Gamma_r/[1 - \exp(-x_{r})],
\end{equation}
with $x_{r} = (E_u - E_d)/T_r$, $r = c,h$. Further, $\Gamma = \Gamma_h = \Omega \Gamma_0$ and $\Gamma_c = \Omega^{-3/2}\Gamma_0$. All the four pairs of rates satisfy the local detailed balance condition~\eref{eq:detailed_balance}. 

In order to evaluate the average power of the solar cell and its fluctuations measured by the constancy~\eref{eq:constancy}, it is useful to construct the tilted rate matrix for the output work and use the large deviation result~\eref{eq:LDF_help}.  If the solar cell transports a single electron from the low to the high chemical potential, it performs work $\Delta \mu = \mu_r - \mu_l$. The work done during the reversed process is $-\Delta \mu$. To construct the tilted rate matrix for work, we thus need to multiply the rate $L_{ru}$ by $\exp(-s_w \Delta\mu)$, the rate $L_{ld}$ by $\exp(s_w \Delta\mu)$, and keep the remaining transition rates unchanged. The resulting tilted rate matrix reads
\begin{equation}
L(s_w) =  \begin{pmatrix}
- L_{dl} - L_{ur} & \exp(s_w \Delta\mu)L_{ld}  & \exp(-s_w \Delta\mu) L_{ru}\\
 L_{dl} & - L_{ld} -  L^c_{ud} - L^h_{ud} &  L^c_{du} + L^h_{du}\\
L_{ur} & L^c_{ud} + L^h_{ud} & - L_{ru} -  L^c_{du} - L^h_{du}
\end{pmatrix}.
\label{eq:solar_cell2}
\end{equation}
Its largest eigenvalue $\lambda_{max}(s_w)$ then determines the scaled cumulant generating function,
\begin{equation}
  \lambda_{max}(s_w) \approx   \frac{1}{t} \log \chi(s_w),
\end{equation}
through Eq.~\eref{eq:LDF_help}. Finally, Eq.~\eref{eq:raw_moments_qin} implies that the average power and constancy can be obtained as
\begin{equation}
    P = -\left.\frac{\partial}{\partial s_w} \lambda_{max}(s_w)\right|_{s_w=0},\quad \Delta_P = 
    \left.\frac{\partial^2}{\partial s_w^2} \lambda_{max}(s_w)\right|_{s_w=0}.
\end{equation}
These formulas can be easily evaluated numerically. Nevertheless, the present model allows to find explicit expressions~\cite{Pietzonka/Seifert:2018} following from the results of Ref.~\cite{Koza1999}.

\section{Concluding remarks} 

Processes of converting the `disordered' thermal energy called heat into an `ordered' work are of vital importance for our society and the whole biosphere. At the macroscale, power output and efficiency of heat engines are universally limited by laws of classical thermodynamics. At the microscale, the second law can be formulated in terms of symmetry relations for fluctuations of entropy and its implications are far from being completely understood. Here, we have reviewed recent findings revealing how and to what extent these fundamental symmetries control fluctuations in performance of microscopic heat engines. Furthermore, we have provided a general overview of theoretical tools of stochastic thermodynamics and discussed basic phenomenology of probability density functions for work and heat.

The review is devoted to small systems subjected to the stochastic Markovian dynamics in contact with one or more equilibrium heat reservoirs, where thermodynamically consistent definitions of work, heat, and entropy for the corresponding individual stochastic trajectories are well established. Applying them to microscopic heat engines, one should distinguish between two situations. Steady-state heat engines operate under time-independent non-equilibrium conditions in contact with several heat reservoirs and deliver a stochastic output work proportional to excitation currents. On the other hand, cyclic heat engines are driven by periodic variation of their energy spectrum and the bath temperature and operate in a time-periodic steady state. Their stochastic output work is determined by the time spent in the individual microstates during the cycle. Mean values of output works for steady-state and cyclic heat engines obey the same limitations imposed by the second law. However, their fluctuations exhibit strikingly different statistics. 

Besides the corresponding probability densities, these differences are revealed in trade-off relations for efficiency, power, and power fluctuations of heat engines derived from thermodynamic uncertainty relations. For steady-state heat engines, they imply that the more deterministic the power, the smaller the efficiency. In contrast, cyclic heat engines can operate arbitrarily close to the Carnot efficiency and deliver a positive and stable output power. 

Unlike the averages of stochastic work $w_{out}$ and heat $q_{in}$, mean value of the stochastic efficiency $\tilde{\eta} = w_{out}/q_{in}$ is not bounded by the second law. In fact, its probability density usually exhibits heavy tails and $\left<\tilde{\eta}\right>$ frequently diverges. More insights into the behavior of stochastic efficiency can be obtained by employing the fluctuation theorem for entropy production that generalizes the second law to individual stochastic trajectories. It implies the fluctuation theorem for heat engines that analogously generalizes the Carnot efficiency and reveals interesting features of statistics of the stochastic efficiency. Namely, if evaluated from heat and work measured over extended periods of time, the most likely value of $\tilde{\eta}$ equals to the standard definition of efficiency $\eta = \left<w_{out}\right>/\left<q_{in}\right>$. Furthermore, the rate function for stochastic efficiency, describing this behavior, attains a special value when $\tilde{\eta}$ equals the Carnot efficiency. For time-reversal symmetric protocol, it corresponds to the least likely efficiency and, for time-asymmetric protocols, to the efficiency which is equally likely for forward and time-reversed driving.

In addition to general derivations, all these fascinating results can be intuitively understood, tested, and their assumptions violated using paradigmatic exactly solvable models of stochastic thermodynamics: driven two-level system and harmonic oscillator. Besides being analytically tractable, these models are also of vital experimental importance as they represent good approximations to the most frequently used experimental setups.

Despite the amount and diversity of the presented results, we believe that there is still much more to come. In particular, the thermodynamic uncertainty relations valid for steady-state heat engines can be applied to an arbitrary transport process not necessarily driven by a temperature gradient. While these machines often cannot be categorized as heat engines, the thermodynamic uncertainty relations should still limit their performance in terms of transport coefficients, their fluctuations, and transport efficiency. Of particular interest are processes driven by information~\cite{Lutz/Ciliberto:2015,Parrondo/etal:2015,Rosinberg/Horowitz:EPL:2016,Paneru/Pak:APX:2020,Paneru/etal:NatCommun:2020} and transport in systems of many interacting particles~\cite{Lips2018}. \textcolor{black}{Another twist in the theory is expected for systems in contact with non-equilibrium reservoirs, sometimes referred to as active heat engines, where the very definitions of work and heat are still subjects of keen discussions~\cite{Krishnamurthy2016, Pietzonka2019,Holubec2020,Holubec2020a,Ekeh2020,Gronchi2021,Fodor_2021,Zakine/etal:Entropy2019, Kumari/etal:PRE2020, Gronchi2021} and thus the results valid for cyclic and steady-state heat engines operating with equilibrium reservoirs cannot be taken for granted.} Fluctuations in coherent open quantum systems have a similar status~\cite{breuer2002theory} as they break validity of most of the classical results~\cite{Agarwalla2018,Halpern2019,Miller2019,Scandi2020,Denzler2021}. Finally, it is not unlikely that other general relations similar to the thermodynamic uncertainty relations will be established and provide new insights into the fundamental laws governing performance of microscopic machines. \textcolor{black}{An example is the article~\cite{saryal2021universal} deriving a new type of inequalities for fluctuations of steady-state currents based on the linear response theory, which appeared shortly after the submission of this review.}

\ack 
We are grateful to our mentor Petr Chvosta for introducing us to the secrets of stochastic modeling in physics. We also thank Zhuolin Ye for commenting on an earlier version of the manuscript.
Financial support by the Czech Science Foundation (Project
No.\ 20-02955J) is gratefully acknowledged. VH also thanks for support by the Humboldt foundation. 

\section*{References}
\bibliographystyle{naturemag}
\bibliography{references_HEreview}

\begin{thebibliography}{100}
\expandafter\ifx\csname url\endcsname\relax
  \def\url#1{\texttt{#1}}\fi
\expandafter\ifx\csname urlprefix\endcsname\relax\def\urlprefix{URL }\fi
\providecommand{\bibinfo}[2]{#2}
\providecommand{\eprint}[2][]{\url{#2}}

\bibitem{Carnot:1824}
\bibinfo{author}{Carnot, S.}
\newblock \emph{\bibinfo{title}{R{\' e}flexions sur la puissance motrice du feu
  et sur les machines propres {\` a} d{\' e}velopper cette puissance}}
  (\bibinfo{publisher}{Bachelier}, \bibinfo{address}{Paris},
  \bibinfo{year}{1824}).

\bibitem{Clausius:1850}
\bibinfo{author}{Clausius, R.}
\newblock \bibinfo{title}{Ueber die bewegende {Kraft} der {W{\" a}rme} und die
  {Gesetze}, welche sich daraus f{\" u}r die {W{\" a}rmelehre} selbst ableiten
  lassen}.
\newblock \emph{\bibinfo{journal}{Ann. Phys.}} \textbf{\bibinfo{volume}{155}},
  \bibinfo{pages}{500--524} (\bibinfo{year}{1850}).
\newblock \urlprefix\url{https://doi.org/10.1002/andp.18501550403}.

\bibitem{Clausius:1854}
\bibinfo{author}{Clausius, R.}
\newblock \bibinfo{title}{Ueber eine ver{\" a}nderte {Form} des zweiten
  {Hauptsatzes} der mechanischen {W{\" a}rmetheorie}}.
\newblock \emph{\bibinfo{journal}{Ann. Phys.}} \textbf{\bibinfo{volume}{169}},
  \bibinfo{pages}{481--506} (\bibinfo{year}{1854}).
\newblock \urlprefix\url{https://doi.org/10.1002/andp.18541691202}.

\bibitem{Thomson:1852}
\bibinfo{author}{Thomson, W.}
\newblock \bibinfo{title}{Ii. on the dynamical theory of heat, with numerical
  results deduced from {Mr. J}oule's equivalent of a thermal unit, and {M.
  R}egnault's observations on steam}.
\newblock \emph{\bibinfo{journal}{The London, Edinburgh, and Dublin
  Philosophical Magazine and Journal of Science}} \textbf{\bibinfo{volume}{4}},
  \bibinfo{pages}{8--21} (\bibinfo{year}{1852}).
\newblock \urlprefix\url{https://doi.org/10.1080/14786445208647064}.

\bibitem{Kubo:book1968}
\bibinfo{author}{Kubo, R.}
\newblock \emph{\bibinfo{title}{Thermodynamics: An advanced course with
  problems and solutions}} (\bibinfo{publisher}{North-Holland Publishing
  Company}, \bibinfo{address}{Amsterdam}, \bibinfo{year}{1968}).
\newblock \bibinfo{note}{In cooperation with Hiroshi Ichimura, Tsunemaru Usui,
  Natsuki Hashitsume}.

\bibitem{callen1998thermodynamics}
\bibinfo{author}{Callen, H.~B.}
\newblock \emph{\bibinfo{title}{Thermodynamics and an Introduction to
  Thermostatistics}} (\bibinfo{publisher}{John Wiley \& Sons},
  \bibinfo{address}{New York}, \bibinfo{year}{1985}).

\bibitem{Holubec/Ryabov:2018}
\bibinfo{author}{Holubec, V.} \& \bibinfo{author}{Ryabov, A.}
\newblock \bibinfo{title}{Cycling tames power fluctuations near optimum
  efficiency}.
\newblock \emph{\bibinfo{journal}{Phys. Rev. Lett.}}
  \textbf{\bibinfo{volume}{121}}, \bibinfo{pages}{120601}
  (\bibinfo{year}{2018}).
\newblock \urlprefix\url{https://doi.org/10.1103/PhysRevLett.121.120601}.

\bibitem{Seifert:2012}
\bibinfo{author}{Seifert, U.}
\newblock \bibinfo{title}{Stochastic thermodynamics, fluctuation theorems and
  molecular machines}.
\newblock \emph{\bibinfo{journal}{Rep. Prog. Phys.}}
  \textbf{\bibinfo{volume}{75}}, \bibinfo{pages}{126001}
  (\bibinfo{year}{2012}).
\newblock \urlprefix\url{https://doi.org/10.1088/0034-4885/75/12/126001}.

\bibitem{Chvosta/etal:2010a}
\bibinfo{author}{Chvosta, P.}, \bibinfo{author}{Holubec, V.},
  \bibinfo{author}{Ryabov, A.}, \bibinfo{author}{Einax, M.} \&
  \bibinfo{author}{Maass, P.}
\newblock \bibinfo{title}{Thermodynamics of two-stroke engine based on
  periodically driven two-level system}.
\newblock \emph{\bibinfo{journal}{Physica E}} \textbf{\bibinfo{volume}{42}},
  \bibinfo{pages}{472--476} (\bibinfo{year}{2010}).
\newblock \urlprefix\url{https://doi.org/10.1016/j.physe.2009.06.031}.

\bibitem{Schmiedl2007}
\bibinfo{author}{Schmiedl, T.} \& \bibinfo{author}{Seifert, U.}
\newblock \bibinfo{title}{Efficiency at maximum power: An analytically solvable
  model for stochastic heat engines}.
\newblock \emph{\bibinfo{journal}{{EPL} (Europhys. Lett.)}}
  \textbf{\bibinfo{volume}{81}}, \bibinfo{pages}{20003} (\bibinfo{year}{2007}).
\newblock \urlprefix\url{https://doi.org/10.1209/0295-5075/81/20003}.

\bibitem{CalvoHernandez/etal:2015}
\bibinfo{author}{Calvo~Hern{\'a}ndez, A.}, \bibinfo{author}{Roco, J. M.~M.},
  \bibinfo{author}{Medina, A.}, \bibinfo{author}{Velasco, S.} \&
  \bibinfo{author}{Guzm{\'a}n-Vargas, L.}
\newblock \bibinfo{title}{The maximum power efficiency 1-{\textsurd}$\tau$:
  Research, education, and bibliometric relevance}.
\newblock \emph{\bibinfo{journal}{Eur. Phys. J. Special Topics}}
  \textbf{\bibinfo{volume}{224}}, \bibinfo{pages}{809--823}
  (\bibinfo{year}{2015}).
\newblock \urlprefix\url{https://doi.org/10.1140/epjst/e2015-02429-4}.

\bibitem{Tu:2021}
\bibinfo{author}{Tu, Z.-C.}
\newblock \bibinfo{title}{Abstract models for heat engines}.
\newblock \emph{\bibinfo{journal}{Front. Phys.}} \textbf{\bibinfo{volume}{16}},
  \bibinfo{pages}{33202} (\bibinfo{year}{2021}).
\newblock \urlprefix\url{https://doi.org/10.1007/s11467-020-1029-6}.

\bibitem{Sekimoto:2010}
\bibinfo{author}{Sekimoto, K.}
\newblock \emph{\bibinfo{title}{Stochastic energetics}}.
\newblock Lecture notes in physics, 799 (\bibinfo{publisher}{Springer},
  \bibinfo{address}{Berlin}, \bibinfo{year}{2010}).
\newblock \urlprefix\url{https://doi.org/10.1007/978-3-642-05411-2}.

\bibitem{Evans/Searles:2002}
\bibinfo{author}{Evans, D.~J.} \& \bibinfo{author}{Searles, D.~J.}
\newblock \bibinfo{title}{The fluctuation theorem}.
\newblock \emph{\bibinfo{journal}{Adv. Phys.}} \textbf{\bibinfo{volume}{51}},
  \bibinfo{pages}{1529--1585} (\bibinfo{year}{2002}).
\newblock \urlprefix\url{https://doi.org/10.1080/00018730210155133}.

\bibitem{Harris/Schuetz:2007}
\bibinfo{author}{Harris, R.~J.} \& \bibinfo{author}{Sch{\"u}tz, G.~M.}
\newblock \bibinfo{title}{Fluctuation theorems for stochastic dynamics}.
\newblock \emph{\bibinfo{journal}{J. Stat. Mech.}}
  \textbf{\bibinfo{volume}{2007}}, \bibinfo{pages}{P07020--P07020}
  (\bibinfo{year}{2007}).
\newblock \urlprefix\url{https://doi.org/10.1088/1742-5468/2007/07/p07020}.

\bibitem{Kurchan:2007}
\bibinfo{author}{Kurchan, J.}
\newblock \bibinfo{title}{Non-equilibrium work relations}.
\newblock \emph{\bibinfo{journal}{J. Stat. Mech.}}
  \textbf{\bibinfo{volume}{2007}}, \bibinfo{pages}{P07005--P07005}
  (\bibinfo{year}{2007}).
\newblock \urlprefix\url{https://doi.org/10.1088/1742-5468/2007/07/p07005}.

\bibitem{Jarzynski:2011}
\bibinfo{author}{Jarzynski, C.}
\newblock \bibinfo{title}{Equalities and inequalities: Irreversibility and the
  second law of thermodynamics at the nanoscale}.
\newblock \emph{\bibinfo{journal}{Annu. Rev. Condens. Matter Phys.}}
  \textbf{\bibinfo{volume}{2}}, \bibinfo{pages}{329--351}
  (\bibinfo{year}{2011}).
\newblock
  \urlprefix\url{https://doi.org/10.1146/annurev-conmatphys-062910-140506}.

\bibitem{Bochkov/Kuzovlev:2013}
\bibinfo{author}{Bochkov, G.~N.} \& \bibinfo{author}{Kuzovlev, Y.~E.}
\newblock \bibinfo{title}{Fluctuation{\textendash}dissipation relations.
  {A}chievements and misunderstandings}.
\newblock \emph{\bibinfo{journal}{Phys.-Usp.}} \textbf{\bibinfo{volume}{56}},
  \bibinfo{pages}{590--602} (\bibinfo{year}{2013}).
\newblock \urlprefix\url{https://doi.org/10.3367/ufne.0183.201306d.0617}.

\bibitem{Marsland/England:2018}
\bibinfo{author}{Marsland, R.} \& \bibinfo{author}{England, J.}
\newblock \bibinfo{title}{Limits of predictions in thermodynamic systems: A
  review}.
\newblock \emph{\bibinfo{journal}{Rep. Prog. Phys.}}
  \textbf{\bibinfo{volume}{81}}, \bibinfo{pages}{016601}
  (\bibinfo{year}{2018}).
\newblock \urlprefix\url{https://doi.org/10.1088/1361-6633/aa9101}.

\bibitem{Seifert:2019}
\bibinfo{author}{Seifert, U.}
\newblock \bibinfo{title}{From stochastic thermodynamics to thermodynamic
  inference}.
\newblock \emph{\bibinfo{journal}{Ann. Rev. Cond. Matt. Phys.}}
  \textbf{\bibinfo{volume}{10}}, \bibinfo{pages}{171--192}
  (\bibinfo{year}{2019}).
\newblock
  \urlprefix\url{https://doi.org/10.1146/annurev-conmatphys-031218-013554}.

\bibitem{Horowitz/Gingrich:2020}
\bibinfo{author}{Horowitz, J.~M.} \& \bibinfo{author}{Gingrich, T.~R.}
\newblock \bibinfo{title}{Thermodynamic uncertainty relations constrain
  non-equilibrium fluctuations}.
\newblock \emph{\bibinfo{journal}{Nat. Phys.}} \textbf{\bibinfo{volume}{16}},
  \bibinfo{pages}{15--20} (\bibinfo{year}{2020}).
\newblock \urlprefix\url{https://doi.org/10.1038/s41567-019-0702-6}.

\bibitem{Parrondo/etal:2015}
\bibinfo{author}{Parrondo, J. M.~R.}, \bibinfo{author}{Horowitz, J.~M.} \&
  \bibinfo{author}{Sagawa, T.}
\newblock \bibinfo{title}{Thermodynamics of information}.
\newblock \emph{\bibinfo{journal}{Nat. Phys.}} \textbf{\bibinfo{volume}{11}},
  \bibinfo{pages}{131--139} (\bibinfo{year}{2015}).
\newblock \urlprefix\url{https://doi.org/10.1038/nphys3230}.

\bibitem{Lutz/Ciliberto:2015}
\bibinfo{author}{Lutz, E.} \& \bibinfo{author}{Ciliberto, S.}
\newblock \bibinfo{title}{Information: From {M}axwell’s demon to
  {L}andauer’s eraser}.
\newblock \emph{\bibinfo{journal}{Phys. Today}} \textbf{\bibinfo{volume}{68}},
  \bibinfo{pages}{30--35} (\bibinfo{year}{2015}).
\newblock \urlprefix\url{https://doi.org/10.1063/PT.3.2912}.

\bibitem{Bustamante/etal:2005}
\bibinfo{author}{Bustamante, C.}, \bibinfo{author}{Liphardt, J.} \&
  \bibinfo{author}{Ritort, F.}
\newblock \bibinfo{title}{The nonequilibrium thermodynamics of small systems}.
\newblock \emph{\bibinfo{journal}{Phys. Today}} \textbf{\bibinfo{volume}{58}},
  \bibinfo{pages}{43--48} (\bibinfo{year}{2005}).
\newblock \urlprefix\url{https://doi.org/10.1063/1.2012462}.

\bibitem{Ritort:2006}
\bibinfo{author}{Ritort, F.}
\newblock \bibinfo{title}{Single-molecule experiments in biological physics:
  Methods and applications}.
\newblock \emph{\bibinfo{journal}{J. Phys. Condens. Matt.}}
  \textbf{\bibinfo{volume}{18}}, \bibinfo{pages}{R531--R583}
  (\bibinfo{year}{2006}).
\newblock \urlprefix\url{https://doi.org/10.1088/0953-8984/18/32/r01}.

\bibitem{Ciliberto/etal:2013}
\bibinfo{author}{Ciliberto, S.}, \bibinfo{author}{Gomez-Solano, R.} \&
  \bibinfo{author}{Petrosyan, A.}
\newblock \bibinfo{title}{Fluctuations, linear response, and currents in
  out-of-equilibrium systems}.
\newblock \emph{\bibinfo{journal}{Annu. Rev. Condens. Matter Phys.}}
  \textbf{\bibinfo{volume}{4}}, \bibinfo{pages}{235--261}
  (\bibinfo{year}{2013}).
\newblock
  \urlprefix\url{https://doi.org/10.1146/annurev-conmatphys-030212-184240}.

\bibitem{Ciliberto:2017}
\bibinfo{author}{Ciliberto, S.}
\newblock \bibinfo{title}{Experiments in stochastic thermodynamics: Short
  history and perspectives}.
\newblock \emph{\bibinfo{journal}{Phys. Rev. X}} \textbf{\bibinfo{volume}{7}},
  \bibinfo{pages}{021051} (\bibinfo{year}{2017}).
\newblock \urlprefix\url{https://link.aps.org/doi/10.1103/PhysRevX.7.021051}.

\bibitem{Benenti/etal:2017}
\bibinfo{author}{Benenti, G.}, \bibinfo{author}{Casati, G.},
  \bibinfo{author}{Saito, K.} \& \bibinfo{author}{Whitney, R.~S.}
\newblock \bibinfo{title}{Fundamental aspects of steady-state conversion of
  heat to work at the nanoscale}.
\newblock \emph{\bibinfo{journal}{Phys. Rep.}} \textbf{\bibinfo{volume}{694}},
  \bibinfo{pages}{1--124} (\bibinfo{year}{2017}).
\newblock \urlprefix\url{https://doi.org/10.1016/j.physrep.2017.05.008}.

\bibitem{Gillespie:1992}
\bibinfo{author}{Gillespie, D.~T.}
\newblock \emph{\bibinfo{title}{Markov Processes}}
  (\bibinfo{publisher}{Academic Press}, \bibinfo{address}{San Diego},
  \bibinfo{year}{1992}).

\bibitem{vanKampen2007}
\bibinfo{author}{{Van Kampen}, N.~G.}
\newblock \emph{\bibinfo{title}{Stochastic processes in physics and chemistry}}
  (\bibinfo{publisher}{Elsevier}, \bibinfo{address}{Amsterdam},
  \bibinfo{year}{2007}).

\bibitem{breuer2002theory}
\bibinfo{author}{Breuer, H.-P.}, \bibinfo{author}{Petruccione, F.}
  \emph{et~al.}
\newblock \emph{\bibinfo{title}{The theory of open quantum systems}}
  (\bibinfo{publisher}{Oxford University Press}, \bibinfo{address}{Oxford},
  \bibinfo{year}{2002}).

\bibitem{Maes2003}
\bibinfo{author}{Maes, C.} \& \bibinfo{author}{Neto{\v{c}}n{\'y}, K.}
\newblock \bibinfo{title}{Time-reversal and entropy}.
\newblock \emph{\bibinfo{journal}{J. Stat. Phys.}}
  \textbf{\bibinfo{volume}{110}}, \bibinfo{pages}{269--310}
  (\bibinfo{year}{2003}).
\newblock \urlprefix\url{https://doi.org/10.1023/A:1021026930129}.

\bibitem{Holubec2011}
\bibinfo{author}{Holubec, V.}, \bibinfo{author}{Chvosta, P.},
  \bibinfo{author}{Einax, M.} \& \bibinfo{author}{Maass, P.}
\newblock \bibinfo{title}{Attempt time {M}onte {C}arlo: An alternative for
  simulation of stochastic jump processes with time-dependent transition
  rates}.
\newblock \emph{\bibinfo{journal}{{EPL} (Europhys. Lett.)}}
  \textbf{\bibinfo{volume}{93}}, \bibinfo{pages}{40003} (\bibinfo{year}{2011}).
\newblock \urlprefix\url{https://doi.org/10.1209/0295-5075/93/40003}.

\bibitem{Risken1996}
\bibinfo{author}{Risken, H.}
\newblock \emph{\bibinfo{title}{The Fokker-Planck Equation}}
  (\bibinfo{publisher}{Springer}, \bibinfo{address}{Berlin, Heidelberg},
  \bibinfo{year}{1996}).

\bibitem{Blickle2012}
\bibinfo{author}{Blickle, V.} \& \bibinfo{author}{Bechinger, C.}
\newblock \bibinfo{title}{Realization of a micrometre-sized stochastic
  heat engine}.
\newblock \emph{\bibinfo{journal}{Nat. Phys.}} \textbf{\bibinfo{volume}{8}},
  \bibinfo{pages}{143--146} (\bibinfo{year}{2012}).
\newblock \urlprefix\url{https://doi.org/10.1038/nphys2163}.

\bibitem{Martinez2016}
\bibinfo{author}{Mart{\'i}nez, I.~A.} \emph{et~al.}
\newblock \bibinfo{title}{Brownian {C}arnot engine}.
\newblock \emph{\bibinfo{journal}{Nat. Phys.}} \textbf{\bibinfo{volume}{12}},
  \bibinfo{pages}{67--70} (\bibinfo{year}{2016}).
\newblock \urlprefix\url{https://doi.org/10.1038/nphys3518}.

\bibitem{Martinez2017}
\bibinfo{author}{Mart{\,i}nez, I.~A.}, \bibinfo{author}{Rold{\,a}n, {\, E}.},
  \bibinfo{author}{Dinis, L.} \& \bibinfo{author}{Rica, R.~A.}
\newblock \bibinfo{title}{Colloidal heat engines: A review}.
\newblock \emph{\bibinfo{journal}{Soft Matter}} \textbf{\bibinfo{volume}{13}},
  \bibinfo{pages}{22--36} (\bibinfo{year}{2017}).
\newblock \urlprefix\url{http://dx.doi.org/10.1039/C6SM00923A}.

\bibitem{Speck:2011}
\bibinfo{author}{Speck, T.}
\newblock \bibinfo{title}{Work distribution for the driven harmonic oscillator
  with time-dependent strength: Exact solution and slow driving}.
\newblock \emph{\bibinfo{journal}{J. Phys. A}} \textbf{\bibinfo{volume}{44}},
  \bibinfo{pages}{305001} (\bibinfo{year}{2011}).
\newblock \urlprefix\url{http://stacks.iop.org/1751-8121/44/i=30/a=305001}.

\bibitem{Holubec2014}
\bibinfo{author}{Holubec, V.}
\newblock \bibinfo{title}{An exactly solvable model of a stochastic heat
  engine: optimization of power, power fluctuations and efficiency}.
\newblock \emph{\bibinfo{journal}{J. Stat. Mech.}}
  \textbf{\bibinfo{volume}{2014}}, \bibinfo{pages}{P05022}
  (\bibinfo{year}{2014}).
\newblock \urlprefix\url{https://doi.org/10.1088/1742-5468/2014/05/p05022}.

\bibitem{Holubec2015}
\bibinfo{author}{Holubec, V.} \& \bibinfo{author}{Ryabov, A.}
\newblock \bibinfo{title}{Efficiency at and near maximum power of
  low-dissipation heat engines}.
\newblock \emph{\bibinfo{journal}{Phys. Rev. E}} \textbf{\bibinfo{volume}{92}},
  \bibinfo{pages}{052125} (\bibinfo{year}{2015}).
\newblock \urlprefix\url{https://link.aps.org/doi/10.1103/PhysRevE.92.052125}.

\bibitem{Ryabov/etal:2013}
\bibinfo{author}{Ryabov, A.}, \bibinfo{author}{Dierl, M.},
  \bibinfo{author}{Chvosta, P.}, \bibinfo{author}{Einax, M.} \&
  \bibinfo{author}{Maass, P.}
\newblock \bibinfo{title}{Work distribution in a time-dependent
  logarithmic--harmonic potential: Exact results and asymptotic analysis}.
\newblock \emph{\bibinfo{journal}{J. Phys. A}} \textbf{\bibinfo{volume}{46}},
  \bibinfo{pages}{075002} (\bibinfo{year}{2013}).
\newblock \urlprefix\url{https://doi.org/10.1088/1751-8113/46/7/075002}.

\bibitem{Kwon/etal:2013}
\bibinfo{author}{Kwon, C.}, \bibinfo{author}{Noh, J.~D.} \&
  \bibinfo{author}{Park, H.}
\newblock \bibinfo{title}{Work fluctuations in a time-dependent harmonic
  potential: Rigorous results beyond the overdamped limit}.
\newblock \emph{\bibinfo{journal}{Phys. Rev. E}} \textbf{\bibinfo{volume}{88}},
  \bibinfo{pages}{062102} (\bibinfo{year}{2013}).

\bibitem{Holubec/etal:2019}
\bibinfo{author}{Holubec, V.}, \bibinfo{author}{Kroy, K.} \&
  \bibinfo{author}{Steffenoni, S.}
\newblock \bibinfo{title}{Physically consistent numerical solver for
  time-dependent {Fokker-Planck} equations}.
\newblock \emph{\bibinfo{journal}{Phys. Rev. E}} \textbf{\bibinfo{volume}{99}},
  \bibinfo{pages}{032117} (\bibinfo{year}{2019}).
\newblock \urlprefix\url{https://doi.org/doi/10.1103/PhysRevE.99.032117}.

\bibitem{Esposito/Parrondo:PRE2015}
\bibinfo{author}{Esposito, M.} \& \bibinfo{author}{Parrondo, J. M.~R.}
\newblock \bibinfo{title}{Stochastic thermodynamics of hidden pumps}.
\newblock \emph{\bibinfo{journal}{Phys. Rev. E}} \textbf{\bibinfo{volume}{91}},
  \bibinfo{pages}{052114} (\bibinfo{year}{2015}).
\newblock \urlprefix\url{https://link.aps.org/doi/10.1103/PhysRevE.91.052114}.

\bibitem{Raz/etal:PRX2016}
\bibinfo{author}{Raz, O.}, \bibinfo{author}{Suba\ifmmode \mbox{\c{s}}\else
  \c{s}\fi{}\ifmmode \imath \else~\i \fi{}, Y.} \& \bibinfo{author}{Jarzynski,
  C.}
\newblock \bibinfo{title}{Mimicking nonequilibrium steady states with
  time-periodic driving}.
\newblock \emph{\bibinfo{journal}{Phys. Rev. X}} \textbf{\bibinfo{volume}{6}},
  \bibinfo{pages}{021022} (\bibinfo{year}{2016}).
\newblock \urlprefix\url{https://link.aps.org/doi/10.1103/PhysRevX.6.021022}.

\bibitem{Brandner/etal:PRX2015}
\bibinfo{author}{Brandner, K.}, \bibinfo{author}{Saito, K.} \&
  \bibinfo{author}{Seifert, U.}
\newblock \bibinfo{title}{Thermodynamics of micro- and nano-systems driven by
  periodic temperature variations}.
\newblock \emph{\bibinfo{journal}{Phys. Rev. X}} \textbf{\bibinfo{volume}{5}},
  \bibinfo{pages}{031019} (\bibinfo{year}{2015}).
\newblock \urlprefix\url{https://link.aps.org/doi/10.1103/PhysRevX.5.031019}.

\bibitem{Barato/Seifert:NJP2017}
\bibinfo{author}{Barato, A.~C.} \& \bibinfo{author}{Seifert, U.}
\newblock \bibinfo{title}{Thermodynamic cost of external control}.
\newblock \emph{\bibinfo{journal}{New. J. Phys.}}
  \textbf{\bibinfo{volume}{19}}, \bibinfo{pages}{073021}
  (\bibinfo{year}{2017}).
\newblock \urlprefix\url{https://doi.org/10.1088/1367-2630/aa77d0}.

\bibitem{Ryabov/etal:2016}
\bibinfo{author}{Ryabov, A.} \emph{et~al.}
\newblock \bibinfo{title}{Transport coefficients for a confined {Brownian}
  ratchet operating between two heat reservoirs}.
\newblock \emph{\bibinfo{journal}{J. Stat. Mech.}}
  \textbf{\bibinfo{volume}{2016}}, \bibinfo{pages}{093202}
  (\bibinfo{year}{2016}).
\newblock \urlprefix\url{https://doi.org/10.1088/1742-5468/2016/09/093202}.

\bibitem{Holubec/etal:2017}
\bibinfo{author}{Holubec, V.} \emph{et~al.}
\newblock \bibinfo{title}{Thermal ratchet effect in confining geometries}.
\newblock \emph{\bibinfo{journal}{Entropy}} \textbf{\bibinfo{volume}{19}}
  (\bibinfo{year}{2017}).
\newblock \urlprefix\url{https://doi.org/10.3390/e19040119}.

\bibitem{JarzynskiCompariosn2007}
\bibinfo{author}{Jarzynski, C.}
\newblock \bibinfo{title}{{Comparison of far-from-equilibrium work relations}}.
\newblock \emph{\bibinfo{journal}{C. R. Physique}}
  \textbf{\bibinfo{volume}{8}}, \bibinfo{pages}{495} (\bibinfo{year}{2007}).
\newblock \urlprefix\url{http://dx.doi.org/10.1016/j.crhy.2007.04.010}.

\bibitem{Reif:1965}
\bibinfo{author}{{Reif}, F.}
\newblock \emph{\bibinfo{title}{{Fundamentals of Statistical and Thermal
  Physics}}} (\bibinfo{publisher}{McGraw-Hill}, \bibinfo{address}{New York},
  \bibinfo{year}{1965}).

\bibitem{Goldstein/etal:2002}
\bibinfo{author}{Goldstein, H.}, \bibinfo{author}{Poole, C.~P.} \&
  \bibinfo{author}{Safko, J.~L.}
\newblock \emph{\bibinfo{title}{{Classical Mechanics}}}
  (\bibinfo{publisher}{Addison-Wesley Longman}, \bibinfo{address}{San
  Francisco}, \bibinfo{year}{2002}), \bibinfo{edition}{3} edn.

\bibitem{HorowitzJarzynskiJSTAT}
\bibinfo{author}{Horowitz, J.} \& \bibinfo{author}{Jarzynski, C.}
\newblock \bibinfo{title}{{Comparison of work fluctuation relations}}.
\newblock \emph{\bibinfo{journal}{J. Stat. Mech.}}
  \textbf{\bibinfo{volume}{2007}}, \bibinfo{pages}{P11002}
  (\bibinfo{year}{2007}).
\newblock \urlprefix\url{http://dx.doi.org/10.1088/1742-5468/2007/11/P11002}.

\bibitem{VilarRubi2008}
\bibinfo{author}{Vilar, J. M.~G.} \& \bibinfo{author}{Rubi, J.~M.}
\newblock \bibinfo{title}{{Failure of the work-{H}amiltonian connection for
  free-energy calculations}}.
\newblock \emph{\bibinfo{journal}{Phys. Rev. Lett.}}
  \textbf{\bibinfo{volume}{100}}, \bibinfo{pages}{020601}
  (\bibinfo{year}{2008}).
\newblock \urlprefix\url{http://dx.doi.org/10.1103/PhysRevLett.100.020601}.

\bibitem{Peliti2008JSTAT}
\bibinfo{author}{Peliti, L.}
\newblock \bibinfo{title}{{On the work–{H}amiltonian connection in
  manipulated systems}}.
\newblock \emph{\bibinfo{journal}{J. Stat. Mech.}}
  \textbf{\bibinfo{volume}{2008}}, \bibinfo{pages}{P05002}
  (\bibinfo{year}{2008}).
\newblock \urlprefix\url{http://dx.doi.org/10.1088/1742-5468/2008/05/P05002}.

\bibitem{HorowitzJarzynski2008}
\bibinfo{author}{Horowitz, J.} \& \bibinfo{author}{Jarzynski, C.}
\newblock \bibinfo{title}{Comment on ``{F}ailure of the work-{H}amiltonian
  connection for free-energy calculations''}.
\newblock \emph{\bibinfo{journal}{Phys. Rev. Lett.}}
  \textbf{\bibinfo{volume}{101}}, \bibinfo{pages}{098901}
  (\bibinfo{year}{2008}).
\newblock \urlprefix\url{http://dx.doi.org/10.1103/PhysRevLett.101.098901}.

\bibitem{ZimanyiSilbey2009}
\bibinfo{author}{Zimanyi, E.~N.} \& \bibinfo{author}{Silbey, R.~J.}
\newblock \bibinfo{title}{{The work-{H}amiltonian connection and the usefulness
  of the Jarzynski equality for free energy calculations}}.
\newblock \emph{\bibinfo{journal}{J. Chem. Phys.}}
  \textbf{\bibinfo{volume}{130}}, \bibinfo{pages}{171102}
  (\bibinfo{year}{2009}).
\newblock \urlprefix\url{http://dx.doi.org/10.1063/1.3132747}.
\newblock \bibinfo{note}{10.1063/1.3132747}.

\bibitem{Talkner/etal:PRE2007}
\bibinfo{author}{Talkner, P.}, \bibinfo{author}{Lutz, E.} \&
  \bibinfo{author}{H\"anggi, P.}
\newblock \bibinfo{title}{Fluctuation theorems: Work is not an observable}.
\newblock \emph{\bibinfo{journal}{Phys. Rev. E}} \textbf{\bibinfo{volume}{75}},
  \bibinfo{pages}{050102} (\bibinfo{year}{2007}).
\newblock \urlprefix\url{https://link.aps.org/doi/10.1103/PhysRevE.75.050102}.

\bibitem{Campisi/etal:RMP2011}
\bibinfo{author}{Campisi, M.}, \bibinfo{author}{H\"anggi, P.} \&
  \bibinfo{author}{Talkner, P.}
\newblock \bibinfo{title}{Colloquium: Quantum fluctuation relations:
  Foundations and applications}.
\newblock \emph{\bibinfo{journal}{Rev. Mod. Phys.}}
  \textbf{\bibinfo{volume}{83}}, \bibinfo{pages}{771--791}
  (\bibinfo{year}{2011}).
\newblock \urlprefix\url{https://link.aps.org/doi/10.1103/RevModPhys.83.771}.

\bibitem{Perarnau-Llobet/etal:PRL2017}
\bibinfo{author}{Perarnau-Llobet, M.}, \bibinfo{author}{B\"aumer, E.},
  \bibinfo{author}{Hovhannisyan, K.~V.}, \bibinfo{author}{Huber, M.} \&
  \bibinfo{author}{Acin, A.}
\newblock \bibinfo{title}{No-go theorem for the characterization of work
  fluctuations in coherent quantum systems}.
\newblock \emph{\bibinfo{journal}{Phys. Rev. Lett.}}
  \textbf{\bibinfo{volume}{118}}, \bibinfo{pages}{070601}
  (\bibinfo{year}{2017}).
\newblock
  \urlprefix\url{https://link.aps.org/doi/10.1103/PhysRevLett.118.070601}.

\bibitem{Hovhannisyan/Imparato:arxiv2021}
\bibinfo{author}{{Hovhannisyan}, K.~V.} \& \bibinfo{author}{{Imparato}, A.}
\newblock \bibinfo{title}{{Energy conservation and Jarzynski equality are
  incompatible for quantum work}}.
\newblock \emph{\bibinfo{journal}{arXiv e-prints}}
  \bibinfo{pages}{arXiv:2104.09364} (\bibinfo{year}{2021}).
\newblock \urlprefix\url{https://arxiv.org/abs/2104.09364}.
\newblock \eprint{2104.09364}.

\bibitem{Holubec2020}
\bibinfo{author}{Holubec, V.}, \bibinfo{author}{Steffenoni, S.},
  \bibinfo{author}{Falasco, G.} \& \bibinfo{author}{Kroy, K.}
\newblock \bibinfo{title}{Active {B}rownian heat engines}.
\newblock \emph{\bibinfo{journal}{Phys. Rev. Research}}
  \textbf{\bibinfo{volume}{2}}, \bibinfo{pages}{043262} (\bibinfo{year}{2020}).
\newblock
  \urlprefix\url{https://link.aps.org/doi/10.1103/PhysRevResearch.2.043262}.

\bibitem{Gronchi2021}
\bibinfo{author}{Gronchi, G.} \& \bibinfo{author}{Puglisi, A.}
\newblock \bibinfo{title}{Optimization of an active heat engine}.
\newblock \emph{\bibinfo{journal}{Phys. Rev. E}}
  \textbf{\bibinfo{volume}{103}}, \bibinfo{pages}{052134}
  (\bibinfo{year}{2021}).
\newblock \urlprefix\url{https://link.aps.org/doi/10.1103/PhysRevE.103.052134}.

\bibitem{Imparato2005}
\bibinfo{author}{Imparato, A.} \& \bibinfo{author}{Peliti, L.}
\newblock \bibinfo{title}{Work probability distribution in single-molecule
  experiments}.
\newblock \emph{\bibinfo{journal}{Europhys. Lett. ({EPL})}}
  \textbf{\bibinfo{volume}{69}}, \bibinfo{pages}{643--649}
  (\bibinfo{year}{2005}).
\newblock \urlprefix\url{https://doi.org/10.1209/epl/i2004-10390-3}.

\bibitem{Subrt/Chvosta:2007}
\bibinfo{author}{{\v{S}}ubrt, E.} \& \bibinfo{author}{Chvosta, P.}
\newblock \bibinfo{title}{Exact analysis of work fluctuations in two-level
  systems}.
\newblock \emph{\bibinfo{journal}{J. Stat. Mech.}}
  \textbf{\bibinfo{volume}{2007}} (\bibinfo{year}{2007}).
\newblock \urlprefix\url{https://doi.org/10.1088/1742-5468/2007/09/p09019}.

\bibitem{holubec2014non}
\bibinfo{author}{Holubec, V.}
\newblock \emph{\bibinfo{title}{Non-equilibrium Energy Transformation
  Processes: Theoretical Description at the Level of Molecular Structures}}
  (\bibinfo{publisher}{Springer}, \bibinfo{address}{New York},
  \bibinfo{year}{2014}).

\bibitem{Touchette:2009}
\bibinfo{author}{Touchette, H.}
\newblock \bibinfo{title}{The large deviation approach to statistical
  mechanics}.
\newblock \emph{\bibinfo{journal}{Phys. Rep.}} \textbf{\bibinfo{volume}{478}},
  \bibinfo{pages}{1--69} (\bibinfo{year}{2009}).
\newblock
  \urlprefix\url{https://www.sciencedirect.com/science/article/pii/S0370157309001410}.

\bibitem{Speck2004}
\bibinfo{author}{Speck, T.} \& \bibinfo{author}{Seifert, U.}
\newblock \bibinfo{title}{Distribution of work in isothermal nonequilibrium
  processes}.
\newblock \emph{\bibinfo{journal}{Phys. Rev. E}} \textbf{\bibinfo{volume}{70}},
  \bibinfo{pages}{066112} (\bibinfo{year}{2004}).
\newblock \urlprefix\url{https://link.aps.org/doi/10.1103/PhysRevE.70.066112}.

\bibitem{Miller2019}
\bibinfo{author}{Miller, H. J.~D.}, \bibinfo{author}{Scandi, M.},
  \bibinfo{author}{Anders, J.} \& \bibinfo{author}{Perarnau-Llobet, M.}
\newblock \bibinfo{title}{Work fluctuations in slow processes: Quantum
  signatures and optimal control}.
\newblock \emph{\bibinfo{journal}{Phys. Rev. Lett.}}
  \textbf{\bibinfo{volume}{123}}, \bibinfo{pages}{230603}
  (\bibinfo{year}{2019}).
\newblock
  \urlprefix\url{https://link.aps.org/doi/10.1103/PhysRevLett.123.230603}.

\bibitem{Scandi2020}
\bibinfo{author}{Scandi, M.}, \bibinfo{author}{Miller, H. J.~D.},
  \bibinfo{author}{Anders, J.} \& \bibinfo{author}{Perarnau-Llobet, M.}
\newblock \bibinfo{title}{Quantum work statistics close to equilibrium}.
\newblock \emph{\bibinfo{journal}{Phys. Rev. Research}}
  \textbf{\bibinfo{volume}{2}}, \bibinfo{pages}{023377} (\bibinfo{year}{2020}).
\newblock
  \urlprefix\url{https://link.aps.org/doi/10.1103/PhysRevResearch.2.023377}.

\bibitem{Jarzynski1997}
\bibinfo{author}{Jarzynski, C.}
\newblock \bibinfo{title}{Nonequilibrium equality for free energy differences}.
\newblock \emph{\bibinfo{journal}{Phys. Rev. Lett.}}
  \textbf{\bibinfo{volume}{78}}, \bibinfo{pages}{2690--2693}
  (\bibinfo{year}{1997}).
\newblock \urlprefix\url{https://link.aps.org/doi/10.1103/PhysRevLett.78.2690}.

\bibitem{Crooks1999}
\bibinfo{author}{Crooks, G.~E.}
\newblock \bibinfo{title}{Entropy production fluctuation theorem and the
  nonequilibrium work relation for free energy differences}.
\newblock \emph{\bibinfo{journal}{Phys. Rev. E}} \textbf{\bibinfo{volume}{60}},
  \bibinfo{pages}{2721--2726} (\bibinfo{year}{1999}).
\newblock \urlprefix\url{https://link.aps.org/doi/10.1103/PhysRevE.60.2721}.

\bibitem{Sinitsyn:2011}
\bibinfo{author}{Sinitsyn, N.~A.}
\newblock \bibinfo{title}{Fluctuation relation for heat engines}.
\newblock \emph{\bibinfo{journal}{J. Phys. A}} \textbf{\bibinfo{volume}{44}},
  \bibinfo{pages}{405001} (\bibinfo{year}{2011}).
\newblock \urlprefix\url{https://doi.org/10.1088/1751-8113/44/40/405001}.

\bibitem{Garcia-Garcia/etal:2010}
\bibinfo{author}{Garc\'{\i}a-Garc\'{\i}a, R.}, \bibinfo{author}{Dom\'{\i}nguez,
  D.}, \bibinfo{author}{Lecomte, V.} \& \bibinfo{author}{Kolton, A.~B.}
\newblock \bibinfo{title}{Unifying approach for fluctuation theorems from joint
  probability distributions}.
\newblock \emph{\bibinfo{journal}{Phys. Rev. E}} \textbf{\bibinfo{volume}{82}},
  \bibinfo{pages}{030104} (\bibinfo{year}{2010}).
\newblock \urlprefix\url{https://link.aps.org/doi/10.1103/PhysRevE.82.030104}.

\bibitem{Verley/etal:2014}
\bibinfo{author}{Verley, G.}, \bibinfo{author}{den Broeck, C.~V.} \&
  \bibinfo{author}{Esposito, M.}
\newblock \bibinfo{title}{Work statistics in stochastically driven systems}.
\newblock \emph{\bibinfo{journal}{New J. Phys.}} \textbf{\bibinfo{volume}{16}},
  \bibinfo{pages}{095001} (\bibinfo{year}{2014}).
\newblock \urlprefix\url{https://doi.org/10.1088/1367-2630/16/9/095001}.

\bibitem{Campisi2014}
\bibinfo{author}{Campisi, M.}
\newblock \bibinfo{title}{Fluctuation relation for quantum heat engines and
  refrigerators}.
\newblock \emph{\bibinfo{journal}{J. Phys. A}} \textbf{\bibinfo{volume}{47}},
  \bibinfo{pages}{245001} (\bibinfo{year}{2014}).
\newblock \urlprefix\url{https://doi.org/10.1088/1751-8113/47/24/245001}.

\bibitem{Gingrich/etal:2014}
\bibinfo{author}{Gingrich, T.~R.}, \bibinfo{author}{Rotskoff, G.~M.},
  \bibinfo{author}{Vaikuntanathan, S.} \& \bibinfo{author}{Geissler, P.~L.}
\newblock \bibinfo{title}{Efficiency and large deviations in time-asymmetric
  stochastic heat engines}.
\newblock \emph{\bibinfo{journal}{New J. Phys.}} \textbf{\bibinfo{volume}{16}},
  \bibinfo{pages}{102003} (\bibinfo{year}{2014}).
\newblock \urlprefix\url{https://doi.org/10.1088/1367-2630/16/10/102003}.

\bibitem{Proesmans2015}
\bibinfo{author}{Proesmans, K.}, \bibinfo{author}{Cleuren, B.} \&
  \bibinfo{author}{den Broeck, C.~V.}
\newblock \bibinfo{title}{Stochastic efficiency for effusion as a thermal
  engine}.
\newblock \emph{\bibinfo{journal}{{EPL} (Europhys. Lett.)}}
  \textbf{\bibinfo{volume}{109}}, \bibinfo{pages}{20004}
  (\bibinfo{year}{2015}).
\newblock \urlprefix\url{https://doi.org/10.1209/0295-5075/109/20004}.

\bibitem{Polettini/etal:2015}
\bibinfo{author}{Polettini, M.}, \bibinfo{author}{Verley, G.} \&
  \bibinfo{author}{Esposito, M.}
\newblock \bibinfo{title}{Efficiency statistics at all times: {C}arnot limit at
  finite power}.
\newblock \emph{\bibinfo{journal}{Phys. Rev. Lett.}}
  \textbf{\bibinfo{volume}{114}}, \bibinfo{pages}{050601}
  (\bibinfo{year}{2015}).
\newblock
  \urlprefix\url{https://link.aps.org/doi/10.1103/PhysRevLett.114.050601}.

\bibitem{Jiang/etal:2015}
\bibinfo{author}{Jiang, J.-H.}, \bibinfo{author}{Agarwalla, B.~K.} \&
  \bibinfo{author}{Segal, D.}
\newblock \bibinfo{title}{Efficiency statistics and bounds for systems with
  broken time-reversal symmetry}.
\newblock \emph{\bibinfo{journal}{Phys. Rev. Lett.}}
  \textbf{\bibinfo{volume}{115}}, \bibinfo{pages}{040601}
  (\bibinfo{year}{2015}).
\newblock
  \urlprefix\url{https://link.aps.org/doi/10.1103/PhysRevLett.115.040601}.

\bibitem{Proesmans/etal:2016b}
\bibinfo{author}{Proesmans, K.}, \bibinfo{author}{Dreher, Y.},
  \bibinfo{author}{Gavrilov, M. c.~v.}, \bibinfo{author}{Bechhoefer, J.} \&
  \bibinfo{author}{Van~den Broeck, C.}
\newblock \bibinfo{title}{Brownian duet: A novel tale of thermodynamic
  efficiency}.
\newblock \emph{\bibinfo{journal}{Phys. Rev. X}} \textbf{\bibinfo{volume}{6}},
  \bibinfo{pages}{041010} (\bibinfo{year}{2016}).
\newblock \urlprefix\url{https://link.aps.org/doi/10.1103/PhysRevX.6.041010}.

\bibitem{Park/etal:PRE:2016}
\bibinfo{author}{Park, J.-M.}, \bibinfo{author}{Chun, H.-M.} \&
  \bibinfo{author}{Noh, J.~D.}
\newblock \bibinfo{title}{Efficiency at maximum power and efficiency
  fluctuations in a linear {B}rownian heat-engine model}.
\newblock \emph{\bibinfo{journal}{Phys. Rev. E}} \textbf{\bibinfo{volume}{94}},
  \bibinfo{pages}{012127} (\bibinfo{year}{2016}).
\newblock \urlprefix\url{https://link.aps.org/doi/10.1103/PhysRevE.94.012127}.

\bibitem{Vroylandt/etal:PRE:2016}
\bibinfo{author}{Vroylandt, H.}, \bibinfo{author}{Bonfils, A.} \&
  \bibinfo{author}{Verley, G.}
\newblock \bibinfo{title}{Efficiency fluctuations of small machines with
  unknown losses}.
\newblock \emph{\bibinfo{journal}{Phys. Rev. E}} \textbf{\bibinfo{volume}{93}},
  \bibinfo{pages}{052123} (\bibinfo{year}{2016}).
\newblock \urlprefix\url{https://link.aps.org/doi/10.1103/PhysRevE.93.052123}.

\bibitem{Proesmans/VanDenBroeck:Chaos:2017}
\bibinfo{author}{Proesmans, K.} \& \bibinfo{author}{Van~den Broeck, C.}
\newblock \bibinfo{title}{The underdamped {B}rownian duet and stochastic linear
  irreversible thermodynamics}.
\newblock \emph{\bibinfo{journal}{Chaos}} \textbf{\bibinfo{volume}{27}},
  \bibinfo{pages}{104601} (\bibinfo{year}{2017}).
\newblock \urlprefix\url{https://doi.org/10.1063/1.5001187}.

\bibitem{Gupta/Sabhapandit:PRE:2017}
\bibinfo{author}{Gupta, D.} \& \bibinfo{author}{Sabhapandit, S.}
\newblock \bibinfo{title}{Stochastic efficiency of an isothermal work-to-work
  converter engine}.
\newblock \emph{\bibinfo{journal}{Phys. Rev. E}} \textbf{\bibinfo{volume}{96}},
  \bibinfo{pages}{042130} (\bibinfo{year}{2017}).
\newblock \urlprefix\url{https://link.aps.org/doi/10.1103/PhysRevE.96.042130}.

\bibitem{Gupta:JSTAT:2018}
\bibinfo{author}{Gupta, D.}
\newblock \bibinfo{title}{Exact distribution for work and stochastic efficiency
  of an isothermal machine}.
\newblock \emph{\bibinfo{journal}{J. Stat. Mech.}}
  \textbf{\bibinfo{volume}{2018}}, \bibinfo{pages}{073201}
  (\bibinfo{year}{2018}).
\newblock \urlprefix\url{https://doi.org/10.1088/1742-5468/aace09}.

\bibitem{Verley2014}
\bibinfo{author}{Verley, G.}, \bibinfo{author}{Esposito, M.},
  \bibinfo{author}{Willaert, T.} \& \bibinfo{author}{Van~den Broeck, C.}
\newblock \bibinfo{title}{The unlikely {C}arnot efficiency}.
\newblock \emph{\bibinfo{journal}{Nat. Commun.}} \textbf{\bibinfo{volume}{5}},
  \bibinfo{pages}{4721} (\bibinfo{year}{2014}).
\newblock \urlprefix\url{https://doi.org/10.1038/ncomms5721}.

\bibitem{Verley/etal:2014b}
\bibinfo{author}{Verley, G.}, \bibinfo{author}{Willaert, T.},
  \bibinfo{author}{Van~den Broeck, C.} \& \bibinfo{author}{Esposito, M.}
\newblock \bibinfo{title}{Universal theory of efficiency fluctuations}.
\newblock \emph{\bibinfo{journal}{Phys. Rev. E}} \textbf{\bibinfo{volume}{90}},
  \bibinfo{pages}{052145} (\bibinfo{year}{2014}).
\newblock \urlprefix\url{https://doi.org/10.1103/PhysRevE.90.052145}.

\bibitem{Manikandan/etal:PRL:2019}
\bibinfo{author}{Manikandan, S.~K.}, \bibinfo{author}{Dabelow, L.},
  \bibinfo{author}{Eichhorn, R.} \& \bibinfo{author}{Krishnamurthy, S.}
\newblock \bibinfo{title}{Efficiency fluctuations in microscopic machines}.
\newblock \emph{\bibinfo{journal}{Phys. Rev. Lett.}}
  \textbf{\bibinfo{volume}{122}}, \bibinfo{pages}{140601}
  (\bibinfo{year}{2019}).
\newblock
  \urlprefix\url{https://link.aps.org/doi/10.1103/PhysRevLett.122.140601}.

\bibitem{Rana/etal:2014}
\bibinfo{author}{Rana, S.}, \bibinfo{author}{Pal, P.~S.},
  \bibinfo{author}{Saha, A.} \& \bibinfo{author}{Jayannavar, A.~M.}
\newblock \bibinfo{title}{Single-particle stochastic heat engine}.
\newblock \emph{\bibinfo{journal}{Phys. Rev. E}} \textbf{\bibinfo{volume}{90}},
  \bibinfo{pages}{042146} (\bibinfo{year}{2014}).
\newblock \urlprefix\url{https://link.aps.org/doi/10.1103/PhysRevE.90.042146}.

\bibitem{Proesmans/VanDenBroeck:NJP:2015}
\bibinfo{author}{Proesmans, K.} \& \bibinfo{author}{den Broeck, C.~V.}
\newblock \bibinfo{title}{Stochastic efficiency: Five case studies}.
\newblock \emph{\bibinfo{journal}{New J. Phys.}} \textbf{\bibinfo{volume}{17}},
  \bibinfo{pages}{065004} (\bibinfo{year}{2015}).
\newblock \urlprefix\url{https://doi.org/10.1088/1367-2630/17/6/065004}.

\bibitem{Rana/etal:PhysicaA:2016}
\bibinfo{author}{Rana, S.}, \bibinfo{author}{Pal, P.}, \bibinfo{author}{Saha,
  A.} \& \bibinfo{author}{Jayannavar, A.}
\newblock \bibinfo{title}{Anomalous {B}rownian refrigerator}.
\newblock \emph{\bibinfo{journal}{Physica A}} \textbf{\bibinfo{volume}{444}},
  \bibinfo{pages}{783--798} (\bibinfo{year}{2016}).
\newblock
  \urlprefix\url{https://www.sciencedirect.com/science/article/pii/S0378437115009541}.

\bibitem{Sune/Imparato:JPhysA:2019}
\bibinfo{author}{Su{\~{n}}{\'{e}}, M.} \& \bibinfo{author}{Imparato, A.}
\newblock \bibinfo{title}{Efficiency fluctuations in steady-state machines}.
\newblock \emph{\bibinfo{journal}{J. Phys. A}} \textbf{\bibinfo{volume}{52}},
  \bibinfo{pages}{045003} (\bibinfo{year}{2019}).
\newblock \urlprefix\url{https://doi.org/10.1088/1751-8121/aaf2f8}.

\bibitem{Cerino/etal:PRE:2015}
\bibinfo{author}{Cerino, L.}, \bibinfo{author}{Puglisi, A.} \&
  \bibinfo{author}{Vulpiani, A.}
\newblock \bibinfo{title}{Kinetic model for the finite-time thermodynamics of
  small heat engines}.
\newblock \emph{\bibinfo{journal}{Phys. Rev. E}} \textbf{\bibinfo{volume}{91}},
  \bibinfo{pages}{032128} (\bibinfo{year}{2015}).
\newblock \urlprefix\url{https://link.aps.org/doi/10.1103/PhysRevE.91.032128}.

\bibitem{Vroylandt/etal:PRL:2020}
\bibinfo{author}{Vroylandt, H.}, \bibinfo{author}{Esposito, M.} \&
  \bibinfo{author}{Verley, G.}
\newblock \bibinfo{title}{Efficiency fluctuations of stochastic machines
  undergoing a phase transition}.
\newblock \emph{\bibinfo{journal}{Phys. Rev. Lett.}}
  \textbf{\bibinfo{volume}{124}}, \bibinfo{pages}{250603}
  (\bibinfo{year}{2020}).
\newblock
  \urlprefix\url{https://link.aps.org/doi/10.1103/PhysRevLett.124.250603}.

\bibitem{Rosinberg/Horowitz:EPL:2016}
\bibinfo{author}{Rosinberg, M.~L.} \& \bibinfo{author}{Horowitz, J.~M.}
\newblock \bibinfo{title}{Continuous information flow fluctuations}.
\newblock \emph{\bibinfo{journal}{Europhys. Lett. ({EPL})}}
  \textbf{\bibinfo{volume}{116}}, \bibinfo{pages}{10007}
  (\bibinfo{year}{2016}).
\newblock \urlprefix\url{https://doi.org/10.1209/0295-5075/116/10007}.

\bibitem{Paneru/etal:NatCommun:2020}
\bibinfo{author}{Paneru, G.}, \bibinfo{author}{Dutta, S.},
  \bibinfo{author}{Sagawa, T.}, \bibinfo{author}{Tlusty, T.} \&
  \bibinfo{author}{Pak, H.~K.}
\newblock \bibinfo{title}{Efficiency fluctuations and noise induced
  refrigerator-to-heater transition in information engines}.
\newblock \emph{\bibinfo{journal}{Nat. Commun.}} \textbf{\bibinfo{volume}{11}},
  \bibinfo{pages}{1012} (\bibinfo{year}{2020}).
\newblock \urlprefix\url{https://doi.org/10.1038/s41467-020-14823-x}.

\bibitem{Paneru/Pak:APX:2020}
\bibinfo{author}{Paneru, G.} \& \bibinfo{author}{Pak, H.~K.}
\newblock \bibinfo{title}{Colloidal engines for innovative tests of information
  thermodynamics}.
\newblock \emph{\bibinfo{journal}{Adv. Phys. X}} \textbf{\bibinfo{volume}{5}},
  \bibinfo{pages}{1823880} (\bibinfo{year}{2020}).
\newblock \urlprefix\url{https://doi.org/10.1080/23746149.2020.1823880}.
\newblock \eprint{https://doi.org/10.1080/23746149.2020.1823880}.

\bibitem{Esposito/etal:2015}
\bibinfo{author}{Esposito, M.}, \bibinfo{author}{Ochoa, M.~A.} \&
  \bibinfo{author}{Galperin, M.}
\newblock \bibinfo{title}{Efficiency fluctuations in quantum thermoelectric
  devices}.
\newblock \emph{\bibinfo{journal}{Phys. Rev. B}} \textbf{\bibinfo{volume}{91}},
  \bibinfo{pages}{115417} (\bibinfo{year}{2015}).
\newblock \urlprefix\url{https://link.aps.org/doi/10.1103/PhysRevB.91.115417}.

\bibitem{Agarwalla/etal:2015}
\bibinfo{author}{Agarwalla, B.~K.}, \bibinfo{author}{Jiang, J.-H.} \&
  \bibinfo{author}{Segal, D.}
\newblock \bibinfo{title}{Full counting statistics of vibrationally assisted
  electronic conduction: Transport and fluctuations of thermoelectric
  efficiency}.
\newblock \emph{\bibinfo{journal}{Phys. Rev. B}} \textbf{\bibinfo{volume}{92}},
  \bibinfo{pages}{245418} (\bibinfo{year}{2015}).
\newblock \urlprefix\url{https://link.aps.org/doi/10.1103/PhysRevB.92.245418}.

\bibitem{Cuetara/Esposito:NJP:2015}
\bibinfo{author}{Cuetara, G.~B.} \& \bibinfo{author}{Esposito, M.}
\newblock \bibinfo{title}{Double quantum dot coupled to a quantum point
  contact: A stochastic thermodynamics approach}.
\newblock \emph{\bibinfo{journal}{New J. Phys.}} \textbf{\bibinfo{volume}{17}},
  \bibinfo{pages}{095005} (\bibinfo{year}{2015}).
\newblock \urlprefix\url{https://doi.org/10.1088/1367-2630/17/9/095005}.

\bibitem{Crepieux/Michelini:JSTAT:2016}
\bibinfo{author}{Cr{\'{e}}pieux, A.} \& \bibinfo{author}{Michelini, F.}
\newblock \bibinfo{title}{Heat-charge mixed noise and thermoelectric efficiency
  fluctuations}.
\newblock \emph{\bibinfo{journal}{J. Stat. Mech.}}
  \textbf{\bibinfo{volume}{2016}}, \bibinfo{pages}{054015}
  (\bibinfo{year}{2016}).
\newblock \urlprefix\url{https://doi.org/10.1088/1742-5468/2016/05/054015}.

\bibitem{Tang/etal:PRB:2018}
\bibinfo{author}{Tang, G.}, \bibinfo{author}{Thingna, J.} \&
  \bibinfo{author}{Wang, J.}
\newblock \bibinfo{title}{Thermodynamics of energy, charge, and spin currents
  in a thermoelectric quantum-dot spin valve}.
\newblock \emph{\bibinfo{journal}{Phys. Rev. B}} \textbf{\bibinfo{volume}{97}},
  \bibinfo{pages}{155430} (\bibinfo{year}{2018}).
\newblock \urlprefix\url{https://link.aps.org/doi/10.1103/PhysRevB.97.155430}.

\bibitem{Denzler/Lutz:PRR:2020}
\bibinfo{author}{Denzler, T.} \& \bibinfo{author}{Lutz, E.}
\newblock \bibinfo{title}{Efficiency fluctuations of a quantum heat engine}.
\newblock \emph{\bibinfo{journal}{Phys. Rev. Research}}
  \textbf{\bibinfo{volume}{2}}, \bibinfo{pages}{032062} (\bibinfo{year}{2020}).
\newblock
  \urlprefix\url{https://link.aps.org/doi/10.1103/PhysRevResearch.2.032062}.

\bibitem{Barato/Seifert:2015}
\bibinfo{author}{Barato, A.~C.} \& \bibinfo{author}{Seifert, U.}
\newblock \bibinfo{title}{Thermodynamic uncertainty relation for biomolecular
  processes}.
\newblock \emph{\bibinfo{journal}{Phys. Rev. Lett.}}
  \textbf{\bibinfo{volume}{114}}, \bibinfo{pages}{158101}
  (\bibinfo{year}{2015}).
\newblock \urlprefix\url{https://doi.org/10.1103/PhysRevLett.114.158101}.

\bibitem{Gingrich/etal:2016}
\bibinfo{author}{Gingrich, T.~R.}, \bibinfo{author}{Horowitz, J.~M.},
  \bibinfo{author}{Perunov, N.} \& \bibinfo{author}{England, J.~L.}
\newblock \bibinfo{title}{Dissipation bounds all steady-state current
  fluctuations}.
\newblock \emph{\bibinfo{journal}{Phys. Rev. Lett.}}
  \textbf{\bibinfo{volume}{116}}, \bibinfo{pages}{120601}
  (\bibinfo{year}{2016}).
\newblock \urlprefix\url{https://doi.org/doi/10.1103/PhysRevLett.116.120601}.

\bibitem{Dechant/Sasa:2018}
\bibinfo{author}{Dechant, A.} \& \bibinfo{author}{ichi Sasa, S.}
\newblock \bibinfo{title}{Current fluctuations and transport efficiency for
  general {L}angevin systems}.
\newblock \emph{\bibinfo{journal}{J. Stat. Mech.}}
  \textbf{\bibinfo{volume}{2018}}, \bibinfo{pages}{063209}
  (\bibinfo{year}{2018}).
\newblock \urlprefix\url{https://doi.org/10.1088%2F1742-5468%2Faac91a}.

\bibitem{Liu/etal:2019}
\bibinfo{author}{Liu, K.}, \bibinfo{author}{Gong, Z.} \& \bibinfo{author}{Ueda,
  M.}
\newblock \bibinfo{title}{Thermodynamic uncertainty relation for arbitrary
  initial states}.
\newblock \emph{\bibinfo{journal}{Phys. Rev. Lett.}}
  \textbf{\bibinfo{volume}{125}}, \bibinfo{pages}{140602}
  (\bibinfo{year}{2020}).
\newblock
  \urlprefix\url{https://link.aps.org/doi/10.1103/PhysRevLett.125.140602}.

\bibitem{Zhang:2019}
\bibinfo{author}{{Zhang}, Y.}
\newblock \bibinfo{title}{{Comment on ``Fluctuation Theorem Uncertainty
  Relation'' and ``Thermodynamic Uncertainty Relations from Exchange
  Fluctuation Theorems''}}.
\newblock \emph{\bibinfo{journal}{arXiv e-prints}}
  \textbf{\bibinfo{volume}{2019}}, \bibinfo{pages}{arXiv:1910.12862}
  (\bibinfo{year}{2019}).
\newblock \urlprefix\url{https://arxiv.org/abs/1910.12862}.

\bibitem{Hasegawa/VanVu:2019}
\bibinfo{author}{Hasegawa, Y.} \& \bibinfo{author}{Van~Vu, T.}
\newblock \bibinfo{title}{Fluctuation theorem uncertainty relation}.
\newblock \emph{\bibinfo{journal}{Phys. Rev. Lett.}}
  \textbf{\bibinfo{volume}{123}}, \bibinfo{pages}{110602}
  (\bibinfo{year}{2019}).
\newblock
  \urlprefix\url{https://link.aps.org/doi/10.1103/PhysRevLett.123.110602}.

\bibitem{Timpanaro/etal:2019}
\bibinfo{author}{Timpanaro, A.~M.}, \bibinfo{author}{Guarnieri, G.},
  \bibinfo{author}{Goold, J.} \& \bibinfo{author}{Landi, G.~T.}
\newblock \bibinfo{title}{Thermodynamic uncertainty relations from exchange
  fluctuation theorems}.
\newblock \emph{\bibinfo{journal}{Phys. Rev. Lett.}}
  \textbf{\bibinfo{volume}{123}}, \bibinfo{pages}{090604}
  (\bibinfo{year}{2019}).
\newblock
  \urlprefix\url{https://link.aps.org/doi/10.1103/PhysRevLett.123.090604}.

\bibitem{Pietzonka/Seifert:2018}
\bibinfo{author}{Pietzonka, P.} \& \bibinfo{author}{Seifert, U.}
\newblock \bibinfo{title}{Universal trade-off between power, efficiency, and
  constancy in steady-state heat engines}.
\newblock \emph{\bibinfo{journal}{Phys. Rev. Lett.}}
  \textbf{\bibinfo{volume}{120}}, \bibinfo{pages}{190602}
  (\bibinfo{year}{2018}).

\bibitem{Pietzonka/etal:2016}
\bibinfo{author}{Pietzonka, P.}, \bibinfo{author}{Barato, A.~C.} \&
  \bibinfo{author}{Seifert, U.}
\newblock \bibinfo{title}{Universal bound on the efficiency of molecular
  motors}.
\newblock \emph{\bibinfo{journal}{J. Stat. Mech.}}
  \textbf{\bibinfo{volume}{2016}}, \bibinfo{pages}{124004}
  (\bibinfo{year}{2016}).
\newblock \urlprefix\url{https://doi.org/10.1088/1742-5468/2016/12/124004}.

\bibitem{Barato/Sefiert:2016}
\bibinfo{author}{Barato, A.~C.} \& \bibinfo{author}{Seifert, U.}
\newblock \bibinfo{title}{Cost and precision of {Brownian} clocks}.
\newblock \emph{\bibinfo{journal}{Phys. Rev. X}} \textbf{\bibinfo{volume}{6}},
  \bibinfo{pages}{041053} (\bibinfo{year}{2016}).
\newblock \urlprefix\url{https://doi.org/10.1103/PhysRevX.6.041053}.

\bibitem{Kheradsoud/eta:2019}
\bibinfo{author}{Kheradsoud, S.} \emph{et~al.}
\newblock \bibinfo{title}{Power, efficiency and fluctuations in a quantum point
  contact as steady-state thermoelectric heat engine}.
\newblock \emph{\bibinfo{journal}{Entropy}} \textbf{\bibinfo{volume}{21}},
  \bibinfo{pages}{777} (\bibinfo{year}{2019}).
\newblock \urlprefix\url{https://www.mdpi.com/1099-4300/21/8/777}.

\bibitem{Koyuk/etal:2018}
\bibinfo{author}{Koyuk, T.}, \bibinfo{author}{Seifert, U.} \&
  \bibinfo{author}{Pietzonka, P.}
\newblock \bibinfo{title}{A generalization of the thermodynamic uncertainty
  relation to periodically driven systems}.
\newblock \emph{\bibinfo{journal}{J. Phys. A}} \textbf{\bibinfo{volume}{52}},
  \bibinfo{pages}{02LT02} (\bibinfo{year}{2018}).
\newblock \urlprefix\url{https://doi.org/10.1088/1751-8121/aaeec4}.

\bibitem{Miller/etal:PRL2021}
\bibinfo{author}{Miller, H. J.~D.}, \bibinfo{author}{Mohammady, M.~H.},
  \bibinfo{author}{Perarnau-Llobet, M.} \& \bibinfo{author}{Guarnieri, G.}
\newblock \bibinfo{title}{Thermodynamic uncertainty relation in slowly driven
  quantum heat engines}.
\newblock \emph{\bibinfo{journal}{Phys. Rev. Lett.}}
  \textbf{\bibinfo{volume}{126}}, \bibinfo{pages}{210603}
  (\bibinfo{year}{2021}).
\newblock
  \urlprefix\url{https://link.aps.org/doi/10.1103/PhysRevLett.126.210603}.

\bibitem{Miller/etal:PRE2021}
\bibinfo{author}{Miller, H. J.~D.}, \bibinfo{author}{Mohammady, M.~H.},
  \bibinfo{author}{Perarnau-Llobet, M.} \& \bibinfo{author}{Guarnieri, G.}
\newblock \bibinfo{title}{Joint statistics of work and entropy production along
  quantum trajectories}.
\newblock \emph{\bibinfo{journal}{Phys. Rev. E}}
  \textbf{\bibinfo{volume}{103}}, \bibinfo{pages}{052138}
  (\bibinfo{year}{2021}).
\newblock \urlprefix\url{https://link.aps.org/doi/10.1103/PhysRevE.103.052138}.

\bibitem{Koyuk/Seifert:2019}
\bibinfo{author}{Koyuk, T.} \& \bibinfo{author}{Seifert, U.}
\newblock \bibinfo{title}{Operationally accessible bounds on fluctuations and
  entropy production in periodically driven systems}.
\newblock \emph{\bibinfo{journal}{Phys. Rev. Lett.}}
  \textbf{\bibinfo{volume}{122}}, \bibinfo{pages}{230601}
  (\bibinfo{year}{2019}).
\newblock
  \urlprefix\url{https://link.aps.org/doi/10.1103/PhysRevLett.122.230601}.

\bibitem{Harunari/etal:2020}
\bibinfo{author}{Harunari, P.~E.}, \bibinfo{author}{Fiore, C.~E.} \&
  \bibinfo{author}{Proesmans, K.}
\newblock \bibinfo{title}{Exact statistics and thermodynamic uncertainty
  relations for a periodically driven electron pump}.
\newblock \emph{\bibinfo{journal}{J. Phys. A}} \textbf{\bibinfo{volume}{53}},
  \bibinfo{pages}{374001} (\bibinfo{year}{2020}).
\newblock \urlprefix\url{https://doi.org/10.1088/1751-8121/aba05e}.

\bibitem{Speck2007}
\bibinfo{author}{Speck, T.} \& \bibinfo{author}{Seifert, U.}
\newblock \bibinfo{title}{The {J}arzynski relation, fluctuation theorems, and
  stochastic thermodynamics for non-{M}arkovian processes}.
\newblock \emph{\bibinfo{journal}{J. Stat. Mech.}}
  \textbf{\bibinfo{volume}{2007}}, \bibinfo{pages}{L09002--L09002}
  (\bibinfo{year}{2007}).
\newblock \urlprefix\url{https://doi.org/10.1088/1742-5468/2007/09/l09002}.

\bibitem{Li/etal:2019}
\bibinfo{author}{Li, J.}, \bibinfo{author}{Horowitz, J.~M.},
  \bibinfo{author}{Gingrich, T.~R.} \& \bibinfo{author}{Fakhri, N.}
\newblock \bibinfo{title}{Quantifying dissipation using fluctuating currents}.
\newblock \emph{\bibinfo{journal}{Nat. Commun.}} \textbf{\bibinfo{volume}{10}},
  \bibinfo{pages}{1666} (\bibinfo{year}{2019}).
\newblock \urlprefix\url{https://doi.org/10.1038/s41467-019-09631-x}.

\bibitem{Manikandan/etal:2020}
\bibinfo{author}{Manikandan, S.~K.}, \bibinfo{author}{Gupta, D.} \&
  \bibinfo{author}{Krishnamurthy, S.}
\newblock \bibinfo{title}{Inferring entropy production from short experiments}.
\newblock \emph{\bibinfo{journal}{Phys. Rev. Lett.}}
  \textbf{\bibinfo{volume}{124}}, \bibinfo{pages}{120603}
  (\bibinfo{year}{2020}).
\newblock
  \urlprefix\url{https://link.aps.org/doi/10.1103/PhysRevLett.124.120603}.

\bibitem{Skinner/Dunkel:2021}
\bibinfo{author}{Skinner, D.~J.} \& \bibinfo{author}{Dunkel, J.}
\newblock \bibinfo{title}{Improved bounds on entropy production in living
  systems}.
\newblock \emph{\bibinfo{journal}{Proc. Natl. Acad. Sci. U.S.A.}}
  \textbf{\bibinfo{volume}{118}} (\bibinfo{year}{2021}).
\newblock \urlprefix\url{https://doi.org/10.1073/pnas.2024300118}.

\bibitem{Schuler/etal:2005}
\bibinfo{author}{Schuler, S.}, \bibinfo{author}{Speck, T.},
  \bibinfo{author}{Tietz, C.}, \bibinfo{author}{Wrachtrup, J.} \&
  \bibinfo{author}{Seifert, U.}
\newblock \bibinfo{title}{Experimental test of the fluctuation theorem for a
  driven two-level system with time-dependent rates}.
\newblock \emph{\bibinfo{journal}{Phys. Rev. Lett.}}
  \textbf{\bibinfo{volume}{94}}, \bibinfo{pages}{180602}
  (\bibinfo{year}{2005}).
\newblock \urlprefix\url{https://doi.org/10.1103/PhysRevLett.94.180602}.

\bibitem{Tietz/etal:2006}
\bibinfo{author}{Tietz, C.}, \bibinfo{author}{Schuler, S.},
  \bibinfo{author}{Speck, T.}, \bibinfo{author}{Seifert, U.} \&
  \bibinfo{author}{Wrachtrup, J.}
\newblock \bibinfo{title}{Measurement of stochastic entropy production}.
\newblock \emph{\bibinfo{journal}{Phys. Rev. Lett.}}
  \textbf{\bibinfo{volume}{97}}, \bibinfo{pages}{050602}
  (\bibinfo{year}{2006}).
\newblock \urlprefix\url{https://doi.org/10.1103/PhysRevLett.97.050602}.

\bibitem{Kung/etal:2012}
\bibinfo{author}{K\"ung, B.} \emph{et~al.}
\newblock \bibinfo{title}{Irreversibility on the level of single-electron
  tunneling}.
\newblock \emph{\bibinfo{journal}{Phys. Rev. X}} \textbf{\bibinfo{volume}{2}},
  \bibinfo{pages}{011001} (\bibinfo{year}{2012}).
\newblock \urlprefix\url{https://link.aps.org/doi/10.1103/PhysRevX.2.011001}.

\bibitem{Koski/etal:2013}
\bibinfo{author}{Koski, J.~V.} \emph{et~al.}
\newblock \bibinfo{title}{Distribution of entropy production in a
  single-electron box}.
\newblock \emph{\bibinfo{journal}{Nat. Phys.}} \textbf{\bibinfo{volume}{9}},
  \bibinfo{pages}{644--648} (\bibinfo{year}{2013}).
\newblock \urlprefix\url{https://doi.org/10.1038/nphys2711}.

\bibitem{Jop/etal:2008}
\bibinfo{author}{Jop, P.}, \bibinfo{author}{Petrosyan, A.} \&
  \bibinfo{author}{Ciliberto, S.}
\newblock \bibinfo{title}{Work and dissipation fluctuations near the stochastic
  resonance of a colloidal particle}.
\newblock \emph{\bibinfo{journal}{Europhys. Lett. ({EPL})}}
  \textbf{\bibinfo{volume}{81}}, \bibinfo{pages}{50005} (\bibinfo{year}{2008}).
\newblock \urlprefix\url{https://doi.org/10.1209/0295-5075/81/50005}.

\bibitem{Ritort:2007}
\bibinfo{author}{Ritort, F.}
\newblock \emph{\bibinfo{title}{Nonequilibrium Fluctuations in Small Systems:
  From Physics to Biology}}, chap.~\bibinfo{chapter}{2},
  \bibinfo{pages}{31--123} (\bibinfo{publisher}{John Wiley \& Sons, Ltd},
  \bibinfo{year}{2007}).
\newblock \urlprefix\url{https://doi.org/10.1002/9780470238080.ch2}.

\bibitem{Chvosta2007}
\bibinfo{author}{Chvosta, P.}, \bibinfo{author}{Reineker, P.} \&
  \bibinfo{author}{Schulz, M.}
\newblock \bibinfo{title}{Probability distribution of work done on a two-level
  system during a nonequilibrium isothermal process}.
\newblock \emph{\bibinfo{journal}{Phys. Rev. E}} \textbf{\bibinfo{volume}{75}},
  \bibinfo{pages}{041124} (\bibinfo{year}{2007}).
\newblock \urlprefix\url{https://link.aps.org/doi/10.1103/PhysRevE.75.041124}.

\bibitem{Chvosta/etal:2010b}
\bibinfo{author}{Chvosta, P.}, \bibinfo{author}{Einax, M.},
  \bibinfo{author}{Holubec, V.}, \bibinfo{author}{Ryabov, A.} \&
  \bibinfo{author}{Maass, P.}
\newblock \bibinfo{title}{Energetics and performance of a microscopic heat
  engine based on exact calculations of work and heat distributions}.
\newblock \emph{\bibinfo{journal}{J. Stat. Mech.}}
  \textbf{\bibinfo{volume}{2010}}, \bibinfo{pages}{P03002}
  (\bibinfo{year}{2010}).
\newblock \urlprefix\url{https://doi.org/10.1088/1742-5468/2010/03/p03002}.

\bibitem{Verley/etal:2013}
\bibinfo{author}{Verley, G.}, \bibinfo{author}{Van~den Broeck, C.} \&
  \bibinfo{author}{Esposito, M.}
\newblock \bibinfo{title}{Modulated two-level system: Exact work statistics}.
\newblock \emph{\bibinfo{journal}{Phys. Rev. E}} \textbf{\bibinfo{volume}{88}},
  \bibinfo{pages}{032137} (\bibinfo{year}{2013}).
\newblock \urlprefix\url{https://link.aps.org/doi/10.1103/PhysRevE.88.032137}.

\bibitem{Barato/Chetrite:2018}
\bibinfo{author}{Barato, A.~C.} \& \bibinfo{author}{Chetrite, R.}
\newblock \bibinfo{title}{Current fluctuations in periodically driven systems}.
\newblock \emph{\bibinfo{journal}{J. Stat. Mech.}}
  \textbf{\bibinfo{volume}{2018}}, \bibinfo{pages}{053207}
  (\bibinfo{year}{2018}).
\newblock \urlprefix\url{https://doi.org/10.1088/1742-5468/aabfc5}.

\bibitem{Mandaiya/Khaymovich:2019}
\bibinfo{author}{Mandaiya, A.} \& \bibinfo{author}{Khaymovich, I.~M.}
\newblock \bibinfo{title}{Time-reversal symmetric crooks and
  {G}allavotti-{C}ohen fluctuation relations in driven classical {M}arkovian
  systems}.
\newblock \emph{\bibinfo{journal}{J. Stat. Mech.}}
  \textbf{\bibinfo{volume}{2019}}, \bibinfo{pages}{054006}
  (\bibinfo{year}{2019}).
\newblock \urlprefix\url{https://doi.org/10.1088/1742-5468/ab11c1}.

\bibitem{Salazar:2020}
\bibinfo{author}{Salazar, D. S.~P.}
\newblock \bibinfo{title}{Work distribution in thermal processes}.
\newblock \emph{\bibinfo{journal}{Phys. Rev. E}}
  \textbf{\bibinfo{volume}{101}}, \bibinfo{pages}{030101}
  (\bibinfo{year}{2020}).
\newblock \urlprefix\url{https://link.aps.org/doi/10.1103/PhysRevE.101.030101}.

\bibitem{Chvosta/etal:2020}
\bibinfo{author}{Chvosta, P.}, \bibinfo{author}{Lips, D.},
  \bibinfo{author}{Holubec, V.}, \bibinfo{author}{Ryabov, A.} \&
  \bibinfo{author}{Maass, P.}
\newblock \bibinfo{title}{Statistics of work performed by optical tweezers with
  general time-variation of their stiffness}.
\newblock \emph{\bibinfo{journal}{J. Phys. A}} \textbf{\bibinfo{volume}{53}},
  \bibinfo{pages}{275001} (\bibinfo{year}{2020}).
\newblock \urlprefix\url{https://doi.org/10.1088/1751-8121/ab95c2}.

\bibitem{Slater:1960}
\bibinfo{author}{Slater, L.~J.}
\newblock \emph{\bibinfo{title}{Confluent Hypergeometric Functions}}
  (\bibinfo{publisher}{Cambridge University Press},
  \bibinfo{address}{Cambridge}, \bibinfo{year}{1960}).

\bibitem{Trepagnier/etal:2004}
\bibinfo{author}{Trepagnier, E.~H.} \emph{et~al.}
\newblock \bibinfo{title}{Experimental test of {Hatano and Sasa's}
  nonequilibrium steady-state equality}.
\newblock \emph{\bibinfo{journal}{Proc. Natl. Acad. Sci. U.S.A.}}
  \textbf{\bibinfo{volume}{101}}, \bibinfo{pages}{15038}
  (\bibinfo{year}{2004}).
\newblock \urlprefix\url{https://doi.org/10.1073/pnas.0406405101}.

\bibitem{Carberry/etal:2004}
\bibinfo{author}{Carberry, D.~M.} \emph{et~al.}
\newblock \bibinfo{title}{Fluctuations and irreversibility: An experimental
  demonstration of a second-law-like theorem using a colloidal particle held in
  an optical trap}.
\newblock \emph{\bibinfo{journal}{Phys. Rev. Lett.}}
  \textbf{\bibinfo{volume}{92}}, \bibinfo{pages}{140601}
  (\bibinfo{year}{2004}).
\newblock \urlprefix\url{https://doi.org/10.1103/PhysRevLett.92.140601}.

\bibitem{Carberry/etal:2007}
\bibinfo{author}{Carberry, D.~M.}, \bibinfo{author}{Baker, M. A.~B.},
  \bibinfo{author}{Wang, G.~M.}, \bibinfo{author}{Sevick, E.~M.} \&
  \bibinfo{author}{Evans, D.~J.}
\newblock \bibinfo{title}{An optical trap experiment to demonstrate fluctuation
  theorems in viscoelastic media}.
\newblock \emph{\bibinfo{journal}{J. Optics A}} \textbf{\bibinfo{volume}{9}},
  \bibinfo{pages}{S204} (\bibinfo{year}{2007}).
\newblock \urlprefix\url{http://stacks.iop.org/1464-4258/9/i=8/a=S13}.

\bibitem{Andrieux/etal:2007}
\bibinfo{author}{Andrieux, D.} \emph{et~al.}
\newblock \bibinfo{title}{Entropy production and time asymmetry in
  nonequilibrium fluctuations}.
\newblock \emph{\bibinfo{journal}{Phys. Rev. Lett.}}
  \textbf{\bibinfo{volume}{98}}, \bibinfo{pages}{150601}
  (\bibinfo{year}{2007}).
\newblock \urlprefix\url{https://doi.org/10.1103/PhysRevLett.98.150601}.

\bibitem{Khan/Sood:2011}
\bibinfo{author}{Khan, M.} \& \bibinfo{author}{Sood, A.~K.}
\newblock \bibinfo{title}{Irreversibility-to-reversibility crossover in
  transient response of an optically trapped particle}.
\newblock \emph{\bibinfo{journal}{Europhys. Lett. ({EPL})}}
  \textbf{\bibinfo{volume}{94}}, \bibinfo{pages}{60003} (\bibinfo{year}{2011}).
\newblock \urlprefix\url{http://stacks.iop.org/0295-5075/94/i=6/a=60003}.

\bibitem{Mestres/etal:2014}
\bibinfo{author}{Mestres, P.}, \bibinfo{author}{Martinez, I.~A.},
  \bibinfo{author}{Ortiz-Ambriz, A.}, \bibinfo{author}{Rica, R.~A.} \&
  \bibinfo{author}{Roldan, E.}
\newblock \bibinfo{title}{Realization of nonequilibrium thermodynamic processes
  using external colored noise}.
\newblock \emph{\bibinfo{journal}{Phys. Rev. E}} \textbf{\bibinfo{volume}{90}},
  \bibinfo{pages}{032116} (\bibinfo{year}{2014}).

\bibitem{vanZon/Cohen:2003}
\bibinfo{author}{van Zon, R.} \& \bibinfo{author}{Cohen, E. G.~D.}
\newblock \bibinfo{title}{Stationary and transient work-fluctuation theorems
  for a dragged {Brownian} particle}.
\newblock \emph{\bibinfo{journal}{Phys. Rev. E}} \textbf{\bibinfo{volume}{67}},
  \bibinfo{pages}{046102} (\bibinfo{year}{2003}).
\newblock \urlprefix\url{https://doi.org/10.1103/PhysRevE.67.046102}.

\bibitem{vanZon/Cohen:2004}
\bibinfo{author}{van Zon, R.} \& \bibinfo{author}{Cohen, E. G.~D.}
\newblock \bibinfo{title}{Extended heat-fluctuation theorems for a system with
  deterministic and stochastic forces}.
\newblock \emph{\bibinfo{journal}{Phys. Rev. E}} \textbf{\bibinfo{volume}{69}},
  \bibinfo{pages}{056121} (\bibinfo{year}{2004}).
\newblock \urlprefix\url{https://doi.org/10.1103/PhysRevE.69.056121}.

\bibitem{Cohen:2008}
\bibinfo{author}{Cohen, E. G.~D.}
\newblock \bibinfo{title}{Properties of nonequilibrium steady states: A path
  integral approach}.
\newblock \emph{\bibinfo{journal}{J. Stat. Mech.}}
  \textbf{\bibinfo{volume}{2008}}, \bibinfo{pages}{P07014}
  (\bibinfo{year}{2008}).
\newblock \urlprefix\url{https://doi.org/10.1088/1742-5468/2008/07/p07014}.

\bibitem{Nickelsen/Engel:2011}
\bibinfo{author}{Nickelsen, D.} \& \bibinfo{author}{Engel, A.}
\newblock \bibinfo{title}{Asymptotics of work distributions: The
  pre-exponential factor}.
\newblock \emph{\bibinfo{journal}{Eur. Phys. J. B}}
  \textbf{\bibinfo{volume}{82}}, \bibinfo{pages}{207} (\bibinfo{year}{2011}).
\newblock \urlprefix\url{https://doi.org/10.1140/epjb/e2011-20133-y}.

\bibitem{Subasi/Jarzynski:2013}
\bibinfo{author}{Suba\c{s}i, Y.} \& \bibinfo{author}{Jarzynski, C.}
\newblock \bibinfo{title}{Microcanonical work and fluctuation relations for an
  open system: An exactly solvable model}.
\newblock \emph{\bibinfo{journal}{Phys. Rev. E}} \textbf{\bibinfo{volume}{88}},
  \bibinfo{pages}{042136} (\bibinfo{year}{2013}).
\newblock \urlprefix\url{https://doi.org/10.1103/PhysRevE.88.042136}.

\bibitem{Kim/etal:2014}
\bibinfo{author}{Kim, K.}, \bibinfo{author}{Kwon, C.} \& \bibinfo{author}{Park,
  H.}
\newblock \bibinfo{title}{Heat fluctuations and initial ensembles}.
\newblock \emph{\bibinfo{journal}{Phys. Rev. E}} \textbf{\bibinfo{volume}{90}},
  \bibinfo{pages}{032117} (\bibinfo{year}{2014}).
\newblock \urlprefix\url{https://doi.org/10.1103/PhysRevE.90.032117}.

\bibitem{Kwon/etal:2011}
\bibinfo{author}{Kwon, C.}, \bibinfo{author}{Noh, J.~D.} \&
  \bibinfo{author}{Park, H.}
\newblock \bibinfo{title}{Nonequilibrium fluctuations for linear diffusion
  dynamics}.
\newblock \emph{\bibinfo{journal}{Phys. Rev. E}} \textbf{\bibinfo{volume}{83}},
  \bibinfo{pages}{061145} (\bibinfo{year}{2011}).
\newblock \urlprefix\url{https://doi.org/10.1103/PhysRevE.83.061145}.

\bibitem{Holubec/Ryabov:2017}
\bibinfo{author}{Holubec, V.} \& \bibinfo{author}{Ryabov, A.}
\newblock \bibinfo{title}{Work and power fluctuations in a critical heat
  engine}.
\newblock \emph{\bibinfo{journal}{Phys. Rev. E}} \textbf{\bibinfo{volume}{96}},
  \bibinfo{pages}{030102} (\bibinfo{year}{2017}).
\newblock \urlprefix\url{https://doi.org/10.1103/PhysRevE.96.030102}.

\bibitem{Hoppenau/Engel:2013}
\bibinfo{author}{Hoppenau, J.} \& \bibinfo{author}{Engel, A.}
\newblock \bibinfo{title}{On the work distribution in quasi-static processes}.
\newblock \emph{\bibinfo{journal}{J. Stat. Mech.}}
  \textbf{\bibinfo{volume}{2013}}, \bibinfo{pages}{P06004}
  (\bibinfo{year}{2013}).

\bibitem{Engel:2009}
\bibinfo{author}{Engel, A.}
\newblock \bibinfo{title}{Asymptotics of work distributions in nonequilibrium
  systems}.
\newblock \emph{\bibinfo{journal}{Phys. Rev. E}} \textbf{\bibinfo{volume}{80}},
  \bibinfo{pages}{021120} (\bibinfo{year}{2009}).

\bibitem{Noh/etal:PRL2013}
\bibinfo{author}{Noh, J.~D.}, \bibinfo{author}{Kwon, C.} \&
  \bibinfo{author}{Park, H.}
\newblock \bibinfo{title}{Multiple dynamic transitions in nonequilibrium work
  fluctuations}.
\newblock \emph{\bibinfo{journal}{Phys. Rev. Lett.}}
  \textbf{\bibinfo{volume}{111}}, \bibinfo{pages}{130601}
  (\bibinfo{year}{2013}).

\bibitem{Holubec/etal:2015}
\bibinfo{author}{Holubec, V.}, \bibinfo{author}{Lips, D.},
  \bibinfo{author}{Ryabov, A.}, \bibinfo{author}{Chvosta, P.} \&
  \bibinfo{author}{Maass, P.}
\newblock \bibinfo{title}{On asymptotic behavior of work distributions for
  driven {Brownian} motion}.
\newblock \emph{\bibinfo{journal}{Eur. Phys. J. B}}
  \textbf{\bibinfo{volume}{88}}, \bibinfo{pages}{340} (\bibinfo{year}{2015}).
\newblock \urlprefix\url{https://doi.org/10.1140/epjb/e2015-60635-x}.

\bibitem{Manikandan/Krishnamurthy:2017}
\bibinfo{author}{Manikandan, S.~K.} \& \bibinfo{author}{Krishnamurthy, S.}
\newblock \bibinfo{title}{Asymptotics of work distributions in a stochastically
  driven system}.
\newblock \emph{\bibinfo{journal}{Eur. Phys. J. B}}
  \textbf{\bibinfo{volume}{90}}, \bibinfo{pages}{258} (\bibinfo{year}{2017}).
\newblock \urlprefix\url{https://doi.org/10.1140/epjb/e2017-80432-9}.

\bibitem{Deza/etal:2009}
\bibinfo{author}{Deza, R.~R.}, \bibinfo{author}{Iz{\'u}s, G.~G.} \&
  \bibinfo{author}{Wio, H.~S.}
\newblock \bibinfo{title}{Fluctuation theorems from non-equilibrium
  {Onsager-Machlup} theory for a {Brownian} particle in a time-dependent
  harmonic potential}.
\newblock \emph{\bibinfo{journal}{Centr. Eur. J. Phys.}}
  \textbf{\bibinfo{volume}{7}}, \bibinfo{pages}{472} (\bibinfo{year}{2009}).

\bibitem{Campisi2016}
\bibinfo{author}{Campisi, M.} \& \bibinfo{author}{Fazio, R.}
\newblock \bibinfo{title}{The power of a critical heat engine}.
\newblock \emph{\bibinfo{journal}{Nat. Commun.}} \textbf{\bibinfo{volume}{7}},
  \bibinfo{pages}{11895} (\bibinfo{year}{2016}).
\newblock \urlprefix\url{https://doi.org/10.1038/ncomms11895}.

\bibitem{Koza1999}
\bibinfo{author}{Koza, Z.}
\newblock \bibinfo{title}{General technique of calculating the drift velocity
  and diffusion coefficient in arbitrary periodic systems}.
\newblock \emph{\bibinfo{journal}{J. Phys. A}} \textbf{\bibinfo{volume}{32}},
  \bibinfo{pages}{7637--7651} (\bibinfo{year}{1999}).
\newblock \urlprefix\url{https://doi.org/10.1088/0305-4470/32/44/303}.

\bibitem{Lips2018}
\bibinfo{author}{Lips, D.}, \bibinfo{author}{Ryabov, A.} \&
  \bibinfo{author}{Maass, P.}
\newblock \bibinfo{title}{Brownian asymmetric simple exclusion process}.
\newblock \emph{\bibinfo{journal}{Phys. Rev. Lett.}}
  \textbf{\bibinfo{volume}{121}}, \bibinfo{pages}{160601}
  (\bibinfo{year}{2018}).
\newblock
  \urlprefix\url{https://link.aps.org/doi/10.1103/PhysRevLett.121.160601}.

\bibitem{Krishnamurthy2016}
\bibinfo{author}{Krishnamurthy, S.}, \bibinfo{author}{Ghosh, S.},
  \bibinfo{author}{Chatterji, D.}, \bibinfo{author}{Ganapathy, R.} \&
  \bibinfo{author}{Sood, A.~K.}
\newblock \bibinfo{title}{A micrometre-sized heat engine operating between
  bacterial reservoirs}.
\newblock \emph{\bibinfo{journal}{Nat. Phys.}} \textbf{\bibinfo{volume}{12}},
  \bibinfo{pages}{1134--1138} (\bibinfo{year}{2016}).
\newblock \urlprefix\url{https://doi.org/10.1038/nphys3870}.

\bibitem{Pietzonka2019}
\bibinfo{author}{Pietzonka, P.}, \bibinfo{author}{Fodor, E.},
  \bibinfo{author}{Lohrmann, C.}, \bibinfo{author}{Cates, M.~E.} \&
  \bibinfo{author}{Seifert, U.}
\newblock \bibinfo{title}{Autonomous engines driven by active matter:
  Energetics and design principles}.
\newblock \emph{\bibinfo{journal}{Phys. Rev. X}} \textbf{\bibinfo{volume}{9}},
  \bibinfo{pages}{041032} (\bibinfo{year}{2019}).
\newblock \urlprefix\url{https://link.aps.org/doi/10.1103/PhysRevX.9.041032}.

\bibitem{Holubec2020a}
\bibinfo{author}{Holubec, V.} \& \bibinfo{author}{Marathe, R.}
\newblock \bibinfo{title}{Underdamped active {B}rownian heat engine}.
\newblock \emph{\bibinfo{journal}{Phys. Rev. E}}
  \textbf{\bibinfo{volume}{102}}, \bibinfo{pages}{060101}
  (\bibinfo{year}{2020}).
\newblock \urlprefix\url{https://link.aps.org/doi/10.1103/PhysRevE.102.060101}.

\bibitem{Ekeh2020}
\bibinfo{author}{Ekeh, T.}, \bibinfo{author}{Cates, M.~E.} \&
  \bibinfo{author}{Fodor, E.}
\newblock \bibinfo{title}{Thermodynamic cycles with active matter}.
\newblock \emph{\bibinfo{journal}{Phys. Rev. E}}
  \textbf{\bibinfo{volume}{102}}, \bibinfo{pages}{010101}
  (\bibinfo{year}{2020}).
\newblock \urlprefix\url{https://link.aps.org/doi/10.1103/PhysRevE.102.010101}.

\bibitem{Fodor_2021}
\bibinfo{author}{Fodor, {\'{E}}.} \& \bibinfo{author}{Cates, M.~E.}
\newblock \bibinfo{title}{Active engines: Thermodynamics moves forward}.
\newblock \emph{\bibinfo{journal}{{EPL} (Europhy. Lett.)}}
  \textbf{\bibinfo{volume}{134}}, \bibinfo{pages}{10003}
  (\bibinfo{year}{2021}).
\newblock \urlprefix\url{https://doi.org/10.1209/0295-5075/134/10003}.

\bibitem{Zakine/etal:Entropy2019}
\bibinfo{author}{Zakine, R.}, \bibinfo{author}{Solon, A.},
  \bibinfo{author}{Gingrich, T.} \& \bibinfo{author}{Van~Wijland, F.}
\newblock \bibinfo{title}{Stochastic {S}tirling engine operating in contact
  with active baths}.
\newblock \emph{\bibinfo{journal}{Entropy}} \textbf{\bibinfo{volume}{19}},
  \bibinfo{pages}{193} (\bibinfo{year}{2017}).
\newblock \urlprefix\url{https://www.mdpi.com/1099-4300/19/5/193}.

\bibitem{Kumari/etal:PRE2020}
\bibinfo{author}{Kumari, A.}, \bibinfo{author}{Pal, P.~S.},
  \bibinfo{author}{Saha, A.} \& \bibinfo{author}{Lahiri, S.}
\newblock \bibinfo{title}{Stochastic heat engine using an active particle}.
\newblock \emph{\bibinfo{journal}{Phys. Rev. E}}
  \textbf{\bibinfo{volume}{101}}, \bibinfo{pages}{032109}
  (\bibinfo{year}{2020}).
\newblock \urlprefix\url{https://link.aps.org/doi/10.1103/PhysRevE.101.032109}.

\bibitem{Agarwalla2018}
\bibinfo{author}{Agarwalla, B.~K.} \& \bibinfo{author}{Segal, D.}
\newblock \bibinfo{title}{Assessing the validity of the thermodynamic
  uncertainty relation in quantum systems}.
\newblock \emph{\bibinfo{journal}{Phys. Rev. B}} \textbf{\bibinfo{volume}{98}},
  \bibinfo{pages}{155438} (\bibinfo{year}{2018}).
\newblock \urlprefix\url{https://link.aps.org/doi/10.1103/PhysRevB.98.155438}.

\bibitem{Halpern2019}
\bibinfo{author}{Yunger~Halpern, N.}, \bibinfo{author}{White, C.~D.},
  \bibinfo{author}{Gopalakrishnan, S.} \& \bibinfo{author}{Refael, G.}
\newblock \bibinfo{title}{Quantum engine based on many-body localization}.
\newblock \emph{\bibinfo{journal}{Phys. Rev. B}} \textbf{\bibinfo{volume}{99}},
  \bibinfo{pages}{024203} (\bibinfo{year}{2019}).
\newblock \urlprefix\url{https://link.aps.org/doi/10.1103/PhysRevB.99.024203}.

\bibitem{Denzler2021}
\bibinfo{author}{Denzler, T.} \& \bibinfo{author}{Lutz, E.}
\newblock \bibinfo{title}{Power fluctuations in a finite-time quantum {C}arnot
  engine}.
\newblock \emph{\bibinfo{journal}{Phys. Rev. Research}}
  \textbf{\bibinfo{volume}{3}}, \bibinfo{pages}{L032041}
  (\bibinfo{year}{2021}).
\newblock
  \urlprefix\url{https://link.aps.org/doi/10.1103/PhysRevResearch.3.L032041}.

\bibitem{saryal2021universal}
\bibinfo{author}{Saryal, S.}, \bibinfo{author}{Gerry, M.},
  \bibinfo{author}{Khait, I.}, \bibinfo{author}{Segal, D.} \&
  \bibinfo{author}{Agarwalla, B.~K.}
\newblock \bibinfo{title}{Universal bounds on fluctuations in continuous
  thermal machines}.
\newblock \emph{\bibinfo{journal}{Phys. Rev. Lett.}}
  \textbf{\bibinfo{volume}{127}}, \bibinfo{pages}{190603}
  (\bibinfo{year}{2021}).
\newblock
  \urlprefix\url{https://link.aps.org/doi/10.1103/PhysRevLett.127.190603}.

\end{thebibliography}


\end{document}